\documentclass[12pt]{article}
\usepackage{amsfonts}
\usepackage[utf8]{inputenc}
\usepackage{bm}
\usepackage{amsmath}
\usepackage{epsfig}
\usepackage{enumitem}

\setlist[enumerate]{%
wide =0.5\parindent,
listparindent=0pt%
}

\newcommand{\bmat}{\left(\begin{array}}
\newcommand{\emat}{\end{array}\right)}

\def\gtrsim{\mathrel{\raise.3ex\hbox{$>$\kern-.75em\lower1ex\hbox{$\sim$}}}}

\def\ov{\overline}
\def\un{\underline}

\def\d{\delta}

\def\r{\rho}

\def\-{\hphantom{-}}
\def\ov{\overline}
\def\s2{\frac{1}{\sqrt2}}

\def\wt{\widetilde}

\def\Dsl{\,\raise.15ex\hbox{/}\mkern-13.5mu D} 

\def\be{\begin{equation}}
\def\ee{\end{equation}}
\def\bea{\begin{eqnarray}}
\def\eea{\end{eqnarray}}

\newcommand{\nn}{\nonumber}

\begin{document}


\pagestyle{plain}

\makeatletter
\@addtoreset{equation}{section}
\makeatother
\renewcommand{\theequation}{\thesection.\arabic{equation}}
\pagestyle{empty}
\begin{center}
\ \

\vskip .5cm

\LARGE{\LARGE\bf Lecture notes: Introduction to the Off-shell Double Copy Program \\[10mm]}
\vskip 0.3cm

\large{Eric Lescano 
 \\[6mm]}

{\small University of Wroclaw, Faculty of Physics and Astronomy, \\ \small\it  Maksa Borna 9, 50-204 Wroclaw,
Poland}

{\small \verb"eric.lescano@uwr.edu.pl"}\\[1cm]

\small{\bf Abstract} \\[0.5cm]\end{center}
 
The present notes are based on a series of lectures prepared for an introductory eight-class course on the modern framework of the off-shell double copy. The course was held from April 16 to May 4, 2026, at Universidad de Buenos Aires (UBA). These lectures, aimed at PhD and master’s students, are self-contained and require only a basic knowledge of classical field theory. The main goal is to review the fundamental concepts of gauge and gravitational theories in order to explore the off-shell frameworks of the single and double copy. In the final part of the course, we explore modern approaches to reinterpreting the single and double copy within T-duality-invariant frameworks.

\newpage

\setcounter{page}{1}
\pagestyle{plain}
\renewcommand{\thefootnote}{\arabic{footnote}}
\setcounter{footnote}{0}
\begin{center}
\includegraphics[width=1\textwidth]{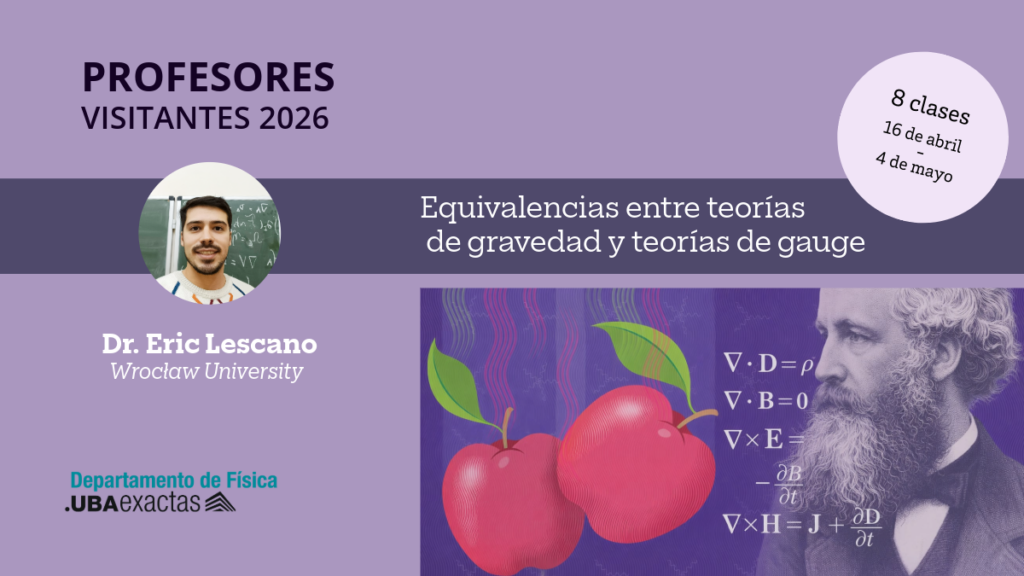}
\end{center}
\tableofcontents
\newpage

\section{Introduction}
The two modern pillars of theoretical physics are gravitational theories, organized by diffeomorphism invariance, and gauge theories, organized by gauge symmetries. Through these lectures we are going to review the basic ingredients to describe these theories: symmetries, tensors, invariants, actions and equations of motion. When we talk about gravitational theories, we typically refer to the Einstein-Hilbert action, but through the course the Weyl action, a higher-derivative model based on the square of certain curvatures, will also take a significative role. When we talk about gauge theories, there is also some ambiguity here: almost always we will refer to Yang-Mills, which can be formulated in its Abelian (Maxwell) or in its non-Abelian frameworks, but from time to time other higher-derivative formulation will emerge. As you can already notice, the higher-derivative contributions in theoretical physics are very important: these contributions can correct theories and modify their leading order behavior.  

The first Lecture of these notes is a quick review to gravity and gauge theories (we suggest references \cite{Part1-1}-\cite{Part1-6} for reviewing the main topics discussed in this part), and in Lecture 2 and Lecture 3 we will study two modern approaches which relate gravitational and gauge theories: the single and double copy \cite{Doublecopy1}-\cite{Doublecopy3}. These frameworks were originally constructed at the level of the amplitudes but, in this course, we will particularly focus on their off-shell formulations (Lagrangian level). This choice will allow us to explore general equivalences without the need of imposing the equations of motion (this is our definition of off-shell though the course). The idea of the single copy is to turn a gravitational theory, like the Einstein-Hilbert theory, into a gauge theory, by identifying the perturbations of the metric tensor with gauge fields. These technique is performed at the level of the equations of motion, allowing one to recast gravitational dynamics in terms of gauge dynamics. We will explore the case of Maxwell theory, obtained from a particular ansatz for the Einstein-Hilbert metric (Kerr-Schild ansatz and its generalization compatible with T-duality invariance). The single copy is extremely powerful since we can turn gravitational solutions into electromagnetic solutions and analyze their behaviors. More fundamentally, the idea of this program is to reverse this engineering and learn about quantum gravity, from the quantum gauge theory.

For this reason, the idea of the double copy is exactly the opposite to the single copy: to start from a gauge theory, let's say Yang-Mills, and to construct a gravitational theory by considering certain identifications. These identifications are done at the level of the action, so one can turn one a gauge action into a gravitational action, perturbatively (in these lectures we will work up to cubic order in fields). Particular focus will be put on the recent approach developed by O. Hohm, F. Jaramillo-Diaz and J. Plefka (HJP) in \cite{HJP}. The double copy procedure of these authors will allow us to construct the Einstein-Hilbert action plus extra matter terms, related to the Kalb-Ramond field and the dilaton. Therefore, by the end of these lectures we will be studying Double Field Theory \cite{DFT1}-\cite{DFT2}, and its relation to the off-shell double copy construction.

\subsection{Outline of the Course}

\begin{itemize}
\item {\bf Lecture 1: } The first lecture is a self-contained review of gravitational and gauge theories: part A discusses the main blocks to construct a principle of action using curvatures (like Einstein-Hilbert or Weyl gravity). In this first approach we keep the fields exact, without perturbing them. In part B we explore perturbation theory. We address general perturbations and the most common gauge fixing: harmonic gauge and TT-gauge. Here we briefly review the inclusion of matter in the formalism. In section C we introduce classical gauge theories following the structure of the Part A: symmetries, actions and equations of motion. In this part re review three different theories: Maxwell (Abelian theory), Yang-Mills (non-Abelian theory) and the so-called DFDF theory.

\item{\bf Lecture 2: }
The second lecture is entirely dedicated to the off-shell formulation of the single and double copy. In part A we explore the Kerr-Schild ansatz in GR, and we use it together with a particular identification of the null vectors in order to construct the Maxwell equations. In part B we explore perturbative Yang-Mills in momentum space, and we identify all the degrees of freedom with gravitational ones, to be able to construct a gravitational action from the Yang-Mills action (off-shell double copy). We will discuss the role of the extra momentum, and the double geometry behind this mechanism. In part C we follow the same strategy as in part B, but our starting point is a higher-derivative gauge theory, whose double copy is related to Weyl gravity.

\item{\bf Lecture 3:}
Finally, in the third lecture we discuss T-duality and its relation to the single and double copy. We introduce supergravity and its T-duality invariant rewriting, in the context of Double Field Theory, and we show how a particular treatment of the single and double copy is possible.

\end{itemize}

After each lecture exercises are included to maximize your learning. The solutions to the exercises are not provided in these notes, but we suggest to write an email in case you need some guidance in a particular resolution.

\newpage

\section{Lecture 1: Introduction to Gravitational and Gauge Theories}
\label{Sec1}

\subsection{Part A: Introduction to Gravity}
\label{A1}

\subsubsection{Differentiable Manifolds}
For this course, we will think of a manifold as a curved $n$-dimensional space. 
If we zoom in to any patch, the manifold looks like $\mathbb{R}^n$. 
Globally, the manifold may have interesting curvature (or topology) so we will later define objects to represent this curvature. 
An important point is that a manifold does not need a notion of metric, used to measure distance, but we will consider differentiable manifolds equipped with a metric.

More precisely, a (topological) manifold $M$ is a space which is locally homeomorphic to $\mathbb{R}^n$. 
This means that for every point $p \in M$, there exists a neighborhood $U \subset M$ and a one-to-one map
\[
\phi : U \to \mathbb{R}^n
\]
whose image is an open subset of $\mathbb{R}^n$. 
The pair $(U,\phi)$ is called a \emph{coordinate chart}, and the collection of such charts that covers $M$ is called an \emph{atlas}. 
The coordinates $x^\mu = \phi^\mu(p)$ are simply labels assigned to points in the patch $U$.

A differentiable manifold is one for which the transition functions between overlapping charts,
\[
x'^\mu = x'^\mu(x),
\]
are smooth ($C^\infty$). This is the minimal structure required to do calculus. 
The manifold itself is not a subset of $\mathbb{R}^n$; it is an abstract space that merely looks like $\mathbb{R}^n$ locally.

A central concept for us will be that of a diffeomorphism. 
A diffeomorphism is a smooth, invertible map
\[
\varphi : M \to M,
\]
with smooth inverse. There are two equivalent ways to interpret such a transformation.

\begin{itemize}
\item {\it Passive viewpoint:} we keep the point $p\in M$ fixed and change the coordinate system used to describe it,
\[
x^\mu \;\longrightarrow\; x'^\mu(x).
\]
This is simply a change of coordinates. Tensor components transform according to the usual Jacobian rules.

\item {\it Active viewpoint:} we keep the coordinate system fixed and move the points,
\[
p \;\longrightarrow\; \varphi(p).
\]
Fields are then pulled back (or pushed forward) accordingly. Infinitesimally, this induces variations of fields generated by a vector field $\xi^\mu(x)$.
\end{itemize}

In these notes we will mostly adopt the active viewpoint when discussing symmetries. 
An infinitesimal diffeomorphism is generated by a vector field $\xi^\mu(x)$ and acts as
\[
x^\mu \;\longrightarrow\; x^\mu + \xi^\mu(x).
\]
The induced variation of any field will be given by its Lie derivative along $\xi^\mu$. 
This provides a unified and operational definition of tensorial transformation laws.

\medskip

For us, the key idea is that coordinates are auxiliary. 
They are convenient labels, but they carry no intrinsic meaning. 
Any physical statement must be invariant under smooth coordinate changes, or equivalently under active diffeomorphisms. 
In gravitational physics this invariance is elevated to a fundamental gauge symmetry.

\medskip

Before introducing tensors systematically through the Lie derivative, we recall the notion of tangent space. 
At each point $p \in M$, we define the tangent space $T_p M$. 
Intuitively, this is the space of possible directions in which one can pass through $p$. 
More formally, a tangent vector at $p$ can be defined as a directional derivative acting on smooth functions $f : M \to \mathbb{R}$. The tangent space will be useful to describe gravitational theories using the vielbein formulation. We will describe this framework later.

With this structure in place, we are ready to define tensors operationally through their infinitesimal transformation under diffeomorphisms, namely through the Lie derivative. 
This will naturally lead us to the question of how derivatives of tensors transform, and to the introduction of the covariant derivative and affine connection as the appropriate tools to preserve tensorial character.

\subsubsection{Infinitesimal Diffeomorphisms and Closure}

We define the Lie derivative as follows,

\bea
L_{\xi} T_{\mu}{}^{\nu} = \xi^{\sigma} \frac{\partial T_{\mu}{}^{\nu}}{\partial x^{\sigma}} + \frac{\partial \xi^{\rho}}{\partial x^{\mu}} T_{\rho}{}^{\nu} - \frac{\partial \xi^{\nu}}{\partial x^{\rho}} T_{\mu}{}^{\rho} + \omega \frac{\partial \xi^{\sigma}}{\partial x^{\sigma}} T_{\mu}{}^{\nu} \, ,\label{infdiftensor}
\eea
where $T_{\mu}{}^{\nu}$ is a (1-1) tensor. For us, this means that indices down transform in a covariant way while indices up transform in a contravariant way. This is our definition for a contravariant transformation (or our definition for a 1-form field, $w_{\mu}$), 
\bea
L_{\xi} w_{\mu} = \xi^{\sigma} \partial_{\sigma} w_{\mu} + \partial_{\mu} \xi^{\rho} w_{\rho} + \omega \partial_{\sigma} \xi^{\sigma} w_{\mu} \, ,\label{1form}
\eea
and, similarly, our definition for a vector/contravariant field is
\bea
L_{\xi} v^{\mu} = \xi^{\sigma} \partial_{\sigma} v^{\mu} - \partial_{\sigma} \xi^{\mu} v^{\sigma} + \omega \partial_{\sigma} \xi^{\sigma} v^{\mu} \, ,\label{vector}
\eea

Of course, general $(p,q)$ tensors can be defined just by including more rotations with respect to each index. Notice that the first term and the last term are always the same, independently of the rank of the tensor. The first term is called "the transport" term, while the last one assigns a weigh $\omega$ to each tensor.  

The Lie derivative satisfies closure
\bea
\Big[\delta_{\xi_1},\delta_{\xi_2} \Big] T_{\mu}{}^{\nu} = \delta_{\xi_{3}} T_{\mu}{}^{\nu}
\eea
through the Lie bracket, defined as follows
\bea
\xi^{\mu}_{3}(x) = \xi^{\rho}_{2} \partial_{\rho} \xi^{\mu}_{1} - (1 \leftrightarrow 2)  \, .
\eea
The closure of a symmetry is a fundamental property that reflects a group structure behind the symmetry (in this case $Diff(M)$).

The demonstration for the generic weight scalar $\omega$ is as follows,
\bea
[\delta_{\xi_1}, \delta_{\xi_2}]\phi & = & \delta_{\xi_1}(\xi_2^{\rho} \partial_{\rho} \phi + \omega \partial_{\rho} \xi_{2}^{\rho}) - (1 \leftrightarrow 2) \nn \\
& = & \xi_{2}^{\rho} \partial_{\rho}(\xi_{1}^{\sigma} \partial_{\sigma} \phi + \omega \partial_{\sigma} \xi_{1}^{\sigma}) - (1 \leftrightarrow 2) \, .
\eea
When we compute closures, neither the parameters nor the constants are transformed. Now we apply Leibniz's rule and we get
\bea
[\delta_{\xi_1}, \delta_{\xi_2}]\phi & = & \xi_{2}^{\rho} \partial_{\rho} \xi_{1}^{\sigma} \partial_{\sigma} \phi + \xi_{2}^{\rho} \xi_{1}^{\sigma} \partial_{\rho} \partial_{\sigma} \phi + w \xi_{2}^{\rho} \partial_{\rho} \partial_{\sigma} \xi_{1}^{\sigma} - (1 \leftrightarrow 2) \, .
\eea
The second term of the RHS cancels out because it is antisymmetric in 1 and 2, and symmetric in $\rho$ and $\sigma$, and the last term can be rewritten as follows:
\bea
[\delta_{\xi_1,\xi_2}]\phi & = & \xi_{2}^{\rho} \partial_{\rho} \xi_{1}^{\sigma} \partial_{\sigma} \phi + w \partial_{\sigma} (\xi_{2}^{\rho} \partial_{\rho} \xi_{1}^{\sigma}) - (1 \leftrightarrow 2)
\eea
which is what we wanted to find (Lie bracket).

Using the notation
\bea
V_{(\mu \nu)} & = & \frac12 V_{\mu \nu} + \frac12 V_{\nu \mu} \, , \\
V_{[\mu \nu]} & = & \frac12 V_{\mu \nu} - \frac12 V_{\nu \mu} \, ,
\eea 
the Lie bracket can be written as $\xi^{\sigma}_{3}=2\xi^{\rho}_{[2} \partial_{\rho} \xi_{1]}^{\sigma}$. 
 
We already know how tensors transform. However, a valid question is what happens to their derivatives. Does the derivative of a scalar field transform into a (0.1) covariant tensor? Does the derivative of a vector transform into a (1,1) rank tensor? In other words, we are asking whether the derivative preserves the tensorial character of the fields.

In the case of the derivative of a scalar field, it does preserve the tensor character. Let's see: consider $\phi$ a scalar field of zero weight, and apply the transformation to the derivative of this field,
\bea
\delta_{\xi} (\partial_{\mu} \phi) = \partial_{\mu} (\delta_{\xi} \phi) = \partial_{\mu} (\xi^{\nu} \partial_{\nu} \phi) = \partial_{\mu} \xi^{\nu} \partial_{\nu} \phi + \xi^{\nu} \partial_{\mu} \partial_{\nu} \phi \, . 
\eea
The second term in the last equality is a transport term, while the first rotates the index, as happens in the covariant transformation. Therefore, it is true that the derivative of a scalar field transforms covariantly.

Let's now see what happens for a vector. We anticipate that for vectors or 1-forms, the tensor character is lost under differentiation. We will show this for the weightless vector $v^{\mu}$.
\bea
\delta_{\xi} (\partial_{\nu} v^{\mu}) = \partial_{\nu} (\xi^{\rho} \partial_{\rho} v^{\mu} - \partial_{\rho} \xi^{\mu} v^{\rho}  ) = \partial_{\nu} \xi^{\rho} \partial_{\rho} v^{\mu} + \xi^{\rho} \partial_{\nu} \partial_{\rho} v^{\mu} - \partial_{\nu} \partial_{\rho} \xi^{\mu} v^{\rho} - \partial_{\rho} \xi^{\mu} \partial_{\nu} v^{\rho} \, . 
\eea
We see, then, that the structure is not covariant, since there is an extra term that has a double derivative acting on the parameter (the third term). Therefore, the derivative of a vector does not transform covariantly. The same applies to a 1-form or tensors of higher rank.

How do we solve this (major) problem? We create a new derivative, the covariant derivative, which will preserve the tensor nature of the tensor we are going to differentiate.

The covariant derivative is defined as follows: given the vector $v^{\mu}$, we define the following derivative:
\bea
\nabla_{\nu} v^{\mu} = \partial_{\nu} v^{\mu} + \Gamma^{\mu}_{\nu \rho} v^{\rho} \, ,
\label{der_cov}
\eea
where $\Gamma^{\mu}_{\nu \rho}$ is a new field. The first thing we need to notice is that $\Gamma^{\mu}_{\nu \rho}$ is, for the moment, an arbitrary quantity that does not transform like a tensor. How do we know this? Because if we analyze the right-hand side of (\ref{der_cov}), the first term transforms non-covariantly, while the left-hand side transforms covariantly. Therefore, the field $\Gamma^{\mu}_{\nu \rho}$ must transform non-covariantly to compensate for this. Thus, the transformation of $\Gamma^{\mu}_{\nu \rho}$ contains a covariant part (as if it were a tensor with respect to its three indices) and a non-covariant part that we can compute explicitly. Let's define the failure of the operator
\bea
\Delta_{\xi} = \delta_{\xi} - {L}_{\xi} 
\eea
which measures only the non-covariant part of each object. Let's apply it to (\ref{der_cov}),
\bea
\Delta_{\xi} (\nabla_{\nu} v^{\mu}) = \Delta_{\xi}(\partial_{\nu} v^{\mu}) + \Delta_{\xi}(\Gamma^{\mu}_{\nu \rho} v^{\rho}) \, .
\eea
The left-hand side is zero, since we decree that the covariant derivative of a tensor transforms as a tensor. The right-hand side has two contributions, one given by the derivative of the vector and the other given by the transformation of the connection.
\bea
0 = \Delta_{\xi}(\partial_{\nu} v^{\mu}) + (\Delta_{\xi} \Gamma^{\mu}_{\nu \rho}) v^{\rho} \, .
\eea
We already calculated the failure of the usual derivative, and it gave us
\bea
0 = - (\partial_{\nu} \partial_{\rho} \xi^{\mu}) v^{\rho} + (\Delta_{\xi} \Gamma^{\mu}_{\nu \rho}) v^{\rho} \, .
\eea
Therefore, we can rule out the connection failure,
\bea
\Delta_{\xi} \Gamma^{\mu}_{\nu \rho} = \partial_{\nu} \partial_{\rho} \xi^{\mu} \, .
\eea
In other words, the complete transformation of the connection is given by
\bea
\Delta_{\xi} \Gamma^{\mu}_{\nu \rho} = L_{\xi}  \Gamma^{\mu}_{\nu \rho} + \partial_{\nu} \partial_{\rho} \xi^{\mu} \, ,
\eea
and due to the non-covariant term, the connection ensures the covariant transformation of the covariant derivative. For now, the connection is a generic field that transforms according to the previous equation. The covariant derivative of a 1-form is defined as follows,
\bea
\nabla_{\nu} w_{\mu} = \partial_{\nu} w_{\mu} - \Gamma^{\rho}_{\nu \mu} w_{\rho} \, \, .
\eea

\subsubsection{Construction of Curvatures}

As we mentioned, a manifold by itself has no notion of distance. To introduce this, we equip it with a metric $g_{\mu\nu}(x)$, a symmetric $(0,2)$ tensor field that assigns to each point a bilinear form on the tangent space. The inverse metric is given by $g^{\mu\nu}(x)$ and satisfies
\bea
g_{\mu \rho} g^{\rho \nu} = \delta_{\mu}^{\nu} \, .
\eea
In general relativity we use a Lorentzian metric, and through these notes we use the signature $(- + + +)$.

The metric defines the invariant line element
\[
ds^2 = g_{\mu\nu}(x)\, dx^\mu dx^\nu,
\]
and allows us to raise and lower indices, define lengths of vectors, and measure angles. Conceptually the differentiable structure precedes the metric structure. 
The Lie derivative, diffeomorphism invariance, and the construction of covariant derivatives can all be formulated independently of any metric. 

However, we will use the metric to fix the affine connection. First, from the covariant derivative of the metric,
\bea
\nabla_{\mu} g_{\nu \rho} = Q_{\mu \nu \rho} \, ,
\eea
we can demand $Q_{\mu \nu \rho}=0$. This tensor, defined by the covariant derivative of the metric, is often called non-metricity. Demanding $Q_{\mu \nu \rho} = 0$ we obtain 
\bea
\partial_{\mu} g_{\nu \rho} - 2 \Gamma^{\sigma}_{\mu (\nu} g_{\rho) \sigma} = 0 \, ,
\eea
and therefore, the most general solution to solve the previous condition is
\bea
\Gamma_{\mu \nu}^{\rho} = \frac12 g^{\rho \sigma} (\partial_{\mu} g_{\nu \sigma} + \partial_{\nu} g_{\mu \sigma} -\partial_{\sigma} g_{\mu \nu}) + \frac12 (T_{\mu \nu}^{\rho} - T_{\mu}{}^{\rho}{}_{\nu} - T_{\nu}{}^{\rho}{}_{\mu} ) \, ,
\eea
where $T_{\mu \nu}^{\rho}=-T_{\nu \mu}^{\rho}$ is called the torsion tensor. 

In GR, we eliminate both the non-metricity and the torsion, leaving the so-called Levi-Civita connection,
\bea
\Gamma_{\mu \nu}^{\rho} = \frac12 g^{\rho \sigma} (\partial_{\mu} g_{\nu \sigma} + \partial_{\nu} g_{\mu \sigma} -\partial_{\sigma} g_{\mu \nu}) \, .
\eea

Since we have fixed our connection, the construction of the curvatures follows like this. First, we compute the commutator of two covariant derivatives,
\bea
\big[\nabla_{\mu},\nabla_{\nu}\big] v^{\rho} & = & R^{\rho}{}_{\sigma \mu \nu} v^{\sigma} \, , \\
\big[\nabla_{\mu},\nabla_{\nu}\big] w_{\rho} & = & - R^{\sigma}{}_{\rho \mu \nu} w_{\sigma} \, .
\eea
From the previous expressions we obtain
\bea
R^{\rho}{}_{\sigma \mu \nu} = 2 \partial_{[\mu} \Gamma^{\rho}_{\nu] \sigma} + 2 \Gamma^{\rho}_{[\mu| \tau} \Gamma_{|\nu] \sigma}^{\tau} \, , 
\eea
which is the definition of the Riemann tensor, in terms of the metric tensor, which emerges directly from our commutator. Since the Riemann tensor transforms covariantly (it is a (1,3) tensor), then we can construct the following covariant (0,2) tensor
\bea
R_{\sigma \nu} = R^{\rho}{}_{\sigma \mu \nu} \delta^{\mu}_{\rho} \, .
\eea
This object is symmetric, and it is known as the Ricci tensor. It will be a very important object, since it describes the dynamics in GR (see next section).

The last curvature that we can define making use of the Ricci tensor and the inverse metric is
\bea
R_{\sigma \nu} g^{\sigma \nu} = R \, ,
\eea
which is a scalar field, known as the scalar curvature or the Ricci scalar. 

Summarizing, we already have defined several tensors: 
\begin{itemize}
    \item $g_{\mu \nu}$ : metric tensor
    \item $Q_{\mu \nu \rho}$: non-metricity (set to zero throughout these notes.)
    \item $T_{\mu \nu}^{\rho}$: torsion tensor (set to zero throughout these notes).
    \item $R^{\rho}{}_{\sigma \mu \nu}$: Riemann tensor (from this one we can obtain the Ricci tensor and the Ricci scalar).
\end{itemize}

{\bf Index symmetries and Bianchi identities for the Riemann tensor}
The index symmetries of the Riemann tensor ($R_{\mu \nu \rho \sigma}=R^{\alpha}{}_{\nu \rho \sigma} g_{\alpha \mu}$) are the following:
\bea
R_{\mu \nu \rho \sigma} & = & - R_{\nu \mu \rho \sigma} \\
R_{\mu \nu \sigma \rho} & = & - R_{\mu \nu \rho \sigma} \\
R_{\mu \nu \rho \sigma} & = & - R_{\rho \sigma \mu \nu} \, .
\eea

When we work with the curvature of a field, instead of the field, we need to remember that there are identities which helps us to eliminate redundancies. For the case of the Riemann tensor, we have an algebraic identity and a differential identity:
\bea
R^{\rho}{}_{[\sigma \mu \nu]} & = & 0 \\
\nabla_{[\lambda} R_{\mu \nu]}{}^{\rho}{}_{\sigma} & = & 0.
\eea
It's part of the problems of this part of the lecture to prove the previous identities.

Due to the previous identities, the number of components of the Riemann tensor is $\frac{D^2(D^2-1)}{12}$. So the Riemann has 20 independent components in $D=4$.
\subsubsection{Gravitational Action Principles}
For constructing a gravitational action principle, we first introduce the following scalar density
\bea
\delta_{\xi} \sqrt{-g} = \xi^{\rho} \partial_{\rho} \sqrt{-g} + \partial^{\rho} \xi_{\rho} \sqrt{-g} \, ,
\eea
where g is the determinant of the metric. This transformation can be easily obtained from the identity $\delta g = g g^{\mu \nu} \delta g_{\mu \nu} \, .$ Using the technology of the previous section we can safely say that the square root of (minus) the determinant of the metric transforms like a scalar with weight $w=1$. This kind of transformation is known as a density, and it is ideal to construct a generic principle of action following this structure,
\bea
S = \int d^4x \sqrt{-g} \, T \, .
\eea
where $T$ is an arbitrary scalar. If the action has the previous form, then
\bea
\delta_{\xi} S = S + \int \partial_{\mu}(\sqrt{-g} \, T \, \xi^{\mu}) \, .
\eea
The last term of the previous variation is a total derivative, so the action is invariant under infinitesimal diffeomorphisms and the Lagrangian
\bea
L = \sqrt{-g} \, T
\eea
is invariant up to total derivatives. Probably, you are thinking that $T=R$ is our ideal candidate for completing our action principle, but $R$ is just one possibility among several. When we use $T=R$, we are formally constructing the Einstein-Hilbert action,
\bea
S_{E-H} = \int d^4x \sqrt{-g} \, R \, .
\eea
But this is not the only possibility that GR offers. For example, we can define
\bea
S_{Riem^2} = \int d^4x \sqrt{-g} R^{\rho}{}_{\sigma \mu \nu} R^{\alpha}{}_{\beta \gamma \delta} g_{\rho \alpha} g^{\sigma \beta} g^{\mu \gamma} g^{\nu \delta}  \, .
\eea
It is pretty obvious that this one is much more complicated than the Einstein-Hilbert action, but in principle it is well-defined if we just focus on the symmetry.

There is a particular action which will be very important for us: Weyl gravity. The Weyl tensor is an important tensor in GR, which can be constructed in the following way,
\bea
\label{Weyl_tensor}
C_{\mu\nu\rho\lambda} & = & R_{\mu\nu\rho\lambda} - \frac{2}{D-2}\left(g_{\mu[\rho}R_{\lambda]\nu} - g_{\nu[\rho}R_{\lambda]\mu}\right)\, \nn \\
& & + \frac{2}{\left(D-1\right)\left(D-2\right)}Rg_{\mu[\rho}g_{\lambda]\nu}.
\eea
where D is the dimension of the space-time. In D=3. the Riemann tensor is fully determined by the Ricci tensor,
\bea
R_{\mu \nu \rho \sigma}|_{D=3} = g_{\mu \rho} R_{\nu \sigma} - g_{\nu \rho} R_{\mu \sigma} - \frac{R}{2} g_{\mu \rho} g_{\nu \sigma} - (\rho \leftrightarrow \sigma) \, ,
\eea
so we can easily see the Weyl tensor vanishes in this particular dimension. The Weyl tensor is the part of the curvature that is not locally determined by matter, so $C_{\mu \nu \rho \sigma}=0$ means that there are no tidal forces not directly sourced by matter (no propagating gravitational degrees of freedom). Therefore, in $D=3$ there is no notion of gravitational waves, since gravity cannot propagate "in between" matter sources.

In $D=4$ the Weyl tensor is given by
\bea
C_{\mu\nu\rho\lambda} & = & R_{\mu\nu\rho\lambda} - \left(g_{\mu[\rho}R_{\lambda]\nu} - g_{\nu[\rho}R_{\lambda]\mu}\right)\, + \frac{1}{3}Rg_{\mu[\rho}g_{\lambda]\nu}
\eea

An interesting example of a solution to GR with vanishing Weyl tensor is FLRW cosmology (the standard model of cosmology to the date). In this scenario, the FLRW metric satisfies $C_{\mu \nu \rho \sigma}=0$ so all curvature comes from the energy density $\rho$ and pressure $p$, and there are no gravitational waves in the background.

The Weyl gravity action is therefore constructed as
\bea
S_{Weyl} = \int d^4x \sqrt{-g} C_{\mu \nu \rho \sigma} C^{\mu \nu \rho \sigma} \, ,
\eea
and it is a conformal invariant model of gravity (to prove it is left as one of the exercises of this Lecture).

Finally, we mention that the number of independent components of the Weyl tensor is $\frac{D(D+1)(D+2)(D-3)}{12}$ (For $D>3$) so in $D=4$ it has $10$ independent components.

\subsubsection{Vielbein Formalism}

We now consider flat vectors $v^a$ defined on the tangent space on every point of the manifold, where the indices $a,b=0,\dots,D-1$ are known as ``flat indices''. Flat vectors transform under Lorentz transformations according to
\bea
\delta_{\Lambda} v^{a} = v^{b} \Lambda_{b}{}^{a}
\eea
where all the contractions are made with a constant flat (inverse) metric $\eta^{ab}$. The Lorentz parameter satisfies $\Lambda_{ab}= - \Lambda_{ba}$. Since we are abandoning the metric formulation, we need to consider a new fundamental field $e_{\mu}{}^{a}$, the vielbein, whose inverse is given by $e^{\mu}{}_{a}$. These objects satisfy,
\bea
e_{\mu}{}^{a} \eta_{ab} e_{\nu}{}^{b} & = & g_{\mu \nu} \, , \\
e^{\mu}{}_{a} \eta^{ab} e^{\nu}{}_{b} & = & g^{\mu \nu} \, ,
\eea
and they transform covariantly under infinitesimal diffeomorphisms and Lorentz transformations,
\bea
\delta_{\xi,\Lambda} e_{\mu}{}^{a} & = & \xi^{\nu} \partial_{\nu} e_{\mu}{}^{a} + \partial_{\mu} \xi^{\nu} e_{\nu}{}^{a} + e_{\mu}{}^{b} \Lambda_{b}{}^{a} \, , \\ 
\delta_{\xi,\Lambda} e^{\mu}{}_{a} & = &  \xi^{\nu}\partial_{\nu}  e^{\mu}{}_{a}  - \partial_{\nu}\xi^{\mu} e^{\nu}{}_{a} + e^{\mu}{}_{b} \Lambda^{b}{}_{a} \, .
\label{inverse0}
\eea
The previous transformations close, and when we have more that one symmetry, we need to check also the mixed closure. The full closure in this case is left as an exercise. 

Analogously to what happens with infinitesimal diffeomorphisms, the transformation of the partial derivative of a flat vector does not match with the transformation of a flat tensor. Considering a generic flat vector $v^a$, we define a flat covariant derivative as
\bea
\nabla_{\mu} v^{a} = \partial_{\mu} v^{a} - w_{\mu}{}^{a}{}_{b} v^{b} \, ,
\eea
where $w_{\mu ab}$ is the spin connection. Imposing
\bea
\nabla_{\mu} e_{\nu}{}^{a} = \partial_{\mu} e_{\nu}{}^{a} - \Gamma_{\mu \nu}^{\rho} e_{\rho}{}^{a} - w_{\mu}{}^{a}{}_{b} e_{\nu}{}^{b} = 0 \, ,
\label{compae}
\eea
we can fully determine the spin connection in terms of the vielbein, $w_{\mu b c} = w_{\mu b c}(e)$. The 2-form version of the Riemann tensor can be written in terms of the vielbein as
\bea
R_{\mu \nu a b} = - 2 \partial_{[\mu} w_{\nu] a b} + 2 w_{[\mu| a}{}^{c} w_{|\nu] c b} \, .
\eea

\subsubsection{Killing Vectors}

Before giving our first example, the Schwarzschild metric, we introduce an important concept: Killing vectors $\xi^\mu$. A Killing vector is defined by
\bea
{\cal L}_{\xi} g_{\mu \nu} = 0 \, ,
\eea
which means that the metric is invariant under the flow generated by $\xi^\mu$. In other words, Killing vectors generate isometries of spacetime.

In General Relativity, defining conserved quantities is subtle due to diffeomorphism invariance. However, in spacetimes admitting Killing vectors, one can associate conserved quantities to these symmetries.

For example, if a spacetime admits a timelike Killing vector field (which in adapted coordinates can be written as $\partial_t$), the metric is independent of time and the spacetime is said to be stationary:
\bea
{\cal L}_{\xi} g_{\mu \nu} = 0 \quad \Longrightarrow \quad \partial_t g_{\mu\nu} = 0 \, .
\eea
In this case, for a particle moving along a geodesic with four-velocity $u^\mu$, the quantity
\bea
E = - \xi_\mu u^\mu
\eea
is conserved along its trajectory, and is interpreted as the energy of the particle.

\subsection{Part B: Perturbation Theory}
\label{B1}
\subsubsection{General Perturbations}
When we talk about perturbations, we refer to a field that it is being expanded around a particular configuration, which is called the background. In gravity, the field that we perturb is the metric, which in general can be perturbed as
\bea
g_{\mu \nu} = \tilde g_{\mu \nu} + h_{\mu \nu} \, ,
\eea
where $\tilde g_{\mu \nu}$ is the background and $h_{\mu \nu}$ is the perturbation. Sometimes we assume that the background is constant, or even flat $\tilde g_{\mu \nu} = \eta_{\mu \nu}$. Let's focus on this last case. The inverse metric needs to satisfy $g^{\mu \nu} g_{\nu \rho} = \delta^{\mu}_{\rho}$ so when we expand around Minkowski with arbitrary perturbations the inverse metric contains an infinite expansion given by
\bea
g^{\mu \nu} = \eta^{\mu \nu} - h^{\mu \nu} + h^{\mu}{}_{\sigma} h^{\sigma \nu} - h^{\mu}{}_{\epsilon} h^{\epsilon}{}_{\sigma} h^{\sigma \nu} + \dots
\eea
where the indices of the previous expression are contracted using the background metric, $h^{\nu}{}_{\sigma} h^{\sigma \rho} = h^{\nu \alpha}\eta_{\alpha \sigma} h^{\sigma \rho}$. If we consider that the perturbation is small ($|h_{\mu \nu}| << 1$ and $|h^{\mu \nu}| << 1$) then one can work at the linearized level. We will show some results for the action and equations of motion in the next section.

\subsubsection{Gauge Fixing and Gravitational Waves}
Let's focus on the Einstein-Hilbert Lagrangian 
\bea
S_{EH} = \int d^4x \sqrt{-g} \, R \, ,
\eea
and  equations of motion
\bea
- \frac12 g_{\mu \nu} R + R_{\mu \nu}  = 0 \, .
\eea
If we consider small perturbations around Minkowski, then the linearized connection is given by
\bea
\Gamma_{\mu \nu}^{(1)\rho} = \frac12 \eta^{\rho \sigma} (\partial_{\mu} h_{\nu \rho} + \partial_{\nu} h_{\mu \rho} - \partial_{\sigma}h_{\mu \nu}) \, ,
\eea
and then the linearized Riemann tensor is given by
\bea
R^{(1)\rho}{}_{\sigma \mu \nu} = \partial_{\mu} \Gamma_{\nu \sigma}^{\rho} - \partial_{\nu} \Gamma_{\mu \sigma}^{\rho} \, .
\eea
In terms of the perturbation we have,
\bea
R^{(1)\rho}{}_{\sigma \mu \nu} & = & \eta^{\rho \alpha} (\partial_{[\mu|} \partial_{\sigma} h_{|\nu] \alpha} - \partial_{[\mu|} \partial_{\alpha} h_{|\nu] \sigma}) \, , \\
R^{(1)}_{\sigma \nu} & = & \frac12 (\partial^{\alpha} \partial_{\sigma} h_{\nu \alpha} + \partial^{\alpha} \partial_{\nu}h_{\sigma \alpha}- \Box h_{\nu \sigma} - \partial_{\nu} \partial_{\sigma} h)  \, , \\
R^{(1)} & = & \partial^{\mu} \partial^{\nu} h_{\mu \nu} - \Box h \, ,
\eea
where $h=h_{\mu}{}^{\mu}$. Now we will obtain the linear expansion for the square root of the determinant. For doing so we use
\bea
g=\textrm{det}(\eta_{\mu \rho} \delta^{\rho}_{\nu} + \eta_{\mu \rho} h^{\rho}{}_{\nu}) = - \textrm{det}(\delta^{\rho}_{\nu} + h^{\rho}{}_{\nu})
\eea
where in the last step we used that the determinant of the Minkowski metric is $-1$. Now we need to recall the identity
\bea
det (1+H) = e^{Tr \ln(1+H)} = -g
\eea
Using $\ln(1+H)=H-\frac12H^2+\frac13 H^3+\dots$ w can define
\bea
Tr\ln(1+H)=Tr H-\frac12 Tr(H^2) + \frac13 Tr(H^3) \, .
\eea
Also recalling $e^{X}=1+X+\frac12 X^2 + \frac16 X^3 + \dots$
we obtain 
\bea
det (1+H) & = & 1 + Tr(H) + \frac12(Tr H)^2 - \frac12 Tr(H^2) + \frac16 (Tr H)^3 + \frac13 Tr(H^3) - \frac12 Tr(H) Tr(H^2) \nn \\ & = & 1 + X = -g \, . 
\eea
Finally, using $\sqrt{1+X} = 1 + \frac12 X - \frac18 X^2 + \frac{1}{16} X^3$ we obtain
\bea
\sqrt{-g} = 1 + \frac12 h + \frac18 h^2 - \frac14 h_{\mu \nu} h^{\mu \nu}   \, . 
\eea
In the last replacement we used $Tr(H)=h$.
So far we have not imposed any gauge fixing, so this is the moment. We start by analyzing which is the symmetry that we will demand on the linearized theory. The symmetry is linearized diffeomorphisms, given by
\bea
\delta_{\xi} h_{\mu \nu} = \partial_{\mu} \xi_{\nu} + \partial_{\nu} \xi_{\mu} \, . 
\label{lindiffeo}
\eea

Now we impose the harmonic gauge (also called De Donder gauge),
\bea
\partial^{\mu} h_{\mu \nu} - \frac12 \partial_{\nu} h = 0 \, ,
\eea
which means that we are using the four components of $\xi$ in (\ref{lindiffeo}) to simplify the degrees of freedom of the metric (originally 10 components, since it is a symmetry tensor). Moreover, if we transform the previous expression we find,
\bea
\delta_{\xi}(\partial^{\mu} h_{\mu \nu} - \frac12 \partial_{\nu} h) = \Box \xi_{\nu} \, ,
\eea
meaning that once the harmonic gauge is imposed, it is preserved by diffeomorphisms that shift the coordinates by vector fields $\xi$ such that $\Box \xi=0$.

Under this gauge the linearized action is just a total derivative, so the leading contribution comes from
\bea
S_{E-H}= - \int d^4x  (R^{(2)}_{\nu \sigma} \eta^{\sigma \nu} - R^{(1)}_{\nu \sigma} h^{\sigma \nu} + \frac12 h R^{(1)}_{\nu \sigma} \eta^{\sigma \nu}) \, ,
\eea
while the equations of motion are given by
\bea
\Box(h_{\mu \nu} - \frac12 \eta_{\mu \nu} h) = 0 \, .
\eea
We define $\bar h_{\mu \nu}=h_{\mu \nu} - \frac12 \eta_{\mu \nu} h$, so the equations turn into the wave equation, with plane wave solution
\bea
\bar h_{\mu \nu} = A_{\mu \nu} e^{i k_{\sigma} x^{\sigma}}
\eea
and $k^2=0$ (waves travel at the speed of light). Typically when we inspect gravitational waves, we use a remanent gauge symmetry to solve the Harmonic gauge (this is called TT-gauge) in the following way,
\bea
\partial^{\mu} h_{\mu \nu} = \partial_{\nu}h=0 \, ,
\eea
and therefore the equations of motion are just $\Box h_{\mu \nu} = 0$. While for the moment we were choosing the diffeomorphism transformations generated by $\xi_{\mu}$ with $\Box \xi_{\mu}=0$, we can also choose
\bea
\xi_{\mu} = C_{\mu} e^{i k_{\sigma} x^{\sigma}}
\eea
to further reduce other 4 degrees of freedom from the metric. Under the TT-gauge (and z-propagation), the polarization takes the form, 
\bea
A_{\mu \nu} = \left(\begin{matrix} 
0 & 0 & 0 & 0 \\
0 & h_{+} & h_{x} & 0 \\
0 & h_{x} & -h_{+} & 0 \\
0 & 0 & 0 & 0
\end{matrix}\right) \ .
\label{generalizedmetric}
\eea
Under the TT-gauge the polarization is traceless $A_{\mu}{}^{\mu} = 0$. Also it satisfies $A_{o \nu}=0$ and
$k^{\mu} A_{\mu \nu}=0$ (the latter is the Fourier transformation of the transverse condition). The remanent $h_{+}$ and $h_{x}$ are the two physical polarizations of gravity.

\subsubsection{Inclusion of Matter}
Let's now go back a little. We already know that we can construct different gravitational theories by defining a Lagrangian, which is a scalar under diffeomorphisms. But so far, these theories do not contain matter. In some sense, we have an empty universe and now it's time to fill it. For doing this, we have two possibilities:
\begin{itemize}
    \item Fields in a curved background: We can couple a matter fields to any of our gravitational fields, which means that we can couple a matter Lagrangian together with our gravitational Lagrangian. If we focus in the E-H Lagrangian, the results is something like
\bea
S_{E-H+matter} = \frac{1}{2\kappa}\int d^4x \sqrt{-g} \Big(R + 2 Lm(\phi)\Big)
\eea    
where now we are explicitly including the coefficient $\kappa= \frac{8 \pi G}{c^4}$ which was omitted so far. Here, $Lm(\phi)$ is a Lagrangian for the matter, and $\phi$ represent an arbitrary matter field. For example, we can couple a massless scalar field,
\bea
Lm(\phi) = -\frac{1}{2} \partial_{\mu} \phi \partial^{\mu} \phi - V(\phi) \ ,
\eea
or more complicated setups, but always respecting that $Lm$ transforms as a scalar.

Let's focus on the massless scalar field, to understand how to compute the dynamics. In principle our fundamental fields our the metric and the scalar field, so now we will have two equations of motion: one coming from the variations with respect to the metric,
\bea
\frac{1}{2\kappa} \sqrt{-g}(R_{\mu \nu} - \frac12 R g_{\mu \nu}) + \frac{ \delta S_m}{\delta g^{\mu \nu}} = 0
\eea
and other one coming from the variation of the action with respect to the scalar field
\bea
\frac{\delta Lm}{\delta \phi} = 0 \, .
\eea
The LHS of the previous equation gives
\bea
\delta_{\phi} Lm = -\partial_{\mu}(\delta \phi) \partial_{\nu} \phi g^{\mu \nu} - \frac{\delta V}{\delta \phi} \, 
\eea
We can promote the derivatives to covariant derivatives, and integrate by parts to obtain
\bea
\delta_{\phi} Lm = \delta \phi \nabla_{\mu} \nabla_{\nu} \phi g^{\mu \nu} - \frac{\delta V}{\delta \phi} = 0\, . 
\eea
In the case $V=0$, we obtain the Klein-Gordon equation $\Box \phi = 0$.

Let's move to the metric equation. We define 
\bea
T_{\mu \nu} & = & -\frac{2}{\sqrt{-g}} \frac{\delta S_{m}}{\delta g^{\mu \nu}}
\eea
so that the full equation of motion with respect to the metric tensor is given by
\bea
R_{\mu \nu} - \frac12 R g_{\mu \nu} = G_{\mu \nu} = \kappa T_{\mu \nu} \, .
\eea
Since we are analyzing the case of the massless scalar field, the energy momentum-tensor is given by
\bea
T_{\mu \nu}= \frac{1}{\kappa} (g_{\mu \nu} L_{m} + 2 \partial_{\mu} \phi \partial_{\nu} \phi + 2 \frac{\delta V}{\delta g^{\mu \nu}}) \, .
\eea
For computing the previous variation we have used $\delta \sqrt{-g}=-\frac12 \sqrt{-g} \delta g^{\mu \nu} g_{\mu \nu}$.

\item Statistical matter on a curved background: The other alternative is to couple statistical matter, meaning particles with a certain dynamics on the manifold. In this sense, one can explore how include a single particle in the curved geometry, and then generalized this idea for simple configurations, like fluid dynamics. This second option is based on relativistic hydrodynamics and, interestingly, in some simple cases, we will be able to formally develop a correspondence between statistical matter dynamics and the field theory description.  
\end{itemize}
For describing statistical matter, the correct starting point is the phase space. For each point $x$ on the manifold, we have introduced its tangent space $\mathbb{P}_{x}$. In this moment, we will define vectors of momenta, $p^\mu$, on this space. In consequence the phase space is a collection $(x,\mathbb{P}_x)$ (this is known as a tangent bundle). The momentum can be considered independent of the position and thus
\bea
\frac{\partial p^{\nu}}{\partial x^{\mu}} = 0 \, .
\label{IndepGR}
\eea
This condition holds in an off-shell formulation of the relativistic kinetic theory which is the scenario that we deal with. Conversely the on-shell condition $p^\mu p_\mu=m^2$ spoils the independence.

The infinitesimal diffeomorphisms of a phase space scalar $v$ with constant weight  $\omega$ can be written as \footnote{Here we follow the conventions of \cite{LescanoMiron}.}
\bea
\delta_{\xi} v = L_{\xi} v + p^{\rho} \frac{\partial \xi^{\sigma}(x)}{\partial x^{\rho}} \frac{\partial v}{\partial p^{\sigma}} \, ,
\label{ediffeos}
\eea
where $L_{\xi}$ is the usual Lie derivative with $\xi^{\mu}=\xi^{\mu}(x)$ (no p dependence).  We may extend (\ref{ediffeos}) to tensors by taking the usual Lie derivative acting on different tensor structures, {\it e.g}. for a (1,1) tensor we have
\bea
L_{\xi} v_{\mu}{}^{\nu} = \xi^{\sigma} \frac{\partial v_{\mu}{}^{\nu}}{\partial x^{\sigma}} + \frac{\partial \xi^{\rho}}{\partial x^{\mu}} v_{\rho}{}^{\nu} - \frac{\partial \xi^{\nu}}{\partial x^{\rho}} v_{\mu}{}^{\rho} + \omega \frac{\partial \xi^{\sigma}}{\partial x^{\sigma}} v_{\mu}{}^{\nu} \, .\label{infdiftensor}
\eea

It is straightforward to check the closure of the transformation (\ref{infdiftensor}),
\bea
\Big[\delta_{\xi_1},\delta_{\xi_2} \Big] v_{\mu}{}^{\nu} = \delta_{\xi_{21}} v_{\mu}{}^{\nu}
\eea
and show that the bracket is given by the Lie bracket, 
\bea
\xi^{\mu}_{12}(x) = \xi^{\rho}_{1} \frac{\partial \xi^{\mu}_{2}}{\partial x^{\rho}} - (1 \leftrightarrow 2)  \, .
\eea
In other words, both the phase space formulation and the space-time formulation share the same bracket.

Since we have tensors acting on the phase space we need to define a natural extension of the covariant derivative in the phase space, namely the Liouville operator $D_{\mu}$. Regarding that we have taken the collection $(x^\mu,p^\mu)$ to be the basis of the phase space, the Liouville operator for an arbitrary tensor reads
\bea
D_{\mu} A^{\rho\lambda}(x,p) = \nabla_{\mu} A^{\rho\lambda}(x,p) -  \Gamma^{\sigma}_{\mu \nu} p^{\nu}\frac{\partial A^{\rho\lambda}(x,p)}{\partial p^{\sigma}} \, ,\label{LiouvilleoperatorGR}
\eea
where $\nabla_\mu$ is the well-known covariant derivative. In particular it satisfies
\bea
D_{\mu} p^{\nu} = D_{\mu} p_{\nu} = 0 \, .
\eea

Finally the diffeomorphism invariant volume element of the phase space is the product of the coordinate and momentum invariant volume elements, namely
\bea
\sqrt g d^{d}{p}\,\sqrt g d^dx=g d^{d}{p} d^dx \, ,
\label{GRvolume}
\eea
with $g$ the determinant of the metric tensor.

The relativistic Boltzmann equation rules the evolution of the one-particle distribution function (1pdf) $f=f[x,p]$, which is a phase space scalar. In its simplest form this equation is 
\bea
p^{\mu} D_{\mu} f = C[f] \, .\label{boltzmanneq}
\eea
The RHS of (\ref{boltzmanneq}) is the collision term which takes into account the non-gravitational interactions between particles. If an equilibrium state is achieved the 1pdf takes its equilibrium form $f=f_{\rm eq}$ and $C[f_{\rm eq}]=0$.

The integration of the first and second moment of the distribution function (with respect to momentum)  give rise to covariant quantities in space-time. For example, 
\bea
\int p^{\mu} f[x,p] \sqrt{g} d^dp = N^{\mu}(x)
\eea
with the usual conservation law or transfer function for the particle current $N^{\mu}$,
\be
\nabla_{\mu}N^{\mu} = 0 \, .
\label{c1}
\ee
If we instead take 
\bea
\int p^{\mu} p^{\nu} f[x,p] \sqrt{g} d^dp = T^{\mu \nu}(x) \, ,
\eea
we can define the energy-momentum tensor, which satisfies the conservation law
\bea
\nabla_{\mu}T^{\mu \nu} = 0 \, 
.
\label{c2}
\eea

Since the energy-momentum tensor and the Einstein tensor have the same properties, we can safely take
\bea
\frac{1}{\kappa} G_{\mu \nu} = \int p_{\mu} p_{\nu} f[x,p] \sqrt{g} d^dp
\eea
which is the Einstein equation, but obtained from the phase space construction. This means that the matter coupled to the manifold does not come from a Lagrangian formulation, but it comes from a statistical distribution. 

The simplest distribution function to consider the generalization of the Maxwell-Boltzmann to curve space (also known as the Maxwell-Juttner distribution), 
\bea
f(x,p) \propto e^{-\beta_{\mu} p^{\mu}} 
\eea
where $\beta_{\mu} = \frac{u_{\mu}}{k_B T}$. The energy-momentum tensor in this case is
\bea
T_{\mu \nu} = (e+p) u_{\mu} u_{\nu} + p g_{\mu \nu} \, ,
\eea
which is known as the energy-momentum tensor of the perfect fluid.

{\bf The perfect fluid-scalar field correspondence} \\
Now we will show that there exist a formal correspondence between coupling a scalar field on a manifold, and coupling particles obeying the Maxwell-Juttner statistics. In the case of a scalar field the energy-momentum tensor is
\bea
T^{\mu \nu}_{\Phi} = \left[\partial^{\mu} \Phi \partial^{\nu} \Phi - \frac12 g^{\mu \nu} \partial^{\rho} \Phi \partial_{\rho} \Phi - V g^{\mu \nu} \right]\, ,
\label{scalar}
\eea

On the other hand, the energy-momentum tensor for an effective perfect fluid reads
\bea
T^{\mu \nu}_{\rm PF} = (e+p) u^{\mu} u^{\nu} + p g^{\mu \nu} \, ,
\label{fluid}
\eea

In order to deduce the scalar-fluid correspondence  we first define the following identification for the velocity,
\bea
u_{\mu} = \frac{\partial_{\mu} \Phi}{\sqrt{|\partial^{\rho} \Phi \partial_{\rho} \Phi|}} \, ,
\eea
with $\partial^{\rho} \Phi \partial_{\rho} \Phi \neq 0$ and $u_{\mu} u^{\mu}=\textrm{sign} (\partial^{\rho} \Phi \partial_{\rho} \Phi)$. Since we use the positive signature, when $\partial^{\rho} \Phi \partial_{\rho} \Phi<0$ the velocity $u_\mu$ defines a time-like vector and the energy density and the pressure of the effective perfect fluid are related to the scalar field through
\bea
e & = & - \frac12 \partial^{\rho} \Phi \partial_{\rho} \Phi + V\, , \\
p & = & - \frac12 \partial^{\rho} \Phi \partial_{\rho} \Phi - V\, .\label{psugra}
\eea

Finally, using the correspondence we are able to find the full action corresponding to a perfect fluid in terms of the scalar field variables, 
\bea
\frac{1}{2 \kappa} \int d^4x \sqrt{-g} (R + 2p(\phi)) \, .
\eea

\subsection{Part C: Introduction to Classical Gauge Theories}
\label{C1}
In the last part of this lecture we will move to gauge theories, where the fundamental degree of freedom is a gauge field $A_{\mu i}$, with $i$ a gauge index (running from 1...N, with N the dimension of the gauge group). During this part, we will forget about the gravitational theory, and we just describe the gauge theory on a curve space. This means that the equation of motion with respect to the metric tensor will give us the effective energy-momentum tensor that the theory is producing.

\subsubsection{Warming up: Maxwell}
Maxwell theory is the simplest case of a gauge theory, because the theory is Abelian ($U(1)$) and we can describe it just with the field $A_{\mu}$. This means that we do not need the extra index since $dim(U(n))=n^2$.

\paragraph{Symmetries and Closure}
The symmetry rules acting on $A_{\mu}$ are
\bea
\delta_{\xi,\lambda} A_{\mu} = L_{\xi} A_{\mu} + \partial_{\mu} \lambda \, .
\eea
When we compute the closure for the A-field we obtain the Lie bracket $\xi_{3}$ plus a gauge parameter
\bea
\lambda_{3} = 2 \xi^{\rho}_{[2} \partial_{\rho} \lambda_{1]} \, . 
\eea
Since the gauge transformation of the $A_{\mu}$ field is given by the derivative of a parameter, then this field transforms in a non-covariant way with respect to gauge transformations. Therefore, we need to construct its curvature, $F_{\mu \nu}$,
\bea
F_{\mu \nu} = \partial_{\mu} A_{\nu} - \partial_{\nu} A_{\mu} \, .
\eea
It is very easy to prove that this curvature is a tensor with respect to both symmetries:
\begin{itemize}
    \item  For the diffeomorphism symmetry we can just promote $\partial_{\mu}\rightarrow \nabla_{\mu}$ since we are not considering torsion. Therefore the curvature is a tensor with respect to diffeomorphisms. 
\item For gauge, we can simple transform an obtain $\delta_{\lambda} F_{\mu \nu} = \partial_{\mu} \partial_{\nu} \lambda - \partial_{\nu} \partial_{\mu} \lambda = 0$. 
\end{itemize}

Just to give you some intuition for the curvature of the A-field, the usual parametrization is given by
\bea
F_{\mu \nu} = \left(\begin{matrix} 
0 & \frac{E_x}{c} & \frac{E_y}{c} & \frac{E_z}{c} \\
-\frac{E_x}{c} & 0 & B_{z} & -B_y \\
-\frac{E_y}{c} & -B_z & 0 & B_x \\
-\frac{E_z}{c} & B_y & -B_x & 0
\end{matrix}\right) \ .
\label{generalizedmetric}
\eea
Let's now introduce the Bianchi identity for $F_{\mu \nu}$. As happened with the Riemann tensor, the F-curvature obeys its own identity, given by
\bea
\nabla_{[\mu} F_{\nu \rho]} = \partial_{[\mu} F_{\nu \rho]} = 0 \, .
\eea
In the first equality we used the fact that there is no torsion on the space-time.

\paragraph{Action and Equations of Motion}
The Maxwell action is given by
\bea
S_{Max} = - \frac14 \int d^4x \sqrt{-g} F_{\mu \nu} F^{\mu \nu} \, .
\eea
The equation of motion with respect to the A-field is given by
\bea
\delta_{A} S_{Max} = - \frac12 \int d^4x \sqrt{-g} \delta_{A} F_{\mu \nu} F^{\mu \nu} \\
= - \frac12 \int \sqrt{-g}  (\nabla_{\mu} \delta A_{\nu}) F^{\mu \nu}
\eea
and using integration by parts we obtain
\bea
\nabla_{\mu} F^{\mu \nu} = 0 \, .
\eea

If we think that the Maxwell theory is coupled as a matter Lagrangian, then the equation of motion with respect to the metric is proportional to the energy-momentum tensor,
\bea
\delta_{g} S_{max} = \frac18 \int d^4x \sqrt{-g} \delta g^{\mu \alpha} g_{\mu \alpha} F_{\nu \beta} F^{\nu \beta}
- \frac12 \int d^4x \sqrt{-g} \delta g^{\mu \alpha} F_{\mu \nu} F_{\alpha \beta} g^{\nu \beta} \, . 
\eea
Therefore 
\bea
T_{\mu \nu} = -2 (F_{\mu}{}^{\beta} F_{\nu \beta} 
- \frac14 g_{\mu \nu} F_{\alpha \beta} F^{\alpha \beta}) \, .
\eea
The previous tensor has no trace, reflecting its scale invariance in $D=4$. Before moving to the non-Abelian case (Yang-Mills), we will briefly discuss a particular case inside Maxwell: the self-dual scenario. Let's define
\bea
\tilde F_{\mu \nu} = \frac{i}{2} \epsilon_{\mu \nu}{}^{\rho \sigma} F_{\rho \sigma} \, ,
\eea
with $\epsilon_{\mu \nu \rho \sigma}$ a Levi-Civita symbol (fully antisymmetric). When $F=\tilde F$ (self-dual Maxwell), the Lagrangian becomes
\bea
-\frac{1}{4} \tilde F_{\mu \nu} F^{\mu \nu} = \nabla_{\mu}K^{\mu} \, .
\eea
This Lagrangian is known as an Abelian Pontryaguin and it is a topological term. Particularly, since it is given by a total derivative, do not affect to the gravitational dynamics.
The explicit form of $K^{\mu}$ can be obtained by explicitly writing
\bea
-\frac14 \tilde F_{\mu \nu} F^{\mu \nu} = - \frac{i}{8} \epsilon^{\mu \nu \rho \sigma} \nabla_{\rho} A_{\sigma} \nabla_{\mu} A_{\nu} \, .
\eea
By integrating by parts this last expression one has,
\bea
K^{\rho} = -\frac{i}{8}  \epsilon^{\rho \mu \nu \sigma} A_{\mu} \partial_{\nu} A_{\sigma} \, .
\eea
So far this is just a curiosity of the self-dual Maxwell theory, but when one promotes the Abelian group to a non-Abelian one, the full Chern-Simons 3 form emerges from $K^{\rho}= -\frac{i}{8}\epsilon^{\rho \mu \nu \sigma} C_{\mu \nu \sigma}$. 

\subsubsection{Yang-Mills}

\paragraph{Symmetries and Closure}
The non-Abelian gauge transformations act on $A_{\mu i}$ in the following way,
\bea
\delta_{\lambda}A_\mu^i & = & \partial_\mu \lambda^i+f^i{}_{jk}\lambda^jA_\mu^k\, , \label{gauge0}
\eea
where $\lambda^{i}$ is an arbitrary parameter. In non-Abelian gauge theory, the transformation
\bea
\delta_{\lambda}v^i & = & f^i{}_{jk}\lambda^j v^k\,  \label{gaugevector}
\eea
defines a gauge vector. This means that the $A_{\mu}{}^{i}$ field transforms in a non-covariant way with respect to non-Abelian transformations, and that's fine. This field transforms as a gauge connection, so we will use it to construct the gauged version of the covariant derivative (for Maxwell theory was not necessary, but now we need it).

Considering an arbitrary gauge vector $v^{i}$, the partial derivative of this vector is not covariant and we need to extend the notion of covariant derivative in order to include the gauge symmetry,
\bea
\nabla_{\mu} v^{i} = \partial_{\mu} v^{i} - f^{i}{}_{jk} A_{\mu}{}^{j} v^{k} \, .
\eea 
Here we use the same notation $\nabla$ for the gauge covariant derivative, so our convention is that $\nabla$ covariantizes the derivative of an object with respect to all the symmetries that the object transforms under (similarly to what we did for the vielbein formalism). Let's see a quick example. Let's consider a quick example: We will explore the covariant derivative of the curvature $F_{\mu \nu}{}^{i}$. This field is now defined as,
\bea
F_{\mu \nu}{}^{i} = 2 \partial_{[\mu} A_{\nu]}{}^{i} - f^{i}{}_{jk} A_{\mu}{}^{j} A_{\nu}{}^{k} \, ,
\eea
and, once again, this curvature transforms covariantly under both diffeomorphisms and non-Abelian gauge transformations
\bea
\delta_{\lambda} F_{\mu \nu i} = f_{ijk} \lambda^{j} F_{\mu \nu}{}^{k} \, ,
\eea
this last statement requires the use of the Jacobi identities,
\bea
f_{il}{}^{m} f_{jk}{}^{l} + f_{jl}{}^{m} f_{ki}{}^{l} + f_{kl}{}^{m} f_{ij}{}^{l} = 0 \, . 
\eea
Now that we have the definition, let's explore the covariant derivative:
\bea
\nabla_{\mu} F_{\nu \rho}{}^{i} = \partial_{\mu} F_{\nu \rho} - 2 \Gamma_{\mu [\nu|}{}^{\sigma} F_{\sigma |\rho]} - f^{i}{}_{j k} A_{\mu}{}^{j} F_{\nu \rho}{}^{k} \, . 
\eea
In the previous expression we clearly observe that the last term is non-covariant with respect to non-Abelian gauge transformations, so it compensates the non-covariance of $\partial_{\mu} F_{\nu \rho}$ under this symmetry. 

Let's now move to the closure of the transformations. Let's consider the full transformation of the A-field,
\bea
\delta_{\xi,\lambda}A_\mu^i & = & \xi^{\rho} \partial_{
\rho
} A_{\mu}{}^{i} + \partial_{\mu} \xi^{\rho} A_{\rho i} + \partial_\mu \lambda^i+f^i{}_{jk}\lambda^jA_\mu^k\, . 
\eea
The brackets for this field now include those from Maxwell plus a gauge bracket proportional to the structure constants given by
\bea
\lambda_{3}{}^{i} = f^{i}{}_{j k} \lambda_{2}^{j} \lambda_{1}^{k} \, . 
\eea
The Bianchi identity for the curvature of the A-field is given by
\bea
\nabla_{[\mu} F_{\nu \rho]}{}^{i} = 0 \, .
\eea
This is the non-Abelian generalization of the Gauss law and the Faraday law, which are encoded in the Bianchi identity for the Maxwell case. Now we have all the ingredients to construct the action and deduce the equations of motion.

\paragraph{Action and Equations of Motion}
The action of Yang-Mills is very similar to the Maxwell action, but now we need to take a trace over the gauge indices,
\bea
S_{YM} = - \frac14 Tr \int d^4x F_{\mu \nu} F^{\mu \nu} = - \frac14 \int d^4x F_{\mu \nu}{}^{i} F^{\mu \nu j} \kappa_{i j} \, , 
\eea
where $k_{i j}$ is a Cartan-Killing metric ($\kappa_{i j} \kappa^{j k}=\delta_{i}^{k}$).
The computation of the equation of motion with respect to the gauge-field is similar to the Maxwell case, but now we have to include the extra term in the definition of the curvature.

\bea
\delta_{A} S_{YM} & = &  - \frac12 \int d^4x \delta_{A} F_{\mu \nu}{}^{i} F^{\mu \nu j} \kappa_{i j} \, , \nn \\
& = & - \int d^4x \delta_{A} (\partial_{\mu} A_{\nu}{}^{i} - f^{i}{}_{k l} A_{\mu}{}^{k} A_{\nu}^{l}) F^{\mu \nu j} \kappa_{i j} \, , \nn \\
& = & - \int d^4x  (\partial_{\mu}  \delta A_{\nu}{}^{i} - 2 f^{i}{}_{k l} \delta A_{\mu}{}^{k} A_{\nu}^{l}) F^{\mu \nu j} \kappa_{i j} \, .
\eea
Therefore the equation of motion is 
\bea
\nabla_{\mu}F^{\mu \nu i} = 0
\eea
where now $\nabla_{\mu}$ contains gauge contributions.

\paragraph{Self-dual Yang-Mills}
Let's explore the self-dual case. Similarly to what happened in Maxwell, now we can obtain a non-commutative Pontryaguin from the action,
\bea
S= -\frac14 Tr \int d^4x \sqrt{-g} F_{\mu \nu} \tilde F^{\mu \nu} = \int d^{4}x \sqrt{-g} \nabla_{\mu} K^{\mu} \, ,
\eea
with 
\bea
K^{\mu} = -\frac{i}{8}\epsilon^{\mu \nu \rho \sigma} (A_{\nu}^{i} \partial_{\rho} A_{\sigma i} - \frac13 A_{\nu}^{i} A_{\rho}^{j} A_{\sigma}^{k} f_{i j k}) \, .
\eea
This is the Chern-Simons current, constructed from the Chern-Simons 3-form
\bea
C_{\nu \rho \sigma} = A_{[\nu}^{i} \partial_{\rho} A_{\sigma] i} - \frac13 A_{\nu}^{i} A_{\rho}^{j} A_{\sigma}^{k} f_{i j k}
\eea
contracted with the epsilon pseudotensor.

Let's try to simplify the prescription of the self-dual case for flat space. We adopt the following light-cone coordinates,
\begin{equation}
    u=\frac{1}{\sqrt{2}}(t-z), \quad 
    v=\frac{1}{\sqrt{2}}(t+z), \quad
    w=\frac{1}{\sqrt{2}}(x+iy), \quad
    \bar{w}=\frac{1}{\sqrt{2}}(x-iy),
\end{equation}
such that the Minkowski metric takes the form\footnote{We recall that we work with mostly-plus metric signature conventions.}
\begin{equation}
    ds^2 = -2(dudv-dwd\bar{w}).
\end{equation}
In these coordinates, the self-duality condition reduces to three independent equations:
\begin{equation}\label{eq:SDLC}
    F_{uw}{}^{i} = 0, \qquad F_{uv}{}^{i} = F_{w\bar{w}}{}^{i}, \qquad F_{v\bar{w}}{}^{i} = 0.
\end{equation}
From now on we will work in the light-cone gauge,
\begin{equation}
    A_u{}^{i} = 0. 
\end{equation}
The first constraint in eq.~\eqref{eq:SDLC} reduces to 
\begin{align}\label{eq:eq1}
    F_{uw}{}^{i} = 0 \quad &\implies \quad \partial_uA_w{}^{i} = 0,
\end{align}
which is trivially solved by setting $A_w{}^{i} = 0$. With this solution, the second constraint in eq.~\eqref{eq:SDLC} yields
\begin{equation}\label{eq:eq2}
    F_{uv}{}^{i} = F_{w\bar{w}}{}^{i} \implies \partial_uA_v{}^{i} = \partial_w A_{\bar{w}}{}^{i}.
\end{equation}
This corresponds to an integrability condition and can be solved by setting
\begin{equation}\label{eq:ICsols}
    A_v{}^{i} = \frac12 \partial_{w}\Psi{}^{i}, \qquad A_{\bar{w}}{}^{i} = \frac12 \partial_{u}\Psi{}^{i},
\end{equation}
where $\Psi^i$ is a gauged scalar field. We therefore find that a self-dual gauge field in light-cone gauge can be written in terms of a single charged scalar,
\begin{equation}\label{eq:SDgaugefield}
    A_{\mu} = \frac12\left(0, \partial_{w}\Psi, 0, \partial_{u}\Psi\right),
\end{equation}
and the final equation in eq.~\eqref{eq:SDLC} provides a constraint for the dynamics of this field,
\begin{equation}\label{eq:LCeom}
    F_{v\bar{w}} = 0 \quad\implies\quad \Box \Psi^{k} - f_{i j k} \partial_u\Psi^{i} \partial_w \Psi^{j} = 0.
\end{equation}
This is the equation of motion for self-dual Yang-Mills (SDYM) in light-cone gauge. 

Let us try to gain some intuition about what these equations describe. First, by constructing polarization vectors in the light-cone coordinates used above, one can show that the self-dual gauge field of eq.~\eqref{eq:SDgaugefield} corresponds to a positive helicity state. Second, by contracting the self-duality condition  with a covariant derivative we find
\begin{equation}\label{eq:YMeqs}
    D^{\mu}F_{\mu\nu} = \frac{i}{2}\epsilon_{\mu\nu\rho\sigma}D^{\mu}F^{\rho\sigma} = 0,
\end{equation}
which vanishes as a consequence of the Bianchi identity. Thus, self-dual gauge fields are automatically solutions to the full Yang-Mills equations. The SDYM equations therefore describe the positive helicity sector of Yang-Mills. The negative helicity sector is captured by the anti-self-dual Yang-Mills equations, which corresponds to the introduction of a negative sign in the self-duality condition.

{\bf Self-dual Gravity Revisited}

The story for self-dual gravity (SDG) in light-cone gauge is broadly analogous to SDYM, albeit more computationally intensive. Here the self-duality condition is 
\begin{equation}\label{eq:SDconditionR}
    R_{\mu\nu\rho\sigma} =  \frac{i}{2}\epsilon_{\mu\nu\gamma\delta}R^{\gamma\delta}_{\;\;\;\rho\sigma},
\end{equation}
where $R_{\mu\nu\rho\sigma}$ is the Riemann tensor. Note that the Riemann tensor is not a 2-form and so this condition at first seems somewhat perplexing. However, it can be easily shown that this constraint descends from the self-duality condition on the curvature 2-form $R_{\mu \nu a b}$. Just like in SDYM, solutions to eq.~\eqref{eq:SDconditionR} are automatically solutions to the vacuum Einstein equations, which can be seen by contracting two of the indices,
\begin{equation}
    R_{\mu\rho} = \frac{i}{2}\epsilon_{\mu}^{\;\;\nu\gamma\delta}R_{\gamma\delta\rho\nu} = 0,
\end{equation}
where the RHS vanishes as a consequence of the Bianchi identity. We now write the metric in the following form
\begin{equation}
    g_{\mu\nu} = \eta_{\mu\nu} + h_{\mu\nu},
\end{equation}
and impose the light-cone gauge via $h_{u\mu} = 0$. We will skip the computation details, but by taking this ansatz in the self-duality condition of eq.~\eqref{eq:SDconditionR}, one finds that the metric can be written in terms of a single scalar field $\phi$,
\begin{equation}
    ds^2 = -2(dudv-dwd\bar{w}) + \kappa(\partial_w^2\phi \,dv^2+ \partial_u^2\phi \,d\bar{w}^2 + 2\partial_u\partial_w\phi \,dvd\bar{w}),
\end{equation}
which satisfies the equation of motion
\begin{equation}\label{eq:LCeomSDG}
    \Box \phi - \kappa\{\partial_u\phi,\partial_w\phi\} = 0,
\end{equation}
where we have introduced the Poisson bracket
\begin{equation}
    \{f,g\} = \partial_w f\partial_u g - \partial_u f\partial_w g.
\end{equation}
Before moving to the last topic of this lecture let us compare eqs.~(\ref{eq:LCeom}) and (\ref{eq:LCeomSDG}). In both cases we obtain a wave equation with a source, where the nonlinear interaction is controlled by a bracket along the same pair of light-cone directions. In SDYM the interaction is governed by the Lie bracket of the gauge algebra, so the nonlinearity is entirely determined by the structure constants of the internal gauge symmetry. In contrast, in SDG the nonlinear term involves the Poisson bracket, which plays the role of a Lie bracket for functions on the $(u,w)$ plane. In this sense, self-dual gravity can be viewed as an analogue of SDYM in which the finite-dimensional gauge algebra is replaced by the infinite-dimensional algebra of area-preserving diffeomorphisms. The deeper relation underlying this analogy is a double copy map between the gauge theory and the gravitational theory. We will return to this idea in the next lecture, where we will discuss it in more general terms.

\subsubsection{DFDF Action}
Now that we have already constructed the actions of both Yang-Mills and self-dual Yang-Mills, we will move to a different non-Abelian gauge theory. At this point we would like to compute something equivalent to Weyl gravity, but for gauge theory. The main difference between Einstein-Hilbert is that in Einstein we are considering a two-derivative Lagrangian ($\sqrt{-g} R$) while in Weyl we have a four-derivative Lagrangian ($\sqrt{-g} C^2$) which is the square of some tensor (the Weyl tensor). Similarly, we can think about some two-derivative candidate (as the Weyl tensor) constructed from the F-curvature, such as
\bea
\nabla_{\mu} F_{\nu \rho} \,.
\eea
There are not many possibilities for constructing a higher-derivative gauge Lagrangian, due to the Bianchi identities and total derivatives. During this course we will study the case $J^2$ with $J_{\nu} = \nabla^{\mu} F_{\mu \nu}$.  

{\bf Action and equations of motion}

Let's consider the following action
\bea
S_{DFDF}= \frac12 \int \sqrt{-g} \nabla_{\mu} F^{\mu \nu i} \nabla^{\rho} F_{\rho \nu i} \, . 
\eea
At first glance we see that the action is invariant under both infinitesimal diffeomorphisms and non-Abelian gauge transformations. The equations of motion with respect to the gauge field can be easily computed by noting that we can safely write $F_{\mu \nu} = 2 \nabla_{[\mu} A_{\nu]}$. Typically, if the object is a connection, the covariant derivative cannot preserve the covariant structure. However, for this particular case the relation is true and it preserves the good behavior of the curvature.
So the variation of the action can be written as
\bea
\delta_{A} S_{DFDF}= 2 \int \sqrt{-g} \nabla_{\mu}  \nabla^{[\mu} \delta A^{\nu] i} \nabla^{\rho} F_{\rho \nu i} \, . 
\eea
We integrate two times the last expression and we obtain,
\bea
\delta_{A} S_{DFDF}= 2 \int \sqrt{-g}  \delta A^{\nu i} \nabla_{[\mu|}  \nabla^{\mu} \nabla^{\rho} F_{\rho |\nu] i} \, . 
\eea
Therefore, the equation of motion for the DFDF theory is given by
\bea
\Box J_{\nu} - \nabla_{\nu}  \nabla^{\mu}  J_{\mu}  = 0 \, .
\eea
If we compare this last equation against the equation of motion of Weyl gravity,
\bea
\nabla^{\rho} \nabla^{\sigma} C_{\mu \rho \nu \sigma} + \frac12 R^{\rho \sigma} C_{\mu \rho \nu \sigma} = 0 \, ,
\eea
in this case we cannot recognize an immediate relation between these theories. However, as we will see in the next section, both theories are also related by the double copy.
\newpage
\subsection{Exercises}

\subsubsection{Part A}
\begin{enumerate}

\item Find the expression for the commutator of two covariant derivatives acting on a vector when the torsion does not vanish. 

\item Show how the Riemann tensor changes under a covariant shift of the connection ($\Gamma \rightarrow \Gamma+T$), and prove that the Riemann tensor still transforms covariantly.

\item Prove that the trace of the LC connection is given by $\Gamma_{\mu \lambda}^{\mu}= \frac{1}{\sqrt{-g}} \partial_{\lambda}{\sqrt{-g}}$. 

\item Prove the identity $\int d^4x \sqrt{-g} \nabla_{\mu} T^{\mu} = \int d^4x \partial_{\mu} (\sqrt{-g} T^{\mu})$ which allow us to partial integrate with the covariant derivative, without considering the measure. 

\item Prove the two Bianchi identities for the Riemann tensor.

\item Prove the "wave identity" for the Ricci tensor, 
\bea
\Box R_{\mu \nu}=2 R_{\mu \rho \nu \sigma} R^{\rho \sigma} - 2 R_{\mu \rho} R^{\rho}{}_{\nu} + \nabla_{\mu} \nabla_{\nu}R.
\nn
\eea

\item Compute the variation of the Ricci scalar with respect to the inverse metric tensor: $\frac{\delta R}{\delta g^{\alpha \beta}} = \frac{\delta}{\delta g^{\alpha \beta}}(R^{\rho}{}_{\sigma \mu \nu} \delta^{\mu}_{\rho} g^{\sigma \nu})$. 

\item Compute the equation of motion for the Einstein-Hilbert Lagrangian. Hint: use the variation of the previous exercise and $\delta g = - g g_{\mu \nu} \delta g^{\mu \nu}$. 

\item Prove that Weyl gravity is Weyl invariant. A Weyl rescaling of for the metric is given by $\delta_{\zeta} g_{\mu \nu} = e^{2\zeta} g_{\mu \nu}$.
    
\item Consider a generic vector $v^{\mu}$ with arbitrary weight and show that the closure $\Big[\delta_{\xi_1},\delta_{\xi_2} \Big] v^{\mu} = \delta_{\xi_{21}} v^{\mu} \, $ holds.

\item Prove the closure of the diffeos plus Lorentz transformations in the vielbein formalism. It is enough to show the closure on the vielbien.

\item Consider a generic flat vector $v^a$ and compute $\delta_{\Lambda}(\partial_{\mu} v^{a})$ and $\delta_{\Lambda}(\nabla_{\mu} v^{a})$. Compare both expressions to obtain $\delta_{\Lambda} w_{\mu a b}$.

\item Use the vielbein compatibility to obtain $w_{abc} = -e^\mu{}_{[a}{} e^\nu{}_{b]}{}\partial_\mu e_{\nu c} +e^\mu{}_{[a}{} e^\nu{}_{c]}{}\partial_\mu e_{\nu b}
+e^\mu{}_{[b}{} e^\nu{}_{c]}{}\partial_\mu e_{\nu a}$ \, .

\item Prove that all the contractions of indices between the Weyl tensor and the inverse metric is zero.

\item The equations of motion for Weyl gravity are given by $B_{\mu \nu}=
\nabla^{\rho} \nabla^{\sigma} C_{\mu \rho \nu \sigma} + \frac12 R^{\rho \sigma} C_{\mu \rho \nu \sigma} = 0 \, , $ where $B_{\mu \nu}$ is called the Bach tensor. Show that every solution of the vacuum Einstein equation ($R_{\mu \nu}=0$) is solution of the vacuum Bach equation (this implies that every solution in Einstein gravity is solution in Weyl gravity).

\item Compute the EOM for: $S = 2 \int \sqrt{-g} (Ric^2 - \frac{1}{3} R^2)$, which is equivalent up to total derivatives with the standard $C^2$ Weyl action. To move from one to another one uses the Gauss Bonnet identity in $D=4$: $Riem^2 = 4 Ric^2 - R^2 + t.d.$ where t.d. is a total derivative.

\end{enumerate}

\subsubsection{Part B}
\begin{enumerate}
\item Show that the linearized Riemann tensor is invariant under linearized diffeomorphisms.

\item Compute the quadratic Einstein-Hilbert action around Minkowski space in terms of the perturbation (harmonic gauge). After integrations by parts you should obtain the Fierz-Pauli Lagrangian: $\frac12 \partial_{\alpha} h_{\mu \nu} \partial^{\alpha} \bar h^{\mu \nu}$. 

\item Considering perturbations around Minkowski, construct the theory $\sqrt{-g}(R+Riem^2)$ up to quadratic order in the perturbation. ¿Is the action invariant under linearized diffeomorphisms? 

\item Considering perturbations around Minkowski, construct the theory $\sqrt{-g} W^2$ up to quadratic order in the perturbation. For doing so consider the action $S_{\textrm{Weyl}} = 2 \int \sqrt{-g} (Ric^2 - \frac{1}{3} R^2)$.

\item Transform the Liouville operator acting on a generic phase space vector, $D_{\mu} v^{\nu}$ and verify explicitly that it transforms as a phase space tensor.

\item Kinematics of the perfect fluid: Define $a^{\mu}=u^{\nu} \nabla_{\nu}{u^{\mu}}$ and consider the decomposition in irreducible tensorial parts: 
\bea
\nabla_{\nu}{u_{\mu}} = w_{\mu \nu} + \sigma_{\mu \nu} + \frac13 \Theta (g_{\mu \nu} + u_{\mu} u_{\nu}) - a_{\mu} u_{\nu} \, \nn
\eea
where $w_{\mu \nu}$ is the vorticity (antisymmetric), $\sigma_{\mu \nu}$ is the shear tensor (symmetric and no trace) and $\Theta$ is the expansion scalar. Using $u^2=-1$, find the explicit form of these quantities in terms of $u$, $\nabla u$, $a$.

\item Considering $\nabla_{\mu} \nabla_{\nu} u_{\lambda} - \nabla_{\nu} \nabla_{\mu} u_{\lambda}=R^{\rho}{}_{\lambda \nu \mu} u_{\rho}$ and the decomposition of the previous exercise, compute the evolution laws:  
\bea
u^{\mu} \nabla_{\mu} \omega_{\lambda \nu} & = & f_{\lambda \nu} \nn \\
u^{\mu} \nabla_{\mu} \sigma_{\lambda \nu} & = & g_{\lambda \nu} \nn \\
u^{\mu} \nabla_{\mu} \Theta & = & h \nn \, .
\eea

\item Consider a dust configuration ($p=0$ and $T^{\mu \nu}=e u^{\mu} u^{\nu}$). Prove that in this case, the evolution law for the expansion is given by the Raychaudhuri equation: $u^{\mu} \nabla_{\mu} \Theta = w^2 - \sigma^2 - \frac13 \Theta^2 - 4 \pi e$.

\item Standard cosmology: Compute the Riemann tensor for the FRW ansatz.

\end{enumerate}
\subsubsection{Part C}
\begin{enumerate}
\item Using the Jacobi identity for $f_{ijk}$, \textit{i.e.}, $
f_{il}{}^{m} f_{jk}{}^{l} + f_{jl}{}^{m} f_{ki}{}^{l} + f_{kl}{}^{m} f_{ij}{}^{l} = 0 \, , $ show that $F_{\mu \nu i}$ transforms covariantly under non-Abelian gauge transformations.

\item Using $f^{i}{}_{jk} f_{i}{}^{mn} = 2 \alpha' \delta_{[j}^{m} \delta_{k]}^{n}$ compute the gauge transformation of the Chern-Simons 3-form $\delta_{\lambda} C^{(g)}_{\mu \nu \rho}$.

\item Consider a generic gauge vector $v^{i}$. Show $\big[\nabla_{\mu},\nabla_{\nu} \big] v^{i} = f_{jk}{}^{i} v^{j} F_{\mu \nu}{}^{k}$.

\item Show that $\nabla_{[\mu} C^{(g)}_{\nu \rho \sigma]}=\frac14 F_{[\mu \nu}{}^{i} F_{\rho \sigma]i}$. 

\item Consider self-dual Yang-Mills in Minkowski space and light-cone gauge. Show that in this gauge the gauge parameters parameters only depends on $v$ and $\bar w$, i.e., $\lambda^{i}= \lambda^{i}(v,\bar w)$.

\item Following the spirit of the previous exercice, compute the transformation of $\delta \Psi$ and prove that this charged scalar transforms non-covariantly. You should obtain that in this gauge $u \partial_{\bar w} \lambda^i = w \partial_{v}\lambda^i$.

\end{enumerate}

\section{Lecture 2: Off-shell Single and Double Copy}
\label{Sec2}

\subsection{Part A: Kerr-Schild Ansatz and Off-shell Single Copy For Maxwell}
\label{A2}
\subsubsection{General Ansatz with Null Vectors}
Our purpose for this section is to obtain the Maxwell equations
¸\bea
\nabla_{\mu} F^{\mu \nu} = 0
\eea
from the general relativity vacuum equation $R_{\mu \nu}=0$. Particularly, our starting point will be general relativity, but the metric will have the following particular ansatz from the very beginning,
\bea
g_{\mu \nu} = ¸\tilde g_{\mu \nu} + \kappa \varphi l^{\rho} l^{\sigma} \tilde{g}_{\rho \mu} \tilde{g}_{\sigma \nu} \, ,
\eea
where $l^{\mu}$ is a null vector,
\bea
l^{\mu} \tilde g_{\mu \nu} l^{\nu} = l^{\mu} g_{\mu \nu} l^{\nu} = 0 \, ,
\eea
$\varphi$ a scalar field and $\kappa$ a constant. We can understand this ansatz as a perturbation of a background solution, $\tilde g_{\mu \nu}$, using the null vectors and the scalar field. The null condition ensures that the inverse metric is finite
\bea
g^{\mu \nu} = \tilde g^{\mu \nu} - \kappa \varphi l^{\mu} l^{\nu} \, . 
\eea
Now we define the background covariant derivative acting on the vector $l^{\nu}$,
\bea
\tilde \nabla_{\mu} l^{\nu} = \partial_{\mu} l^{\nu} + \tilde \Gamma_{\mu \rho}^{\nu} l^{\rho}
\eea
and also defining $l_{\mu} = l^{\nu} \tilde g_{\mu \nu}$ we can write the full connection as
\bea
\Gamma_{\mu \nu}^{\rho} = \tilde \Gamma_{\mu \nu}^{\rho} + \kappa \Gamma_{\mu \nu}^{(1)\rho} + \kappa^2 \Gamma_{\mu \nu}^{(2)\rho} 
\eea
where 
\bea
\Gamma_{\mu \nu}^{(1)\rho} = \tilde g^{\rho \sigma} \tilde \nabla_{(\mu}(\phi l_{\nu)} l_{\sigma}) - \frac12 \tilde g^{\rho \sigma} \tilde \nabla_{\sigma} (\phi l_{\mu} l_{\nu}) \, ,
\eea
and
\bea
\Gamma_{\mu \nu}^{(2)\rho} = \frac12 \phi l^{\rho} l^{\sigma} \tilde \nabla_{\sigma}(\phi l_{\mu} l_{\nu}) \, .
\eea
As reader can easily verify, the contributions $\Gamma_{\mu \nu}^{(1)\rho}$ and $\Gamma_{\mu \nu}^{(2)\rho} $ are tensorial shifts in the LC connection. This means that we can easily write the full Riemann tensor by defining
\bea
\Delta \Gamma_{\mu \nu}^{\rho} = \Gamma_{\mu \nu}^{(1)\rho} + \Gamma_{\mu \nu}^{(2)\rho} \, ,
\eea
and therefore
\bea
R^{\rho}{}_{\epsilon \mu \nu} = \tilde R^{\rho}{}_{\epsilon \mu \nu} + 2 \tilde \nabla_{[\mu} \Delta \Gamma_{\nu] \epsilon}^{\rho} + 2 \Delta \Gamma_{[\mu \alpha}^{\rho} \Delta \Gamma_{\nu] \epsilon}^{\alpha} \, .
\eea
By taking the trace in the previous relation, we have a relation for the equation of motion,
\bea
R_{\epsilon \nu} = \tilde R_{\epsilon \nu} + 2 \tilde \nabla_{[\mu} \Delta \Gamma_{\nu] \epsilon}^{\mu} + 2 \Delta \Gamma_{[\mu \alpha}^{\mu} \Delta \Gamma_{\nu] \epsilon}^{\alpha} \, .
\eea
So far, we have only imposed the null condition. If we analyze the gravitational EOM we find
\bea
R_{\epsilon \nu} = \tilde R_{\epsilon \nu} + \kappa R_{\epsilon \nu}^{(1)} + \kappa^2 R_{\epsilon \nu}^{(2)} + \kappa^3 R_{\epsilon \nu}^{(3)} = 0 \, .
\eea
\subsubsection{The Kerr-Schild Ansatz}
The idea of the Kerr-Schild ansatz is to find an extra relation, apart from the null condition, an simplify the equation of motion to just $\tilde R_{\mu \nu} = R^{(1)}_{\mu \nu}=0$. 

In order to find this magic condition, we first explore the equation $R=0$ which we use to write $R_{\mu \nu}=0$. The expression for the Ricci scalar is 
\bea
R = \tilde R + R^{(1)} + R^{(2)} = 0 \, . 
\eea
where the quadratic part is given by
\bea
R^{(2)} = \frac12 \varphi^2 l^{\nu} \tilde \nabla_{\nu}l^{\rho} l^{\mu} \tilde \nabla_{\mu}l_{\rho} \, .  
\eea
While apriori the powers of the Ricci scalar could higher than the Ricci due to the extra inverse metric $R=R_{\mu \nu} (\tilde g^{\mu \nu} - \kappa \varphi l^{\mu} l^{\nu})$, it is straightforward to prove that the Ricci scalar is at most quadratic. Therefore, at this point we could impose the condition 
\bea
l^{\nu} \tilde \nabla_{\nu}l^{\rho} l^{\mu} \tilde \nabla_{\mu}l_{\rho} = 0
\eea
or simply
\bea
l^{\nu} \tilde \nabla_{\nu}l^{\rho} = 0 \, .
\label{geodesic}
\eea
With these ideas in mind, let's explore now the cubic contributions of the Ricci tensor to decide what to impose. The important thing here is that we would like to include an extra condition to make the equation linear in the $kappa$ parameter, but we would not like to eliminate all the dynamics and just have the trivial $\tilde R_{\mu \nu}=0$. We want to keep some non-trivial physics, but simple.

The cubic contributions to the Ricci tensor are given by
\bea
R^{(3)}_{\nu \sigma} = - \frac12 \varphi^3 l^{\sigma} \tilde \nabla_{\sigma}l^{\rho} l^{\mu} \tilde \nabla_{\mu}l_{\rho} l_{\nu} l_{\sigma} \, , 
\eea
so both conditions eliminate this contribution. Let's move to the quadratic terms,
\bea
R^{(2)}_{\nu \sigma} & = & \frac12 \tilde \nabla_{\mu}{\tilde \nabla_{\rho}{\varphi}} \varphi l^{\mu} l^{\rho} l_{\nu} l_{\sigma} + \frac12 \tilde \nabla_{\mu}{\varphi} \tilde \nabla_{\rho}{l^{\rho}} \varphi l^{\mu} l_{\nu} l_{\sigma} + \tilde \nabla_{\mu}{\varphi} \tilde \nabla_{\rho}{l_{\nu}} \varphi l^{\mu} l^{\rho} l_{\sigma} \nn \\ && + \tilde \nabla_{\mu}{\varphi} \tilde \nabla_{\rho}{l_{\sigma}} \varphi l^{\mu} l^{\rho} l_{\nu} + \frac12 \tilde \nabla_{\mu}{\tilde \nabla_{\rho}{l_{\nu}}} \varphi \varphi l^{\mu} l^{\rho} l_{\sigma} + \frac12 \tilde \nabla_{\mu}{l^{\mu}} \tilde \nabla_{\rho}{l_{\nu}} \varphi \varphi l^{\rho} l_{\sigma} \nn \\ && + \frac12 \tilde \nabla_{\mu}{l_{\nu}} \tilde \nabla_{\gamma}{l_{\sigma}} \varphi \varphi l^{\mu} l^{\gamma} + \frac12 \tilde \nabla_{\mu}{\tilde \nabla_{\rho}{l_{\sigma}}} \varphi \varphi l^{\mu} l^{\rho} l_{\nu} + \frac12 \tilde \nabla_{\mu}{l^{\mu}} \tilde \nabla_{\rho}{l_{\sigma}} \varphi \varphi l^{\rho} l_{\nu} \nn \\ && - \frac12 \tilde \nabla_{\nu}{\tilde \nabla_{\mu}{l^{\rho}}} \varphi \varphi l^{\mu} l_{\sigma} l_{\rho} - \frac12 \tilde \nabla_{\nu}{l^{\mu}} \tilde \nabla_{\rho}{l_{\mu}} \varphi \varphi l^{\rho} l_{\sigma} - \frac12 \tilde \nabla_{\mu}{\varphi} \tilde \nabla_{\rho}{l^{\mu}} \varphi l^{\rho} l_{\nu} l_{\sigma} \nn \\ && - \frac12 \tilde \nabla_{\mu}{l^{\rho}} \tilde \nabla_{\rho}{l^{\mu}} \varphi \varphi l_{\nu} l_{\sigma} + \frac12 \tilde \nabla_{\mu}{l^{\rho}} \tilde \nabla^{\mu}{l_{\rho}} \varphi \varphi l_{\nu} l_{\sigma} \, .
\eea
Some of the terms can still be simplified by using $\tilde \nabla_{\nu}{\tilde \nabla_{\mu}{l^{\rho}}} l_{\rho} = - {\tilde \nabla_{\mu}{l^{\rho}}} \tilde \nabla l_{\rho}$ but still we see that the most efficient way to simplify this expression is to impose the relation (\ref{geodesic}), which is called the geodesic condition. 

It is important to notice at this point that we do not need $R^{(2)}_{\mu \nu} =0$ but
\bea
R^{(2)}_{\mu \nu} \propto R^{(1)}_{\mu \nu} \, .
\eea
When we impose the geodesic condition and simplify $R^{(2)}_{\nu \sigma}$ we obtain,
\bea
R^{(2)}_{\nu \sigma} & = & \frac12 \tilde \nabla_{\mu}{\tilde \nabla_{\rho}{\varphi}} \varphi   l^{\mu} l^{\rho} l_{\nu} l_{\sigma} + \frac12 \tilde \nabla_{\mu}{\varphi} \tilde \nabla_{\rho}{l^{\rho}} \varphi   l^{\mu} l_{\nu} l_{\sigma} \nn \\ && - \frac12 \tilde \nabla_{\mu}{l^{\rho}} \tilde \nabla_{\rho}{l^{\mu}} \varphi \varphi   l_{\nu} l_{\sigma} + \frac12 \tilde \nabla_{\mu}{l_{\lambda}} \tilde \nabla^{\mu}{l^{\lambda}} \varphi \varphi   l_{\nu} l_{\sigma}
\eea
while the linear EOM is 
\bea
R^{(1)}_{\nu \sigma} & = & \frac12 \tilde \nabla_{\mu}{\tilde \nabla_{\nu}{\varphi}}  l^{\mu} l_{\sigma} + \frac12 \tilde \nabla_{\mu}{\tilde \nabla_{\nu}{l_{\sigma}}} \varphi  l^{\mu} + \frac12 \tilde \nabla_{\mu}{\tilde \nabla_{\nu}{l^{\mu}}} \varphi  l_{\sigma} + \frac12 \tilde \nabla_{\mu}{\tilde \nabla_{\sigma}{\varphi}}  l^{\mu} l_{\nu} \nn \\ && + \frac12 \tilde \nabla_{\sigma}{\varphi} \tilde \nabla_{\mu}{l^{\mu}}  l_{\nu} + \frac12 \tilde \nabla_{\mu}{\tilde \nabla_{\sigma}{l_{\nu}}} \varphi  l^{\mu} + \frac12 \tilde \nabla_{\mu}{\varphi} \tilde \nabla_{\sigma}{l_{\nu}}  l^{\mu} + \frac12 \tilde \nabla_{\sigma}{l_{\nu}} \tilde \nabla_{\rho}{l^{\rho}} \varphi  \nn \\ && + \frac12 \tilde \nabla_{\mu}{\tilde \nabla_{\sigma}{l^{\mu}}} \varphi  l_{\nu} + \frac12 \tilde \nabla_{\mu}{\varphi} \tilde \nabla_{\sigma}{l^{\mu}}  l_{\nu} + \frac12 \tilde \nabla_{\sigma}{l^{\mu}} \tilde \nabla_{\mu}{l_{\nu}} \varphi  - \frac12 \tilde \Box{\varphi}  l_{\nu} l_{\sigma} \nn \\ && - \tilde \nabla_{\mu}{\varphi} \tilde \nabla^{\mu}{l_{\nu}}  l_{\sigma} - \tilde \nabla_{\mu}{\varphi} \tilde \nabla^{\mu}{l_{\sigma}}  l_{\nu} - \frac12 \tilde \Box{l_{\nu}} \varphi  l_{\sigma} - \tilde \nabla^{\gamma}{l_{\nu}} \tilde \nabla_{\gamma}{l_{\sigma}} \varphi  \nn \\ && - \frac12 \tilde \Box{l_{\sigma}} \varphi  l_{\nu} + \frac12 \tilde \nabla_{\mu}{\varphi} \tilde \nabla_{\nu}{l^{\mu}}  l_{\sigma} + \frac12 \tilde \nabla_{\mu}{\varphi} \tilde \nabla_{\nu}{l_{\sigma}}  l^{\mu} + \frac12 \tilde \nabla_{\nu}{\varphi} \tilde \nabla_{\mu}{l^{\mu}}  l_{\sigma} \nn \\ && + \frac12 \tilde \nabla_{\nu}{l_{\sigma}} \tilde \nabla_{\rho}{l^{\rho}} \varphi  + \frac12 \tilde \nabla_{\nu}{l^{\mu}} \tilde \nabla_{\mu}{l_{\sigma}} \varphi \, ,
\eea
At this point it is straightforward to prove that
\bea
R^{(2)}_{\nu \sigma}= \varphi l_{(\sigma|} l^{\rho} R^{(1)}_{\rho |\nu)} \, , 
\eea
and therefore now we know that the Kerr-Schild ansatz produce linear equations in $\kappa$, a very powerful result in GR. There are several famous solutions that can be cast in the Kerr-Schild form: Schwarzschild, Kerr and Kerr-Newman black holes, pp-waves solutions and Taub-Nut, for example, are common GR solutions that can be cast in the Kerr-Schild form. 

\subsubsection{Killing Vectors and the Double Copy}
We assume that the geometry admits one Killing vector $\xi^{\mu}$ such that the Lie derivative ${L}_{\xi}$ acting on an exact field vanishes,
\bea
L_{\xi} T_{\mu_{1} \mu_{2} \mu_{3} \dots} =0 \, 
\label{Lie}
\eea
where $T_{\mu_{1} \mu_{2} \mu_{3} \dots}$ is an arbitrary tensor. Moreover we choose a coordinate system where $\xi^{\mu}$ is covariantly constant, \textit{i.e.},
\bea
\nabla_{o\mu}\xi_{\nu} = \nabla_{o[\mu}\xi_{\nu]} = 0 \, ,
\eea
and then condition (\ref{Lie}) is
\bea
\xi^{\mu} \nabla_{o \mu} T_{\mu_{1} \mu_{2} \mu_{3} \dots} = 0 \, .
\eea
We normalize the null vectors to satisfy,
\bea
\xi^{\mu} l_{\mu} = 1 \, .
\eea

In order to obtain the single copy we first contract the linear equation of motion for the metric tensor with $\xi^{\mu}$,
\bea
\xi^{\nu} R^{(1)}_{\nu \sigma} & = &   \frac12 \tilde \nabla_{\mu}{\tilde \nabla_{\sigma}{\varphi}}  l^{\mu}  + \frac12 \tilde \nabla_{\sigma}{\varphi} \tilde \nabla_{\mu}{l^{\mu}}    \nn \\ && + \frac12 \tilde \nabla_{\mu}{\tilde \nabla_{\sigma}{l^{\mu}}} \varphi   + \frac12 \tilde \nabla_{\mu}{\varphi} \tilde \nabla_{\sigma}{l^{\mu}}   - \frac12 \tilde \Box{\varphi}   l_{\sigma} \nn \\ && - \tilde \nabla_{\mu}{\varphi} \tilde \nabla^{\mu}{l_{\sigma}}  - \frac12 \tilde \Box{l_{\sigma}} \varphi  = 0 \, .
\eea

\subsubsection{The Maxwell Equation Appears!}
Let's group the terms of the previous expression as
\bea
\xi^{\nu} R^{(1)}_{\nu \sigma} & = &   \frac12 \tilde \nabla^{\mu}{\tilde \nabla_{\sigma}({\varphi}}  l_{\mu}) - \frac12 \tilde \Box ({\varphi}   l_{\sigma}) = 0 \, .
\eea
Now we simply make the formal identification \bea
A_{\mu} = \varphi l_{\mu} \, ,
\eea
and we obtain
\bea
\xi^{\nu} R^{(1)}_{\nu \sigma}
& = &   \frac12 \tilde \nabla^{\mu}\tilde \nabla_{\sigma}(A_{\mu}) - \frac12 \tilde \Box (A_{\sigma}) 
= \frac12 \tilde \nabla^{\mu} F_{\sigma \mu} = 0 \, .
\eea
The previous equation proves that any solution to the vacuum Maxwell equation can be translated to a gravitational Kerr-Schild background with an isometry, given by the Killing vector $\xi^{\mu}$. 

In the next part, we will reverse the idea of the single copy. We will start from a gauge theory (our examples will cover Maxwell, Yang-Mills and DFDF), and we will obtain different formulations of gravity.

\subsection{Part B: Off-shell Double Copy for Yang-Mills}
\label{B2}
Now we will reverse the logic of the last section and, more importantly, we will:
\begin{itemize}
\item extend the gauge theory to Yang-Mills. 
\item match the gauge theory with the gravitational theory at the level of the action.
\end{itemize}

The fundamental interactions in nature, as far as currently known, are described by Yang–Mills gauge theories and, at first sight, gauge theory and gravitational theories have actions with very different structures and their perturbative expansions behave in very different ways. In particular, when general relativity is treated as a quantum field theory and expanded perturbatively around flat space-time, the resulting interactions are notoriously complicated and the theory is non-renormalizable. This stands in sharp contrast with Yang–Mills theory, whose perturbation theory is by now well understood and under excellent theoretical control.

Despite these differences, remarkable evidence accumulated over the past decades suggests that gauge theory and gravity may be far more closely related than their Lagrangians initially suggest. One of the most striking manifestations of this relationship is the Bern–Carrasco–Johansson (BCJ) double copy \cite{Doublecopy1}, which expresses scattering amplitudes in gravity as a kind of “square’’ of Yang–Mills amplitudes. The key observation behind this construction is that Yang–Mills amplitudes can be reorganized in a special way. In such a representation, the contributions associated with the gauge group—known as color factors—appear on the same footing as the purely kinematical numerator factors, which depend on momenta and polarization data \cite{Part2-5}-\cite{Part2-22}. 

Remarkably, these kinematic numerators can be arranged to satisfy algebraic relations analogous to the Jacobi identities. This property is known as color–kinematics duality. Once this duality is achieved, a simple but powerful operation becomes possible: replacing the color factors by another copy of the kinematic numerators produces scattering amplitudes in gravity. In other words, gravity amplitudes can be obtained as an on-shell “double copy’’ of the gauge-theory building blocks. From the traditional perspective of starting with the Einstein–Hilbert Lagrangian and applying textbook perturbation theory, the same gravity amplitudes are vastly more complicated to compute.

Formal proofs of the double-copy construction exist at tree level, using several complementary approaches. At loop level, while a general proof is still lacking, a large body of evidence has accumulated through explicit constructions of multi-loop amplitudes. The double copy is now a central element of the modern scattering-amplitude program, which often departs from the traditional textbook formulation of quantum field theory. 

Because of this on-shell perspective, one might not expect the double copy to manifest itself directly at the level of an action. Nevertheless, recent work suggests that such a formulation may indeed exist. In particular, in 2021 Hohm, Jaramillo-Díaz, and Plefka \cite{HJP} showed that it is possible to reproduce the full quadratic Einstein–Hilbert Lagrangian, as well as a gauged version of the cubic action, within a framework inspired by the double-copy structure. In this section we will review these results and discuss how they provide a promising step toward understanding the double copy directly at the level of field-theory actions. 

\subsubsection{HJP: Quadratic Theory}

We start from the action for Yang-Mills theory in $D$ dimensions, 
 \be\label{actionYM}
  S_{\rm YM} = -\frac{1}{4 }\int d^D x\,\kappa_{ij} \,F^{\mu\nu \,i} F_{\mu\nu}{}^{j}\;, 
 \ee
where $
  F_{\mu\nu}{}^{i} = \partial_{\mu}A_{\nu}{}^{i} -\partial_{\nu}A_{\mu}{}^{i}  + g_{YM} f^{i}{}_{jk} A_{\mu}{}^{j} A_{\nu}{}^{k}\;, $.   
The quadratic order in fields action, after 
integration by parts, is given by 
 \be
  S_{\rm YM}^{(2)} = \frac{1}{2}\int d^Dx\,\kappa_{ij}\,A^{\mu i} \big(\square A_{\mu}{}^{j} - \partial_{\mu}\partial^{\nu} A_{\nu}{}^{j}\big)\;. 
 \ee
This action is the Maxwell action, so in principle, we will try to reverse the logic of the previous section to try to obtain Einstein-Hilbert. In order to construct the double copy map we need to pass to momentum space. Defining 
\bea
A_{\mu}{}^{i}(k) & \equiv & \frac{1}{(2\pi)^{D/2}}\int d^D x A_{\mu}{}^{i}(x)e^{ikx} \\
A_{\mu}{}^{i}(x) & \equiv & \frac{1}{(2\pi)^{D/2}}\int d^D k A_{\mu}{}^{i}(k)e^{-ikx} \, ,
\eea
the quadratic action can be written as 
 \be
  S_{\rm YM}^{(2)} = -\frac{1}{2} \int_{k}\, d^{D}k \kappa_{ij}\,k^2 \,\Pi^{\mu\nu}(k) A_{\mu}{}^{i}(-k) A_{\nu}{}^{j}(k)\; , 
 \ee
where
 \be\label{PiProjector}
  \Pi^{\mu\nu}(k) \equiv  \eta^{\mu\nu}-\frac{k^{\mu}k^{\nu}}{k^2}\;. 
 \ee 
 The only extra equation that you will need to explicitly prove the quadratic action is
 \bea
 \int d^dx e^{-i(k+p)x}=(2\pi)^{D} \delta^{(D)}(k+p) \, .
\eea 
The operator $\Pi^{(\mu \nu)}$ obeys the following identities:  
 \be\label{projIdentity}
  \Pi^{\mu\nu}(k)k_{\nu} \equiv 0\;, \qquad \Pi^{\mu\nu}\Pi_{\nu\rho} \equiv  \Pi^{\mu}{}_{\rho}\;. 
 \ee
 The first relation implies gauge invariance under 
 \be
   \delta A_{\mu}{}^{i}(k) = k_{\mu} \lambda^{i}(k)  \;, 
 \ee  
where the gauge parameter $\lambda^i(k)$ is an arbitrary function. This transformation will be soon related to linearized diffeomorphism in a suitable way, so the resulting gravity theory will be invariant under linearized diffeomorphisms.

Let us  now turn to the double copy construction: We will replace the color indices $a$ 
by a second set of spacetime indices denoted by a bar, 
$i\rightarrow \bar{\mu}$, corresponding to a second set of spacetime 
momenta $\bar{k}^{\bar{\mu}}$:  
 \be\label{DcFields}
  A_{\mu}{}^{i}(k) \; \rightarrow \; e_{\mu\bar{\mu}}(k,\bar{k})\;. 
 \ee
 At first glance one might be tempted to identify the second set of spacetime indices with the first one, but let's keep it arbitrary, for the moment. Since the gauge group is arbitrary, let's assume that the dimension of the gauge group is adequate to perform this identification.
 
To complete the double copy map at this order we need to define a substitution rule for the Cartan-Killing metric 
$\kappa_{ij}$. Motivated by the YM $\times$ YM structure we choose:
 \be\label{DCCartan}
  \kappa_{ab} \; \rightarrow \; \frac{1}{2}\, \bar{\Pi}^{\bar{\mu}\bar{\nu}}(\bar{k})\,, 
 \ee
where $\bar{\Pi}^{\bar{\mu}\bar{\nu}}$  is defined as in (\ref{PiProjector}), but with all momenta replaced by barred momenta and all indices replaced by 
barred indices.

The quadratic gravity action following from this double copy (DC) prescription then reads
 \be\label{quadraticDoubleCopy}
  S_{\rm DC}^{(2)} = 
  -\frac{1}{4} \int_{k,\bar{k}} k^2\,   \Pi^{\mu\nu}(k)  \bar{\Pi}^{\bar{\mu}\bar{\nu}}(\bar{k}) e_{\mu\bar{\mu}}(-k,-\bar{k}) 
  e_{\nu\bar{\nu}}(k,\bar{k})\,.  
 \ee
Let's observe that now the field $e_{\mu\bar{\mu}}$ depends on doubled momenta $K\equiv (k,\bar{k})$. Moreover, the momenta $k$ and $\bar{k}$ enter the action on the same footing, except that we have chosen the factor in front to be $k^2$ rather than $\bar{k}^2$, but this asymmetry is resolved with the level-matching constraint
 \be\label{levelmatching}
  k^2 = \bar{k}^2\;, 
 \ee 
where two copies of the same flat  space-time metric are used to take the square. In order to lead to the Einstein-Hilbert action, we thus have to assume that the doubled momenta are subject to this constraint (which does have more general 
solutions than the trivial $k=\bar{k}$ for which the theory reduces to a standard linearized  gravity theory). 
We also note that, due to the first identity in (\ref{projIdentity}),  the action is manifestly gauge invariant under  
 \be\label{deltagaugee}
  \delta e_{\mu\bar{\nu}} = k_{\mu}\bar{\lambda}_{\bar{\nu}}  + \bar{k}_{\bar{\nu}}\lambda_{\mu} \;, 
 \ee
with two independent gauge parameters $\lambda_{\mu}$ and $\bar{\lambda}_{\bar{\mu}}$ that depend on doubled momenta 
$K\equiv (k,\bar{k})$, subject to (\ref{levelmatching}). Here it is important to mention that we are not obtaining the symmetry rule from the identification itself. We will comment on this when we analyze the cubic contributions.  
 
We will now show that (\ref{quadraticDoubleCopy}) is indeed equivalent to linearized gravity. 
Writing out the projectors with (\ref{PiProjector}) and using the level-matching constraint (\ref{levelmatching}) the action reads 
  \begin{align}\label{YMsquaredConstrained}
     S_{\rm DC}^{(2)} = -\frac{1}{4} \int_{k,\bar{k}} \Big(&  k^2 e^{\mu\bar{\nu}}e_{\mu\bar{\nu}}-k^{\mu} k^{\rho} e_{\mu\bar{\nu}}e_{\rho}{}^{\bar{\nu}}
     - \bar{k}^{\bar{\nu}}\bar{k}^{\bar{\sigma}} e_{\mu\bar{\nu}} e^{\mu}{}_{\bar{\sigma}} \nn\\ &
     +\frac{1}{k^2} k^{\mu}k^{\rho}\bar{k}^{\bar{\nu}}\bar{k}^{\bar{\sigma}} e_{\mu\bar{\nu}} e_{\rho\bar{\sigma}}
     \Big)\,.
  \end{align}
In order to compare with the standard Einstein-Hilbert action we have to Fourier transform to (doubled) position space. 
This is straightforward except for the last term in (\ref{YMsquaredConstrained}), which due to the factor $\frac{1}{k^2}$ would yield a 
non-local term. This problem is resolved by introducing  an auxiliary scalar field $\phi(k,\bar k)$ (the dilaton):
   \begin{align}\label{localDFTquad}
     S_{\rm DC}^{(2)} = -\frac{1}{4} \int_{k,\bar{k}}\Big(& k^2 e^{\mu\bar{\nu}}e_{\mu\bar{\nu}}-k^{\mu} k^{\rho} e_{\mu\bar{\nu}}e_{\rho}{}^{\bar{\nu}}
     - \bar{k}^{\bar{\nu}}\bar{k}^{\bar{\sigma}} e_{\mu\bar{\nu}} e^{\mu}{}_{\bar{\sigma}} \nn\\ &
          -k^2 \phi^2 +2\phi\, k^{\mu}\bar{k}^{\bar{\nu}} e_{\mu\bar{\nu}}
     \Big)\;. 
  \end{align}
Integrating out  $\phi$ by solving its own field equations, 
 \be\label{PhiSolution}
  \phi = \frac{1}{k^2}k^{\mu}\bar{k}^{\bar{\nu}} e_{\mu\bar{\nu}}\;, 
 \ee
and back-substituting into the action we recover the non-local (\ref{YMsquaredConstrained}). 
Alternatively, without integrating out fields, one may redefine the dilaton as 
$\phi \to \phi' =  \phi - \frac{1}{k^2}k^{\mu}\bar{k}^{\bar{\nu}} e_{\mu\bar{\nu}}$,  
which decouples $\phi'$  from $e_{\mu\bar\nu}$. 
The action  (\ref{localDFTquad}) is of course  still gauge invariant, 
with a gauge transformation for $\phi$ that is determined by the variation of (\ref{PhiSolution}): 
 \be\label{deltaGaugephi}
  \delta\phi = k_{\mu} \lambda^{\mu} + \bar{k}_{\bar{\mu}} \bar{\lambda}^{\bar{\mu}}\;, 
 \ee
where we used (\ref{levelmatching}). 
With the action in the form (\ref{localDFTquad})  it is then straightforward  to Fourier transform to a local action in doubled position space: 
  \begin{align}\label{localDFTquadPosition}
     S_{\rm DC}^{(2)} = \frac{1}{4}& \int d^Dx \,d^D\bar{x} \Big( e^{\mu\bar{\nu}}\square e_{\mu\bar{\nu}}+\partial^{\mu}  e_{\mu\bar{\nu}}\,\partial^{\rho}  e_{\rho}{}^{\bar{\nu}} \nn\\ &
     + \bar{\partial}^{\bar{\nu}} e_{\mu\bar{\nu}}\, \bar{\partial}^{\bar{\sigma}}e^{\mu}{}_{\bar{\sigma}}
          - \phi\square \phi +2\phi \partial^{\mu}\bar{\partial}^{\bar{\nu}} e_{\mu\bar{\nu}}
     \Big) ,
  \end{align}
where $\partial_{\mu} =\frac{\partial}{\partial x^{\mu}}$ and $\bar{\partial}_{\bar{\mu}}=\frac{\partial}{\partial \bar{x}^{\bar{\mu}}}$ are the partial derivatives 
corresponding to the coordinates that are dual to $k^{\mu}$ and $\bar{k}^{\bar{\mu}}$ and hence by (\ref{levelmatching}) subject to the 
constraint 
 \be\label{positionlevelmatching}
  \square \equiv \partial^{\mu} \partial_{\mu} = \bar{\partial}^{\bar{\mu}}\bar{\partial}_{\bar{\mu}}\;. 
 \ee
The gauge transformations (\ref{deltagaugee}) and (\ref{deltaGaugephi}) translate in doubled position space to 
 \be
 \begin{split}
  \delta e_{\mu\bar{\nu}} &= \partial_{\mu}\bar{\lambda}_{\bar{\nu}} + \bar{\partial}_{\bar{\nu}}\lambda_{\mu}\;, \\
  \delta \phi &=  \partial_{\mu} \lambda^{\mu} + \bar{\partial}_{\bar{\mu}} \bar{\lambda}^{\bar{\mu}}\;, 
 \end{split}
 \ee
under which (\ref{localDFTquadPosition}) is invariant, modulo the constraint (\ref{positionlevelmatching}). 
The action (\ref{localDFTquadPosition}) defines precisely the standard quadratic Einstein-Hilbert action
upon setting $x=\bar{x}$. Let's explicitly show it. Let's remember the quadratic Einstein-Hilbert action,
\bea
S_{E-H} \propto \int d^4x  (R^{(2)}_{\nu \sigma} \eta^{\sigma \nu}  +  R^{(1)}_{\nu \sigma} (\frac12 h \eta^{\sigma \nu} -  h^{\sigma \nu})) \, ,
\eea
were $\propto$ means that we are omitting a minus global sign. Let's compute the first term from the previous Lagrangian considering
\bea
R^{(2)\rho}{}_{\sigma \mu \nu} & = & 2 \partial_{[\mu} \Gamma^{(2)\rho}_{\nu] \sigma} + 2 \Gamma^{(1)\rho}_{\lambda [\mu} \Gamma^{(1) \lambda}_{\nu] \sigma} \, .
\eea
The contraction $R^{(2)}_{\sigma \nu} \eta^{\sigma \nu}$ gives
\bea
- \frac12  h \partial^{\beta} \partial^{\nu} h_{\nu \beta} + \frac14  h \Box_{\beta}h + \frac12 h_{\nu \alpha} \partial_{\lambda}\partial^{\alpha} h^{\nu \lambda} - \frac14 h_{\nu \lambda} \Box h^{\nu \lambda} \, .
\eea
On the other hand, $R^{(1)}_{\nu \sigma} (\frac12 h \eta^{\sigma \nu} -  h^{\sigma \nu})$ gives
\bea
R^{(1)}_{\nu \sigma} (\frac12 h \eta^{\sigma \nu} -  h^{\sigma \nu}) =   h \partial^{\beta} \partial^{\nu} h_{\nu \beta} - \frac12  h \Box h - h_{\nu \alpha} \partial_{\lambda}\partial^{\alpha} h^{\nu \lambda} + \frac12 h_{\nu \lambda} \Box h^{\nu \lambda}
\eea
The sum of the previous quantities gives the following Lagrangian,
\bea
L^{(2)}_{E-H} =  \frac12 h \partial^{\beta} \partial^{\nu} h_{\nu \beta} - \frac14  h \Box h - \frac12 h_{\nu \alpha} \partial_{\lambda}\partial^{\alpha} h^{\nu \lambda} + \frac14 h_{\nu \lambda} \Box h^{\nu \lambda} \, .
\eea
We compare the previous Lagrangian with
 \begin{align}
     L_{\rm DC}^{(2)} = \frac{1}{4}&  \Big( e^{\mu{\nu}}\square e_{\mu{\nu}}+\partial^{\mu}  e_{\mu{\nu}}\,\partial^{\rho}  e_{\rho}{}^{{\nu}}
     + {\partial}^{{\nu}} e_{\mu{\nu}}\, {\partial}^{{\sigma}}e^{\mu}{}_{{\sigma}}
          - \phi\square \phi +2\phi \partial^{\mu}{\partial}^{{\nu}} e_{\mu{\nu}}
     \Big) ,
  \end{align}
  where $e_{\mu \nu}=h_{\mu \nu}$ and $\phi=h$. Explicitly (and using partial integration),
   \begin{align}
     L_{\rm DC}^{(2)} = \frac{1}{4}&  \Big( h^{\mu{\nu}}\square h_{\mu{\nu}}- 2   h_{\mu{\nu}}\, \partial^{\mu} \partial^{\rho}  h_{\rho}{}^{{\nu}}
          - h\square h +2h \partial^{\mu}{\partial}^{{\nu}} h_{\mu{\nu}}
     \Big) .
  \end{align}
  Therefore, we can see that the quadratic Einstein-Hilbert Lagrangian can be obtain by considering the off-shell double copy prescription of Hohm, Jaramillo-Diaz and Plefka. In the next section we will partially extend this results to cubic order.

\subsubsection{HJP: Cubic (Gauge-Fixed) Theory}
We now turn to the cubic Yang-Mills theory  and we will extend the double copy construction to 
the cubic action of EH under a particular gauge fixing condition. The cubic part of the Yang-Mills action (\ref{actionYM}) reads 
  \be
   S^{(3)}_{\rm YM} = - g_{YM}\int d^Dx\, f_{ijk}\, \partial^{\mu}A^{\nu a} \,A_{\mu}{}^{b}\, A_{\nu}{}^c\;. 
  \ee
Upon Fourier transforming to momentum space this becomes 
  \be
   S^{(3)}_{\rm YM} = \frac{i g_{YM}}{(2\pi)^{{D}/{2}}} \int_{k_1,k_2,k_3 }
 \!\!\!\!\!\!\!\!\!\!\!\!\! \delta(k_1+k_2+k_3) f_{ijk} k_1^{\mu} A_{1}^{\nu\,i} A_{2\mu}{}^{j} A_{3 \nu}{}^k\,, 
  \ee
where we use the short-hand notation $A_i \equiv A(k_i)$, and we 
performed the $x$-integration, 
introducing  the delta function. It is convenient to write this more symmetrically as 
  \begin{align}
S^{(3)}_{\rm YM}=-&\frac{ig_{YM}}{6(2\pi)^{{D}/{2}} }\int_{k_1,k_2,k_3 } 
 \!\!\!\!\!\!\!\!\!\!  \d (k_{1}+k_{2}+k_{3})\\ &\qquad  \times f_{ijk}\, \Pi^{\mu \nu \rho}(k_{1},k_{2},k_{3}) 
\, A_{1\mu}{}^i A_{2\nu}{}^j A_{3\rho}{}^k\, ,\nn
\end{align}
where we defined 
\begin{equation}\label{3indexPi}
\Pi^{\mu \nu \rho}(k_{1},k_{2},k_{3})\equiv \eta^{\mu \nu} k_{12}^{\r}+\eta^{\nu \rho}k^{\mu}_{23}+\eta^{\rho \mu}k^{\nu}_{31}\, ,
\end{equation}
with $k_{ij}\equiv k_{i}-k_{j}$.

Our task now is to give  the double copy prescription that extends (\ref{DcFields}), (\ref{DCCartan}) to the cubic theory. 
The natural substitution rule is 
\begin{equation}\label{eq:precripcubic}
f_{abc} \; \to \;  \tfrac{i}{4} \,
\ov\Pi^{\ov\mu\ov\nu\ov\rho}(\ov k_{1},\ov k_{2},\ov k_{3})\;. 
\end{equation}
 
Together with  $g_{YM}\rightarrow \frac{1}{2}$,  this gives the cubic action 
\begin{align}
&S^{(3)}_{\rm DC}=\frac{1}{48(2\pi)^{D/2}}\int  {dK_{1}}{dK_{2}}{dK_{3}}\, \d (K_{1}+K_{2}+K_{3})\nn \\
&\qquad \times  \ov\Pi^{\ov\mu\ov\nu\ov\rho}(\ov k_{1},\ov k_{2},\ov k_{3}) \, \Pi^{\mu\nu\rho}(k_{1},k_{2},k_{3})
\nn \, 
e_{1\,\mu\ov\mu}\, e_{2\,\nu\ov\nu}\, e_{3\,\rho\ov\rho}\, , 
\end{align}
where we use the short-hand notation $e_{i\, \mu\ov\mu}\equiv e_{\mu\ov\mu}(K_{i})$, with $K\equiv (k,\bar{k})$ for doubled momenta, 
and  $dK\equiv d^{2D}K$. 
Writing out $\Pi^{\mu\nu\rho}$ and  $\bar{\Pi}^{\bar{\mu}\bar{\nu}\bar{\rho}}$ yields nine terms which, upon relabeling momentum variables  and  indices, reduce to
two terms, and then writing out $k_{ij}=k_i-k_j$ 
 the action becomes
\begin{align}
S^{(3)}_{\rm DC}&=\frac{1}{8(2\pi)^{D/2}}\int  {dK_{1}}{dK_{2}}{dK_{3}}\, \d (K_{1}+K_{2}+K_{3}) \nn \\
&\times\, e_{1\, \mu\ov\mu}\Big[-k_{2}^{\mu}\, e_{2\, \rho\ov\rho}\,\ov k^{\ov\mu}_{3}\, e_{3}^{\rho\ov\rho}+k^{\mu}_{2}\, e_{2\, \nu\ov\rho}\, \ov k^{\ov\rho}_{3}\, e_{3}^{\nu\ov\mu}+k_{2}^{\rho}\, e^{\mu\ov\rho}_{2}\, \ov k^{\ov\mu}_{3}\, e_{3\, \rho\ov\rho} 
+k_{2}^{\mu}\, \ov k_{2}^{\ov\mu}\, e_{2\, \rho\ov\rho}\, e_{3}^{\rho\ov\rho} \nn \\ & -k_{2\, \rho}\, e_{2}^{\mu\ov\rho}\, \ov k_{3\, \ov\rho}\, e_{3}^{\rho\ov\mu}-k_{2}^{\rho}\, \ov k^{\ov\mu}_{2}\, e_{2}^{\mu\ov\rho}\, e_{3\rho\ov\rho}\Big]\;.
\end{align}
Fourier transforming  to position space and integrating by parts, we finally obtain 
\begin{equation}\label{eq:DCBAD}
\begin{split}
S^{(3)}_{\rm DC}=\frac{1}{8}\int d^D x \,d^D\ov x\;  e_{\mu\ov\mu}\, \Big[& \,2\partial^{\mu}e_{\rho\ov\rho}\,\ov \partial^{\ov\mu}e^{\r\ov\rho}-2\partial^{\mu}e_{\nu\ov\rho}\,\ov \partial^{\ov\rho}e^{\nu\ov\mu}-2\partial^{\rho}e^{\mu\ov\rho}\,\ov \partial^{\ov\mu}e_{\rho\ov\rho} \nn \\ &
 +\partial^{\rho}e_{\rho\ov\rho}\,\ov \partial^{\ov\rho}e^{\mu\ov\mu}+\ov \partial_{\ov\rho}e^{\mu\ov\rho}\,\partial_{\rho}e^{\rho\ov\mu}\, \Big]\;.
\end{split}
\end{equation}
In the following we will prove that this action agrees precisely with the cubic Einstein-Hilbert action upon imposing a gauge fixing condition. For simplicity, we will use the TT gauge ($\partial_{\mu}h^{\mu \nu}=h=0$), so the HJP Lagrangian is given by
\bea
L^{(3)}_{HJP} = \frac14 \partial^{\mu} h_{\rho \beta} \partial^{\alpha} h^{\rho \beta} h_{\mu \alpha} - \frac12 \partial^{\mu}h_{\nu \beta} \partial^{\beta} h^{\nu \alpha} h_{\mu \alpha} \, .
\eea
The full cubic EH Lagrangian is given by
\bea
L^{(3)}_{EH} & = & 2 \partial^{\sigma}{}_{\gamma}{h_{\sigma \lambda}} h_{\nu}{}^{\gamma} h^{\nu \lambda} - \Box{h_{\gamma \lambda}} h_{\nu}{}^{\gamma} h^{\nu \lambda} + 2 \partial^{\nu}{h_{\nu}{}^{\sigma}} \partial_{\gamma}{h_{\sigma \lambda}} h^{\gamma \lambda} 
\nn \\ && 
+ \partial_{\mu \nu}{h_{\rho \sigma}} h^{\mu \rho} h^{\nu \sigma} - \partial_{\mu \nu}{h_{\rho \sigma}} h^{\mu \nu} h^{\rho \sigma} - \partial^{\nu}{h_{\nu}{}^{\sigma}} \partial_{\sigma}{h_{\gamma \lambda}} h^{\gamma \lambda} 
\nn \\ && + \partial^{\nu}{h_{\nu \gamma}} \partial^{\sigma}{h_{\sigma \lambda}} h^{\gamma \lambda} - \partial_{\rho \sigma}{h} h_{\nu}^{\rho} h^{\nu \sigma} - \partial_{\rho}{h} \partial^{\nu}{h_{\nu \sigma}} h^{\rho \sigma} \nn \\ && 
- \partial^{\nu}{h} \partial_{\rho}{h_{\nu \sigma}} h^{\rho \sigma} - \frac{3}{4} \partial_{\gamma}{h^{\nu \sigma}} \partial_{\lambda}{h_{\nu \sigma}} h^{\gamma \lambda} - \frac{3}{2} \partial^{\nu}{h^{\sigma}{}_{\gamma}} \partial_{\nu}{h_{\sigma \lambda}} h^{\gamma \lambda} 
\nn \\ && 
+ \frac12 \partial^{\nu}{h} \partial_{\nu}{h_{\rho \sigma}} h^{\rho \sigma} + \frac14 \partial_{\mu}{h} \partial_{\nu}{h} h^{\mu \nu} + \frac12 \partial^{\nu}{h^{\sigma}{}_{\gamma}} \partial_{\sigma}{h_{\nu \lambda}} h^{\gamma \lambda} 
\nn \\ &&
+ \partial^{\nu}{h^{\sigma}{}_{\gamma}} \partial_{\lambda}{h_{\nu \sigma}} h^{\gamma \lambda} - \partial^{\nu}{}_{\rho}{h_{\nu \sigma}} h h^{\rho \sigma} + \frac12 \Box{h_{\rho \sigma}} h h^{\rho \sigma} 
\nn \\ &&
- \frac12 \partial^{\nu}{h_{\nu}{}^{\sigma}} \partial^{\lambda}{h_{\sigma \lambda}} h + \frac12 \partial_{\mu \nu}{h} h h^{\mu \nu} + \frac12 \partial^{\nu}{h} \partial^{\sigma}{h_{\nu \sigma}} h 
\nn \\ &&
+ \frac{3}{8} \partial^{\nu}{h^{\sigma \lambda}} \partial_{\nu}{h_{\sigma \lambda}} h - \frac{1}{8} \partial^{\nu}{h} \partial_{\nu}{h} h - \frac14 \partial^{\nu}{h^{\sigma \lambda}} \partial_{\sigma}{h_{\nu \lambda}} h 
\nn \\ &&
+ \frac{1}{8} \partial^{\nu \sigma}{h_{\nu \sigma}} h^{2} - \frac{1}{8} \Box{h} h^{2} - \frac14 \partial^{\nu \sigma}{h_{\nu \sigma}} h_{\gamma \lambda} h^{\gamma \lambda} 
\nn \\ &&
+ \frac14 \Box{h} h_{\rho \sigma} h^{\rho \sigma}
\eea

and in the TT-gauge the previous expression simplifies and we obtain
\bea
L^{(3)}_{EH}= \frac14 \partial^{\mu} h_{\rho \beta} \partial^{\alpha} h^{\rho \beta} h_{\mu \alpha} - \frac12 \partial^{\mu}h_{\nu \beta} \partial^{\beta} h^{\nu \alpha} h_{\mu \alpha} - \frac14 \Box h^{\sigma}{}_{\gamma} h_{\sigma \lambda} h^{\gamma \lambda} \, .
\eea  
We can see an extra term in the Einstein-Hilbert Lagrangian, proportional to the linearized Ricci scalar
\bea
- \frac14 \Box h^{\sigma}{}_{\gamma} h_{\sigma \lambda} h^{\gamma \lambda} = \frac{1}{4} R^{(1)\sigma}{}_{\gamma} h_{\sigma \lambda} h^{\gamma \lambda} \, .
\eea
Since the quadratic Lagrangian is given by
\bea
L^{(2)}_{EH} = R^{(1)}_{\mu \nu} h^{\mu \nu}
\eea
we can eliminate all the cubic terms proportional to the linearized Ricci tensor by using a redefinition in the perturbation. Therefore, using the TT-gauge, we can explicitly prove the equivalence between Einstein-Hilbert and Yang-Mills (using the off-shell double copy map) up to cubic order in perturbations.

In the next section we will use the same HJP technique, but our starting point will be a higher-derivative theory (DFDF) which its double copy map is related to Weyl gravity.

\subsection{Part C: Off-shell Double Copy for DFDF}
\label{C1}
\subsubsection{LMR: Quadratic Theory}
Now we proceed to study the DFDF theory under the off-shell double copy prescription.
It was demonstrated that, at the level of amplitudes, the double copy of  \cite{Part2-23}-\cite{Part2-24} 
\be
\label{actionhd}
{\cal L} = \frac{1}{2}\kappa_{ab}D_{\mu}F^{\mu\nu a}D_{\rho}F^{\rho}{}_{\nu}{}^{b}\, ,
\ee
its related to Weyl gravity. The main idea of this section is to explore if this resulting double copy presents a structure that can be interpreted from the double geometry framework and in an off-shell program \cite{Part2-25}-\cite{Part2-27}.

Considering the gauge covariant derivative defined as
\be
D_{\rho}F_{\mu\nu}{}^{a} = \partial_{\rho}F_{\mu\nu}{}^{a} + g_{\rm{YM}}f^{a}{}_{bc}A_{\rho}{}^{b}F_{\mu\nu}{}^{c}.
\ee
The expansion of the action \eqref{actionhd} up to quadratic terms and its subsequent integration by parts becomes
\be
\label{quadratic_actionhd}
S^{(2)}_{\rm{HD}} = \frac{1}{2}\int d^{D}x\kappa_{ab}\square A^{\mu a}\left(\square A_{\mu}{}^{b} - \partial_{\mu}\partial^{\nu}A_{\nu}{}^{b}\right)\, .
\ee
This expression is interesting because it contains the quadratic expansion of the pure Yang-Mills action. Going to momentum space it becomes
\be
\label{HD_momentum}
S_{\rm{HD}}^{(2)} = - \frac{1}{2}\int_{k}\,\kappa_{ab}\,k^{4} \,\Pi^{\mu\nu}(k) A_{\mu}{}^{a}(-k) A_{\nu}{}^{b}(k)\; ,  
\ee
After imposing the double copy relations \eqref{DcFields} and \eqref{DCCartan} on the previous action, we obtain
\be
\label{quadraticHD_DoubleCopy}
S_{\rm{HD/DC}}^{(2)} = - \frac{1}{4}\int_{k,\bar{k}} k^4\,\Pi^{\mu\nu}(k)  \bar{\Pi}^{\bar{\mu}\bar{\nu}}(\bar{k}) e_{\mu\bar{\mu}}(-k,-\bar{k})e_{\nu\bar{\nu}}(k,\bar{k}), 
\ee
which, not surprisingly, takes the same form as the Yang-Mills case except for the additional $k^{2}$ contribution. However, this additional contribution avoids the emergence of non-local terms in the action as we can observe after expanding the projectors
\begin{align}
S_{\rm{HD/DC}}^{(2)} = & - \frac{1}{4}\int_{k,\bar{k}}  k^{2}\Big(k^{2} e^{\mu\bar{\nu}}e_{\mu\bar{\nu}} - k^{\mu} k^{\rho} e_{\mu\bar{\nu}}e_{\rho}{}^{\bar{\nu}}  \nn \\ 
& - \bar{k}^{\bar{\nu}}\bar{k}^{\bar{\sigma}} e_{\mu\bar{\nu}}e^{\mu}{}_{\bar{\sigma}} + \frac{1}{k^{2}}k^{\mu}\bar{k}^{\bar{\nu}}k^{\rho}\bar{k}^{\bar{\sigma}}e_{\mu\bar{\nu}}e_{\rho\bar{\sigma}}\Big)\;, 
\end{align}
and hence the introduction of auxiliary fields is not necessary. Transforming the last expression to doubled position space we find the following higher-derivative contributions
\begin{align}
\label{DFDF_DFT}
S_{\rm{HD/DC}}^{(2)} = & - \frac{1}{4}\int d^{D}x d^{D}\bar{x}\Big[\Box e^{\mu\bar{\nu}}\Box e_{\mu\bar{\nu}} - \Box e^{\mu\bar{\nu}}\partial_{\mu}\partial^{\rho}e_{\rho\bar{\nu}}\, \nn \\ 
& - \Box e^{\mu\bar{\nu}}\bar{\partial}_{\bar{\nu}}\bar{\partial}^{\bar{\sigma}}e_{\mu\bar{\sigma}} + \partial^{\mu}\bar{\partial}^{\bar{\nu}}e_{\mu\bar{\nu}}\partial^{\rho}\bar{\partial}^{\bar{\sigma}}e_{\rho\bar{\sigma}}\Big] .
\end{align}
This action can be understood as a higher-derivative extension of DFT with conformal symmetry in the double space (in a classical sense). To clarify this point, we are going to explore the conformal gravity action
up to quadratic order for $g_{\mu\nu}=\eta_{\mu\nu} + h_{\mu\nu}$; we obtain
\bea
\label{quadratic_weyl}
S_{\rm{CG}}^{(2)} & = & \frac{D-3}{D-2}\int d^{D}x\left[\left(\Box h^{\mu\nu}\Box h_{\mu\nu} - 2\Box h^{\mu\nu}\partial_{\mu}\partial^{\rho}h_{\rho\nu} \right. \right. \nn \\  
&+& \left. \left. \partial^{\mu}\partial^{\nu}h_{\mu\nu}\partial^{\rho}\partial^{\lambda}h_{\rho\lambda}\right) - \frac{1}{D-1}\left(\Box h - \partial_{\mu}\partial_{\nu}h^{\mu\nu}\right)^{2}\right] \, . 
\eea

The second term in \eqref{quadratic_weyl} can be removed by imposing a particular gauge fixing condition $\Box h = \partial_{\mu}\partial_{\nu}h^{\mu\nu}$, related to dilatation symmetry $\delta h_{\mu\nu} = -2\lambda_{D}\eta_{\mu\nu}$. It is straightforward to prove that, after setting $x=\bar{x}$ and properly rescaling the metric (and/or considering a particular volume for the double space when we integrate), in the pure gravity case ($e_{\mu\bar{\nu}}\sim h_{\mu\nu}$) the action \eqref{DFDF_DFT} reduces to \eqref{quadratic_weyl}.

While so far the results are promising, we would like to obtain the full quadratic action beyond any gauge fixing condition. In order to address this limitation we will couple a charged scalar field to the initial gauge theory. The full gauge model that we will consider is given by \cite{Part2-23}-\cite{Part2-24},
\bea
{\cal L} & = & a_{1}\kappa_{ab}D_{\mu}F^{\mu\nu a}D_{\rho}F^{\rho}{}_{\nu}{}^{b} + a_{2}\kappa^{\alpha\beta}D_{\mu}\phi_{\alpha}D^{\mu}\phi_{\beta}\,  \nn \\ 
& & + a_{3}f_{abc}F_{\mu}{}^{\nu a}F_{\nu}{}^{\lambda b}F_{\lambda}{}^{\mu c} + a_{4}C^{\alpha}{}_{ab}\phi_{\alpha}F_{\mu\nu}{}^{a} F^{\mu\nu b}\, \nn \\
& & + a_{5}d^{\alpha\beta\gamma}\phi_{\alpha}\phi_{\beta}\phi_{\gamma}\, ,
\label{fullgauge_sect_2}
\eea
where the $a_{i}$ are real coefficients to be determined. The scalar field transforms under the action of the same gauge group but in the real representation, denoted by color indices $\alpha,\beta,\dots$, and it is allowed to interact with the gauge field $A_{\mu}{}^{a}$. This model is related to the DC of Weyl gravity at the level of scattering amplitudes. The coefficients $C_{\alpha ab}$ play the role of Clebsch–Gordan coefficients, mediating the coupling between the scalar and the gauge sector, while $d_{\alpha \beta \gamma}$ is a totally symmetric tensor. These objects are implicitly defined through the relations
\bea
C^{\alpha ab}C^{\alpha cd} & = & f^{ace}f^{ebd} + f^{ade}f^{ecb}\, ,  \nn \\
C^{\alpha ab}d^{\alpha\beta\gamma} & = & (T_{R}^{a})^{\beta\alpha}(T_{R}^{b})^{\alpha\gamma} + C^{\beta ac}C^{\gamma cb} + (a\leftrightarrow b)\, , \nn
\eea
which ensure consistency with the gauge symmetry and determine the structure of the interactions in terms of the underlying group theory.
Starting from the theory proposed in \eqref{fullgauge_sect_2}, the only source of quadratic contributions comes from the kinetic terms of the two fundamental fields. Specifically, we have:
\be
{\cal L}^{(2)} = 4a_{1}\kappa_{ab}\partial_{\mu}\partial^{[\mu}A^{\nu]a}\partial^{\rho}\partial_{[\rho}A_{\nu]}{}^{b} + a_{2}\kappa^{\alpha\beta}\partial_{\mu}\phi_{\alpha}\partial^{\mu}\phi_{\beta}\, .
\ee
We integrate by parts and then transform this action into momentum space, yielding:
\bea
S_{\mathrm{DC}}^{(2)} & = & - \int d^{D}k\left[a_{1}k^{4}\kappa_{ab}\Pi^{\mu\nu}(k)A_{\mu}{}^{a}(k)A_{\nu}^{b}(-k)\right.\, \nn \\ 
& & \ \ \ \ \ \ \ \ \ \ \ \ \ \left. + a_{2}k^{2}\kappa^{\alpha\beta}\phi_{\alpha}(k)\phi_{\beta}(-k)\right]\, .
\eea
At this stage, it is necessary to identify the objects associated with the kinetic term of the scalar field to complete the DC prescription for our model. This leads to
\bea
\label{id1}
\phi_{\alpha}(k) & \longrightarrow & k_{\mu}e^{\mu\bar{\nu}}(k,\bar{k}) - \bar{k}^{\bar{\nu}}\Phi(k,\bar{k})\, , \\
\kappa^{\alpha \beta} & \longrightarrow & \frac{\bar{k}_{\bar{\mu}}\bar{k}_{\bar{\nu}}}{k^{2}}\, ,
\label{id2}
\eea
where $\Phi(k,\bar{k})$ is the original scalar field incorporated by HJP. The identification of the scalar field is strongly indicated by the missing quadratic terms which are essential to completing the quadratic contributions of conformal gravity without relying on a specific gauge.

The quadratic DC Lagrangian, after performing all the identifications, is given by
\bea
\label{DFDF_DFT}
S_{\rm{DC}}^{(2)} & = & - \frac{1}{2}\int d^{D}x d^{D}\bar{x}\left[a_{1}\left(\Box e^{\mu\bar{\nu}}\Box e_{\mu\bar{\nu}} - \Box e^{\mu\bar{\nu}}\partial_{\mu}\partial^{\rho}e_{\rho\bar{\nu}}\right.\right.\, \nn \\ 
& & - \left. \Box e^{\mu\bar{\nu}}\bar{\partial}_{\bar{\nu}}\bar{\partial}^{\bar{\sigma}}e_{\mu\bar{\sigma}} + \partial^{\mu}\bar{\partial}^{\bar{\nu}}e_{\mu\bar{\nu}}\partial^{\rho}\bar{\partial}^{\bar{\sigma}}e_{\rho\bar{\sigma}}\right)\, \nn \\ 
& & \left. - 2a_{2}\left(\partial_{\mu}\partial_{\bar{\nu}}e^{\mu\bar{\nu}} - \Box\Phi\right)^{2}\right]\, .
\eea
This action now includes new contributions coming from the DC of the scalar field $\phi_{\alpha}$. We consider then the pure gravity limit, which demands to identify the coordinates $x$ and $\bar{x}$. The previous action becomes
\bea
S_{\rm{DC}}^{(2)} & = & - \frac{1}{2}\int d^{D}x\Big[a_{1}\left(\Box h^{\mu\nu}\Box h_{\mu\nu} - 2\Box h^{\mu\nu}\partial_{\mu}\partial^{\rho}h_{\rho\nu} \right. \Big. \nn \\  
& & \left. \left. + \ \partial^{\mu}\partial^{\nu}h_{\mu\nu}\partial^{\rho}\partial^{\lambda}h_{\rho\lambda}\right) - 2a_{2}\left(\Box h - \partial_{\mu}\partial_{\nu}h^{\mu\nu}\right)^{2}\right]\, \nn ,
\eea
which present the same structure that the quadratic contributions of the Weyl gravity action. In fact, if we demand both actions to be the same, the coefficients $a_1,a_2$ are fixed as 
\bea
a_1 & = & -2\left(\frac{D-3}{D-2}\right) \, , \\
a_2 & = & -\frac{1}{(D-1)}\left(\frac{D-3}{D-2}\right)\, .
\label{coeffquadratic}
\eea

It is notable, though expected, that the action vanishes in $D=3$, as in three dimensions, the Weyl tensor is identically zero. The new result presented here is the emergence of terms with a non-zero coefficient $a_{2}$, which arise from the scalar field dynamics. 

\subsubsection{LR: Cubic Theory}

Once the quadratic contributions of our model have been obtained and the coefficients $a_{1}$ and $a_{2}$ have been calculated, ensuring that, in the pure gravity limit, the model reduces to Weyl gravity, we are ready to proceed with the calculation of the cubic theory. Now, we will use the full Lagrangian introduced in \eqref{fullgauge_sect_2}
\bea
{\cal L} & = & a_{1}\kappa_{ab}D_{\mu}F^{\mu\nu a}D_{\rho}F^{\rho}{}_{\nu}{}^{b} + a_{2}\kappa^{\alpha\beta}D_{\mu}\phi_{\alpha}D^{\mu}\phi_{\beta}\,  \nn \\ 
& & + a_{3}f_{abc}F_{\mu}{}^{\nu a}F_{\nu}{}^{\lambda b}F_{\lambda}{}^{\mu c} + a_{4}C^{\alpha}{}_{ab}\phi_{\alpha}F_{\mu\nu}{}^{a} F^{\mu\nu b}\, \nn \\
& & + a_{5}d^{\alpha\beta\gamma}\phi_{\alpha}\phi_{\beta}\phi_{\gamma}\, .
\eea
In this Lagrangian, $f_{abc}$ are the structure constants of the gauge algebra, corresponding to the adjoint representation in which the gauge field $A_{\mu}{}^{a}$ transforms. The scalar field $\phi^{\alpha}$ is charged under a real representation $R$ of the gauge group, whose generators are denoted by $(T_{R}^{a})^{\alpha \beta}$, and are assumed to be real and antisymmetric in its last two indices. The indices $\alpha, \beta, \gamma$ corresponds to the real representation of the gauge group and they have their own Cartan-Killing metric $\kappa_{\alpha \beta}$. The gauge generators satisfy
\bea
[T_{a},T_{b}] & = & i f_{ab}{}^{c} T_{c} \, , \\
\{T_{a},T_{b}\} & = & \frac{1}{N} \delta_{ab} + d_{ab}{}^{c} T_{c} \, ,
\eea
while the trace of three generators is given by
\bea
\textrm{Tr}(T_{a} T_{b} T_{c}) = \frac14(d_{abc}+if_{abc}) \, .
\label{trace}
\eea

The interaction terms involving the scalar field introduce additional group-theoretic tensors. The coefficients $C_{\alpha ab}$ play the role of Clebsch–Gordan coefficients, mediating the coupling between the scalar and the gauge sector, while $d_{\alpha \beta \gamma}$ is a totally symmetric structure constant. These objects are implicitly defined through the relations
\bea
C^{\alpha ab} C_{\alpha}{}^{cd} & = & f^{ace}f_{e}{}^{bd} + f^{ade} f_{e}{}^{cb}\, ,  \nn \\
C_{\alpha}{}^{ab} d^{\alpha\beta\gamma} & = & (T_{R}^{a})^{\beta\alpha}(T_{R}^{b})_{\alpha}{}^{\gamma} + C^{\beta a}{}_{c} C^{\gamma cb} + (a\leftrightarrow b)\, , \nn
\eea
which ensure consistency with the gauge symmetry and determine the structure of the interactions in terms of the underlying group theory.

The gauge covariant derivatives are defined as
\bea
D_{\rho}F_{\mu\nu}{}^{a} & = & \partial_{\rho}F_{\mu\nu}{}^{a} + gf^{a}{}_{bc}A_{\rho}{}^{b}F_{\mu\nu}{}^{c}\, , \nn \\
D_{\mu}\phi^{\alpha} & = & \partial_{\mu}\phi^{\alpha} - ig (T_{R}^{a})^{\alpha \beta} A_{\mu a} \phi_{\beta}\, \nn ,
\eea
with the Yang-Mills curvature defined in the standard way $F_{\mu\nu}{}^{a} = 2\partial_{[\mu}A_{\nu]}{}^{a} + gf^{a}{}_{bc}A_{\mu}{}^{b}A_{\nu}{}^{c}$.

Our goal in this section is to construct the gravitational cubic contributions arising from the DC prescription of this model, determining the remaining coefficients $a_3$, $a_4$, and $a_5$ in the same way as we did for the quadratic terms. While analyzing the full Lagrangian is a very ambitious task, we will focus on the case with vanishing generalized dilaton 
\be
\Phi(k,\bar{k}) = \frac{1}{k^{2}}k^{\mu}\bar{k}_{\bar{\nu}}e_{\mu}{}^{\bar{\nu}} = 0 \, .
\ee
 
To construct a minimal proposal for the cubic Lagrangian of CDFT, we identify the following quantities
\bea
\label{idf}
f_{abc} & \rightarrow & i \frac{a_6}{a_1}\left(\eta^{\bar \mu \bar \nu} \bar k_{12}^{\rho} + \eta^{\bar \nu \bar \rho}\bar k_{23}{}^{\bar \mu} + \eta^{\bar \rho \bar \mu} \bar k_{31}^{\bar \nu}\right)\, \\
\label{idC}
C_{\alpha ab} & \rightarrow &  i \bar{k}_{1}^{\bar{\mu}}\eta^{\bar{\nu}\bar{\rho}} \, \\
\label{idomega}
(T_{R}^{a})^{\alpha\beta}A_{\mu a} & \rightarrow & i \omega_{\mu \bar{\nu}\bar{\rho}}(k,\bar{k})\, , \\
\label{idd}
d_{\alpha \beta \gamma} & \rightarrow &  \frac{i}{\bar k^2} \bar k_{1}^{(\bar \mu} \bar k_{2}^{\bar \nu} k_{3}^{\bar \rho)}\, ,
\eea
where $k_{ij}=k_{i}-k_{j}$. In (\ref{idomega}) $\omega_{\mu\bar{\nu}\bar{\rho}}$ illustrates a natural correspondence between a gauge connection and a spin connection for the generalized vielbein. In the above identifications, the only object requiring an additional coefficient to fix its cubic contributions is $f_{abc}$. In this case, the coefficient $a_1$ is inherited from the cubic contribution of the $DFDF$ term, which needs to be canceled. The identifications (\ref{idf})-(\ref{idd}) can be further extended to a more general form in agreement with the index constrains that every object carries, which it would be convenient in a more general set up thanks to extra coefficients. For our proposes we will show that these identifications are enough to reproduce Weyl gravity in a particular (pure gravitational) limit. 
 
We observe that the Weyl tensor is non-trivial in $D>3$ and therefore $a_1^{-1}$ is well defined. The identification (\ref{idf}) follows the same structure as in HJP, but the inclusion of the $\frac{a_1}{a_6}$ coefficients introduces a global factor in this identification, which is crucial for reproducing Weyl gravity in arbitrary dimensions, as we will demonstrate in the next part of this section.

The minimal cubic DC Lagrangian with $\Phi=0$ is given by
\footnotesize
\bea
\label{Cubic_HD_DFT}
&& S_{\rm{DC}}^{(3)}|_{\Phi=0} = \int d^{D}x \ d^{D}\bar{x} (-\frac{a_6}{2}) \left[\Box\partial^{\rho}\bar{\partial}^{\bar{\sigma}}e_{\mu\bar{\nu}}e^{\mu\bar{\nu}}e_{\rho\bar{\sigma}} - \Box\bar{\partial}^{\bar{\sigma}}e_{\mu\bar{\nu}}\partial^{\rho}e^{\mu\bar{\nu}}e_{\rho\bar{\sigma}} + \Box\bar{\partial}^{\bar{\sigma}}e_{\mu\bar{\nu}}\partial^{\mu}e^{\rho\bar{\nu}}e_{\rho\bar{\sigma}} - \Box\bar{\partial}^{\bar{\sigma}}e_{\mu\bar{\nu}}e^{\rho\bar{\nu}}\partial^{\mu}e_{\rho\bar{\sigma}}\right.\, \nn \\
& &  - \Box e_{\mu\bar{\nu}}\partial^{\mu}\bar{\partial}^{\bar{\sigma}}e^{\rho\bar{\nu}}e_{\rho\bar{\sigma}} + \Box e_{\mu\bar{\nu}}\bar{\partial}^{\bar{\sigma}}e^{\rho\bar{\nu}}\partial^{\mu}e_{\rho\bar{\sigma}} - \partial^{\mu}\partial^{\rho}\partial^{\lambda}\bar{\partial}^{\bar{\sigma}}e_{\mu\bar{\nu}}e_{\rho}{}^{\bar{\nu}}e_{\lambda\bar{\sigma}} + \partial^{\mu}\partial^{\rho}\bar{\partial}^{\bar{\sigma}}e_{\mu\bar{\nu}}\partial^{\lambda}e_{\rho}{}^{\bar{\nu}}e_{\lambda\bar{\sigma}}\, \nn \\
& &  - \partial^{\rho}\partial^{\lambda}\bar{\partial}^{\bar{\sigma}}e_{\mu\bar{\nu}}\partial^{\mu}e_{\rho}{}^{\bar{\nu}}e_{\lambda\bar{\sigma}} + \partial^{\rho}\partial^{\lambda}\bar{\partial}^{\bar{\sigma}}e_{\mu\bar{\nu}}e_{\rho}{}^{\bar{\nu}}\partial^{\mu}e_{\lambda\bar{\sigma}} + \partial^{\rho}\partial^{\lambda}e_{\mu\bar{\nu}}\partial^{\mu}\bar{\partial}^{\bar{\sigma}}e_{\rho}{}^{\bar{\nu}}e_{\lambda\bar{\sigma}} - \partial^{\rho}\partial^{\lambda}e_{\mu\bar{\nu}}\bar{\partial}^{\bar{\sigma}}e_{\rho}{}^{\bar{\nu}}\partial^{\mu}e_{\lambda\bar{\sigma}}\, \nn \\
& & \left. - \bar{\partial}^{\bar{\nu}}\bar{\partial}^{\bar{\sigma}}\bar{\partial}^{\bar{\kappa}}\partial^{\rho}e_{\mu\bar{\nu}}e^{\mu}{}_{\bar{\sigma}}e_{\rho\bar{\kappa}} + \bar{\partial}^{\bar{\nu}}\bar{\partial}^{\bar{\sigma}}\bar{\partial}^{\bar{\kappa}}e_{\mu\bar{\nu}}\partial^{\rho}e^{\mu}{}_{\bar{\sigma}}e_{\rho\bar{\kappa}} + \bar{\partial}^{\bar{\nu}}\bar{\partial}^{\bar{\sigma}}e_{\mu\bar{\nu}}\bar{\partial}^{\bar{\kappa}}\partial^{\mu}e^{\rho}{}_{\bar{\sigma}}e_{\rho\bar{\kappa}} - \bar{\partial}^{\bar{\nu}}\bar{\partial}^{\bar{\sigma}}e_{\mu\bar{\nu}}\bar{\partial}^{\bar{\kappa}}e^{\rho}{}_{\bar{\sigma}}\partial^{\mu}e_{\rho\bar{\kappa}}\, \right]\, \nn \\ &&
 + a_{2} \partial^{\sigma}\partial_{\lambda} e^{\lambda \bar \epsilon} \omega^{(1)}_{\sigma \bar \epsilon \bar \tau} \partial_{\epsilon} e^{\epsilon \bar \tau} + 3 a_3 \left[\partial_{\mu} e^{\nu \bar \rho} \partial_{\bar \rho}\partial_{\nu}e^{\lambda}{}_{\bar \epsilon} \partial_{\lambda}e^{\mu \bar \epsilon} - \partial_{\mu} e^{\nu \bar \rho} \partial_{\nu}e^{\lambda}{}_{\bar \epsilon} \partial_{\bar \rho}\partial_{\lambda}e^{\mu \bar \epsilon} \right]  \nn \\ &&
 - 3 a_{3} \left[\partial_{\mu} \partial_{\bar \epsilon} e^{\nu \bar \rho} \partial_{\nu}e^{\lambda}{}_{\bar \rho} \partial^{\mu}e_{\lambda}{}^{\bar \epsilon} - \partial_{\mu} e^{\nu \bar \rho} \partial_{\bar \epsilon} \partial_{\nu}e^{\lambda}{}_{\bar \rho} \partial^{\mu}e_{\lambda}{}^{\bar \epsilon}\right] - 3 a_{3} \left[\partial_{\mu} e^{\nu \bar \rho} \partial_{\nu}\partial_{\bar \rho}e^{\lambda \bar \sigma} \partial^{\mu}e_{\lambda \bar \sigma} - \partial_{\mu} e^{\nu \bar \rho} \partial_{\nu}e^{\lambda \bar \sigma} \partial^{\mu}\partial_{\bar \rho}e_{\lambda \bar \sigma}\right]  
\nn \\ &&  
- 3 a_{3} \left[\partial_{\mu} e^{\nu \bar \rho} \partial_{\nu}e^{\lambda \bar \epsilon} \partial^{\mu}\partial_{\bar \epsilon}e_{\lambda \bar \rho} - \partial_{\mu} \partial_{\bar \epsilon} e^{\nu \bar \rho} \partial_{\nu}e^{\lambda \bar \epsilon} \partial^{\mu}e_{\lambda \bar \rho}\right] + a_{4} \partial_{\bar \mu}\partial_{\epsilon} e^{\epsilon \bar \mu} \partial_{[\sigma} e_{\lambda]}{}^{\bar \nu} \partial^{\lambda}e^{\sigma}{}_{\bar \nu} \, ,
\label{fullLagrangian}
\eea    
\normalsize
where the notation $\omega^{(1)}_{\mu \bar \mu \bar \nu}$ refers to the first order perturbation of the connection. It is possible to see that the Lagrangian \eqref{fullgauge_sect_2} produces terms proportional to the generalized dilaton associated to the coefficient $a_{5}$, and therefore this coefficient does not appear in this Lagrangian. To determine the remaining coefficients, we will compare (\ref{fullLagrangian}) with the cubic contributions of Weyl gravity. 

We will perform this analysis under the assumption of a vanishing wave equation $\Box h_{\mu \nu}=0$ and the traceless condition $h=0$. Given these conditions, the generalized dilaton will not contribute to the gravitational terms after parametrization. Furthermore, our method cannot fix the coefficient $a_5$, as its contributions depend solely on the generalized dilaton, which vanishes in this case.

In the next section, we will demonstrate that this method is sufficiently robust to uniquely determine the coefficients $a_3$, $a_4$ and $a_6$.

\subsubsection{Fixing the Coefficients when $\Phi=0$}

The cubic contributions for Weyl gravity are given by
\be
S_{\rm{CG}}^{(3)} = \int d^{D}x \left[\frac{1}{2}hC^{(1)}_{\mu\nu\rho\lambda}C^{(1)\mu\nu\rho\lambda} - 4h^{\mu\nu}C^{(1)}_{\mu\rho\lambda\sigma}C^{(1)}_{\nu}{}^{\rho\lambda\sigma} + 2C^{(2)}_{\mu\nu\rho\lambda}C^{(1)\mu\nu\rho\lambda}\right]\, .
\ee
Imposing the conditions  
\be
\Box h_{\mu \nu}=h=0 \, ,
\label{conditions}
\ee
which are fully compatible with $\Phi=0$, and performing a series of integrations by parts, the cubic Lagrangian can be reduced to the following two contributions,
\bea
L^{(3)} & = &  f(D)\partial_{\delta}\partial_{\lambda}h^{\epsilon \lambda}\partial^{\nu}h_{\nu \sigma}\partial_{\epsilon}h^{\sigma \delta}\, , \nn \\ 
& & + g(D)\partial_{\mu}\partial_{\sigma}h^{\mu}{}_{\epsilon} \partial_{\delta}\partial_{\lambda}h^{\sigma \epsilon}h^{\delta \lambda}\, ,
\eea
where 
\bea
f(D) & = &\frac{-2(D-3)(11D+15)}{5(D-2)(D-1)} \, , \nn \\
g(D) & = & \frac{(D-3)(34D+5)}{5(D-2)(D-1)} \, . 
\eea
As before, we find that $D=3$ is the only physically meaningful root of both $f(D)$ and $g(D)$, consistent with the physical interpretation of the Weyl tensor. When the double conformal model (\ref{fullLagrangian}) is parametrized and restricted to the pure gravitational limit, it takes the following form
\bea
L^{(3)}_{DC} = && -(\frac{a_6}{4} + 3 a_3) \partial_{\delta} \partial_{\lambda}h^{\epsilon \lambda} \partial^{\nu}h_{\nu \sigma} \partial_{\epsilon}h^{\sigma \delta} \nn \\ && + \frac{15}{4} a_6 \partial_{\mu} \partial_{ \sigma}h^{\mu}{}_{ \epsilon} \partial_{\delta}\partial_{\lambda}{h^{\sigma \epsilon}} h^{\delta \lambda} \nn \\ &&+(-a_6+ \frac{a_4}{2}) \partial_{\mu}\partial_{ \epsilon}h^{\mu \epsilon}\partial_{\sigma}h_{\lambda \nu} \partial^{\nu}h^{\sigma \lambda}) \, .
\eea
Therefore, the coefficients are uniquely fixed as
\bea
a_3= - \frac{1}{45} g(D) - \frac{1}{3}f(D) \, , \\
a_4 = 2 a_{6} = \frac{8}{15} g(D) \, \, .  
\label{cubicfix}
\eea
The preceding identifications determine the double action (\ref{fullLagrangian}), which represents the full cubic CDFT with a vanishing generalized dilaton. In \eqref{cubicfix}, we observe the necessity of the $a_{6}$ coefficient in identifying the structure constants. Additionally, we note that both the $F^{3}$ term and the interaction term between the gauge field and the scalar field are essential for reproducing Weyl gravity in this limit.

\newpage

\subsection{Exercises}

\subsubsection{Part A}
\begin{enumerate}
\item Consider the relaxed form of the Kerr-Schild ansatz with a null vector $l^{\mu}$, but without imposing the geodesic equation. Prove that the connection is shifted by a tensor, by explicitly obtaining the tensor $\tilde \Gamma_{\mu \nu}^{\rho}$.

\item Prove that in the relaxed form of the Kerr-Schild de Ricci tensor contains cubic contributions which are non-vanishing, but the cubic contributions of the Ricci scalar.

\item Prove that the quadratic terms in the EOM are proportional to the linear terms, i.e.,
\bea
R^{(2)}_{\nu \sigma}= \varphi l_{(\sigma|} l^{\rho} R^{(1)}_{\rho |\nu)} \, . \nn
\eea

\item Let's explore the family of gravitational solutions that can be obtained by considering a relaxed Kerr-Schild ansatz (null condition) but also satisfying $\nabla_{\mu}l^{\nu}=0$ and $l^{\mu} \nabla_{\mu} \varphi=0$, which is stronger than geodesic condition. Find the explicit form of the linear equation of motion contracted with the Killing vector.

\item One example of the previous family of solutions is the pp-wave in Brinkmann (light-cone) coordinates $(u,v,x,y)$ where the propagation is in the $v$ direction and the full metric is
\bea
ds^2 = 2 du dv + dx^2 + dy^2 + \varphi(u,x,y) du^2 \, . \nn
\eea
The first three terms are a Minkowski metric (constant background) and its inverse metric, $\eta^{\mu \nu}$ has the following non-vanishing components:
\begin{equation}
\eta^{uv} = \eta^{vu} = 1, \qquad \eta^{xx} = \eta^{yy} = 1 . \nn
\end{equation}
Using $l^{\mu}=(0,1,0,0)$ and $l_{\mu}=(1,0,0,0)$, prove that the equation of motion contracted with the Killing vector reduces to a 2D Laplace equation, 
\bea
\partial_{x}^2 \varphi + \partial_{y}^2 \varphi = 0 \, .
\eea

\item Find the condition that a Kerr-Schild solution requires to satisfy $F^2=0$ after imposing the single copy map. Does the pp-wave satisfy this condition?
\end{enumerate}

\subsubsection{Part B}
\begin{enumerate}
\item Show that the quadratic YM action can be written as 
 \be
  S_{\rm YM}^{(2)} = -\frac{1}{2} \int_{k}\, d^{D}k \kappa_{ij}\,k^2 \,\Pi^{\mu\nu}(k) A_{\mu}{}^{i}(-k) A_{\nu}{}^{j}(k)\; , \nn
 \ee
 in momentum space.

 \item Expand the double copy action
  \be
  S_{\rm DC}^{(2)} = 
  -\frac{1}{4} \int_{k,\bar{k}} k^2\,   \Pi^{\mu\nu}(k)  \bar{\Pi}^{\bar{\mu}\bar{\nu}}(\bar{k}) e_{\mu\bar{\mu}}(-k,-\bar{k}) 
  e_{\nu\bar{\nu}}(k,\bar{k})\,, \nn  
 \ee
and transform it back to configuration space to finally obtain (\ref{YMsquaredConstrained}).
\item Let's include an extra scalar field in the identification of $\phi$. In other words, use $\phi = h + a \varphi$ (a arbitrary) and show that the quadratic double copy procedure of HJP produce a kinetic term for the new scalar field $\varphi$ if we impose the TT-gauge at this order.

\item While the natural identification for the structure constant is the 3-vertex operator,
\begin{equation}
f_{abc} \; \to \;  \tfrac{i}{4} \,
\ov\Pi^{\ov\mu\ov\nu\ov\rho}(\ov k_{1},\ov k_{2},\ov k_{3})\;. 
\end{equation}
there are other possibilities for this identification. Assign arbitrary coefficients to them and construct the most general identification for the structure constant. 

\item Prove that it is possible to match cubic Einstein-Hilbert from cubic Yang-Mills by imposing the double copy map and the Harmonic gauge.
\end{enumerate}

\subsubsection{Part C}
\begin{enumerate}
\item Show that the quadratic DFDF action can be written as 
 \be
   S_{\rm{HD}}^{(2)} = - \frac{1}{2}\int_{k}\,\kappa_{ab}\,k^{4} \,\Pi^{\mu\nu}(k) A_{\mu}{}^{a}(-k) A_{\nu}{}^{b}(k)\; , \nn
 \ee
 in momentum space.

\item Expand the double copy action
\be
S_{\rm{HD/DC}}^{(2)} = - \frac{1}{4}\int_{k,\bar{k}} k^4\,\Pi^{\mu\nu}(k)  \bar{\Pi}^{\bar{\mu}\bar{\nu}}(\bar{k}) e_{\mu\bar{\mu}}(-k,-\bar{k})e_{\nu\bar{\nu}}(k,\bar{k}), \nn 
\ee
and transform it back to configuration space to finally obtain (\ref{DFDF_DFT}).

\item Let's include an extra scalar field in the identification of $\Phi$. In other words, use $\phi = h + a \varphi$ (a arbitrary) and show kind of higher-derivative Lagrangian produces the quadratic double copy procedure of LR.

\item While the natural identification for the symmetric structure constant is given by,
\begin{equation}
d_{abc} \; \to \;  \tfrac{i}{k^2} \,
\ov k_{1}^{(\mu} \ov k_{2}^{\nu} \ov k_{3}^{\rho)}\;. 
\end{equation}
there are other possibilities for this identification. Assign arbitrary coefficients to them and construct the most general identification for the symmetric structure constant.
\end{enumerate}

\newpage

\section{Lecture 3: T-duality Invariant Frameworks and Their Relation to the Single and Double Copy}
\label{Sec3}
The idea of these lectures is to show the formal bridge that exist between gauge theories and gravitational theories. In the lecture II, part A, we use the Kerr-Schild ansatz in Einstein-Hilbert gravity and after multiplying by a particular killing vector, we manage to obtain the Maxwell equation. In lecture II part B, we reverse the logic, and starting by Maxwell (quadratic Yang-Mills) we obtain quadratic Einstein-Hilbert gravity in its TT-gauge form, and starting from cubic Yang-Mills we obtained cubic Einstein-Hilbert under the same gauge. However, behind the double copy map there is more than gravity. There is matter. In this section we will show that
\bea
- \frac{1}{4} Tr \int_{x} (F_{\mu \nu} F^{\mu \nu})^{(2,3)} \rightarrow \int_{x} \sqrt{-g}(R + 4 \partial_{\mu} \phi \partial^{\mu}\phi - \frac{1}{12} H_{\mu \nu \rho} H^{\mu \nu \rho})^{(2,3)} 
\label{NSNS}
\eea
where $\phi$ and $H_{\mu \nu \rho}$ can be interpreted as matter fields. Particularly, $\phi$ is a scalar field called "the dilaton" and $H_{\mu \nu \rho}$ is the curvature of the two-form $b_{\mu \nu}=-b_{\nu \mu}$ (called the Kalb-Ramond field) defined as
\bea
H_{\mu \nu \rho} = 3 \partial_{[\mu} b_{\nu \rho]} \, .
\eea
The previous action can be obtained as the low energy limit of close string theory, and it is known as the NSNS supergravity (even if there are no fermions involved, we call it supergravity). In the next parts we will explore this theory, its T-duality rewriting and its relation to the single and double copy program. 

\subsection{Part A: Introduction to Supergravity}
\label{A2}
\subsubsection{Action Principle}
The low-energy effective action depends on the string formulation under consideration. The superstrings (type II A or type II B) contain bosons and fermions in their spectrum and this is the main difference with respect to the bosonic string that only contains bosons. Schematically, both type-II effective Lagrangians can be written as,
\bea
L_{\rm{type-II}} = L_{\rm NS-NS} + L_{\rm R-R} + L_{\rm NS-R} + L_{\rm R-NS} \, .
\label{typeII0}
\eea
The first term, $L_{\rm NS-NS}$ , encodes the dynamics of the full Lagrangian that we have described perturbatively in (\ref{NSNS}) for $D=10$ (critical dimension). The second term of (\ref{typeII0}), $L_{\rm R-R}$, describes the dynamics of a 1-form $C_{1}$ and a 3-form $C_{3}$ for type IIA superstring and a zero-form $C_{0}$, a two form $C_{2}$ and a four-form $C_{4}$ for type IIB superstring. Both $L_{\rm NS-NS}$ and $L_{\rm R-R}$ are bosonic Lagrangians, meaning that they depend on bosonic quantities. On the other hand, both $L_{\rm NS-R}$ and $L_{\rm R-NS}$ depend on fermionic degrees of freedom, related to the bosonic spectrum through ${\cal N}=2$ supersymmetry. The Lagrangian (\ref{typeII0}) is our first (and last) example of a truly supergravity Lagrangian, since it contains a dynamical metric tensor and fermionic degrees of freedom, both of them related by supersymmetry. Indeed, the low energy limit of all superstring theory formulations is given by a supergravity. Furthermore, supergravity theories in arbitrary dimensions can be studied independently of string theory. It turns out that the first higher-derivative corrections for the type II superstrings consist of eight-derivative terms (for example $Riem^4$), so neither $\alpha'$ nor $\alpha'^2$ corrections are present in this theory. 

Last but not least, we present heterotic string theory. This string is a hybrid formulation. To construct it we have to consider a left-right decomposition for the closed string oscillation. This is analogous to a traveling wave decomposition, where left and right are the directions of propagation of the waves on the closed string. The heterotic string contains a $D=26$ bosonic string formulation on its left part, compactified on a $16$-dimensional torus (for dimensional consistency), and a $D=10$ superstring formulation on its right side. The 10-dimensional bosonic and massless degrees of freedom are a metric tensor, a 2-form, a dilaton and a non-Abelian gauge field $A_{\mu i}$. The index $i$ is a gauge index that runs from $1,\dots,n$ with $n$ the dimension of the gauge group, $E_{8}\times E_{8}$ or $SO(32)$, depending on the heterotic formulation.

This superstring theory is invariant under ${\cal N}=1$ supersymmetry so it contains fermionic degrees of freedom besides the former fields. The effective action is given by a heterotic supergravity,      
\bea
S_{\rm{het}} = \int d^{D}x \sqrt{-g} e^{-2 \phi} (R + 4 \partial_{\mu} \phi \partial^{\mu}\phi - \frac{1}{12} \bar{H}_{\mu \nu \rho} \bar{H}^{\mu \nu \rho} - \frac14 F_{\mu \nu}{}^{i} F^{\mu \nu}{}_{i} + L_{f}) \, ,
\label{het0}
\eea
where $L_{f}$ are two-derivative fermionic terms and\footnote{We use the convention $T_{[\mu \nu]} = \frac12(T_{\mu \nu} - T_{\nu \mu})$ and $T_{(\mu \nu)} = \frac12(T_{\mu \nu} + T_{\nu \mu})$.}
\bea
F_{\mu \nu}{}^{i} = 2 \partial_{[\mu} A_{\nu]}{}^{i} - f^{i}{}_{jk} A_{\mu}{}^{j} A_{\nu}{}^{k} \, ,
\eea
is the gauge field curvature. In the previous expression gauge indices are contracted using a constant Cartan-Killing metric $\kappa_{i j}$ with inverse $\kappa^{i j}$. The first higher-derivative correction to this action principle is a correction of order $\alpha'$. We will discuss the form of these terms in the next section.  

An important difference with respect to the previous formulations is that the curvature of the $b$-field contains extra terms,
\be
\bar H_{\mu\nu\rho}=3\left(\partial_{[\mu}b_{\nu\rho]}-C_{\mu\nu\rho}^{(g)}\right)\, , \label{barH}
\ee 
where $C_{\mu\nu\rho}^{(g)}$ is a so-called Chern-Simons 3-form, defined as
\be
C_{\mu\nu\rho}^{(g)}=A^i_{[\mu}\partial_\nu A_{\rho]i}-\frac13 f_{ijk}A_\mu^i A_\nu^jA_\rho^k \, . 
\ee
The inclusion of these extra contributions will be clear when we inspect the invariance of the action.  

\subsubsection{Symmetry Transformations in Bosonic/Type II Supergravity}
Here we discuss the symmetry rules which leave the universal action  invariant,
\bea
S = \int d^{D}x \sqrt{-g} e^{-2 \phi} (R + 4 \partial_{\mu} \phi \partial^{\mu}\phi - \frac{1}{12} H_{\mu \nu \rho} H^{\mu \nu \rho}) \, .
\label{univ}
\eea
This action is common to all close string formulations, and it is invariant under:
\begin{itemize}
\item {\bf Infinitesimal diffeomorphisms:} The fundamental fields transform as \footnote{We use the convention that contracted indices are not part of the (anti-)symmetrizers.}
\bea
\delta_{\xi} g_{\mu \nu} & = & L_{\xi} g_{\mu \nu} = \xi^{\rho} (\partial_{\rho} g_{\mu \nu}) + 2(\partial_{(\mu} \xi^{\rho}) g_{\rho \nu)} \, , \\
\delta_{\xi} b_{\mu \nu} & = & L_{\xi} b_{\mu \nu} = \xi^{\rho} (\partial_{\rho} b_{\mu \nu}) + 2(\partial_{[\mu} \xi^{\rho}) b_{\rho \nu]} \, , \\
\delta_{\xi} \phi & = & L_{\xi} \phi = \xi^{\rho} (\partial_{\rho} \phi) \, .
\eea

Each term of the Lagrangian
\be
L = e^{-2\phi}(R + 4 \partial_\mu \phi \partial^{\mu}\phi - \frac{1}{12} H_{\mu \nu \lambda} H^{\mu \nu \lambda})
\label{Lag}
\ee
transforms as a scalar (with $\omega=0$). Then the universal action (\ref{univ}) is invariant up to total derivatives.

\item {\bf Abelian gauge transformations:} This transformation only acts on the $b$-field due to its 2-form nature,
\be
\delta_{\zeta} b_{\mu \nu} = 2 \partial_{[\mu} \zeta_{\nu]} \, ,
\ee
where $\zeta$ is an arbitrary parameter. Since 
\bea
\delta_{\zeta} H_{\mu \nu \rho} = 3 \partial_{[\mu}(\delta b_{\nu \rho]}) = 6 \partial_{[\mu}(\partial_{\nu} \zeta_{\rho]}) = 0 \, ,
\eea
the Lagrangian (\ref{Lag}) is gauge invariant.

\subsubsection{Symmetry Transformations in Heterotic Supergravity}

The low-energy heterotic effective Lagrangian,
\be
L_{\rm{het}} = e^{-2\phi}(R + 4 \partial_\mu \phi \partial^{\mu}\phi - \frac{1}{12} \bar H_{\mu \nu \lambda} \bar H^{\mu \nu \lambda} - \frac14 F_{\mu \nu}{}^i F^{\mu \nu}{}_{i}) \, ,
\label{LagHet}
\ee
is invariant under the symmetries previously described since $A_{\mu i}$ transforms as a 1-form under infinitesimal diffeomorphisms, \textit{i.e.},
\bea
\delta_{\xi} A_{\mu i} = L_{\xi} A_{\mu i} = \xi^{\rho} (\partial_{\rho} A_{\mu i}) + (\partial_{\mu} \xi^{\rho}) A_{\rho i} \, ,
\eea
and is invariant under Abelian gauge transformations. The Lagrangian (\ref{LagHet}) is also invariant under non-Abelian gauge transformations.

\item{\bf Non-Abelian gauge transformations:}
The non-Abelian gauge transformations act on $A_{\mu i}$ and $b_{\mu \nu}$ in the following way,
\bea
\delta_{\lambda}A_\mu^i & = & \partial_\mu \lambda^i+f^i{}_{jk}\lambda^jA_\mu^k\, , \\ \delta_{\lambda} b_{\mu\nu} & = & -(\partial_{[\mu} \lambda^i) A_{\nu]i}\, , \label{gauge0}
\eea
where $\lambda^{i}$ is an arbitrary parameter. Considering an arbitrary gauge vector $v^{i}$, the partial derivative of this vector is not covariant and we need to define a covariant derivative for this symmetry in the following way,
\bea
\nabla_{\mu} v^{i} = \partial_{\mu} v^{i} - f^{i}{}_{jk} A_{\mu}{}^{j} v^{k} \, .
\eea 
Here we use the same notation $\nabla$ for the gauge covariant derivative as in the first lecture, so our convention is that $\nabla$ covariantizes the derivative of an object with respect to all the symmetries that the object transforms under. On the other hand,  $F_{\mu \nu i}$ transforms covariantly under non-Abelian gauge transformations
\bea
\delta_{\lambda} F_{\mu \nu i} = f_{ijk} \lambda^{j} F_{\mu \nu}{}^{k} \, , 
\eea
unlike $A_{\mu i}$ whose transformation is not covariant due to the presence of the $\partial_{\mu}\lambda^i$ term. Moreover, we need $\delta \bar H_{\mu \nu \rho}=0$ to ensure the non-Abelian gauge invariance of the effective action. Now it is clear the importance of the Chern-Simons terms in (\ref{barH}),
\bea
\delta_{\lambda} C^{(g)}_{\mu \nu \rho} = \partial_{[\mu}(A_{\nu}{}^{i} \partial_{\rho]} \lambda_{i}) \, .
\eea
The previous transformation exactly cancels the $3 \partial_{[\mu}(\delta_{\lambda} b_{\nu \rho]})$ contribution and therefore (\ref{het0}) is non-Abelian gauge invariant. 
\end{itemize}

\subsubsection{Introduction to Double Field Theory}

As we mentioned in the previous lecture, the different formulations of string theory are defined on a $D$-dimensional manifold. From a phenomenological point of view, this $D$-dimensional target space is often divided
into an external non-compact space-time $M_{D-N}$ and an internal compact space $M_{N}$,
\bea
M_{D} = M_{D-N} \times M_{N} \, ,
\eea
where the simplest case is to consider an internal toroidal manifold. In this scenario, the resulting theory is invariant under different symmetries. One of these symmetries is T-duality or, more precisely, $O(N,N)$ invariance. This symmetry group appears after compactifying the universal NS-NS sector on a N-dimensional torus and it is an exact symmetry of string theory. 

The main idea of Double Field Theory (DFT) is to rewrite a supergravity theory as a T-duality invariant theory \textit{before} compactification. In this sense, DFT represents an $O(D,D)$ invariant theory, where $D$ is the dimension of the target space of the embedded supergravity. All the DFT fields and parameters are $O(D,D)$ multiplets or, in other words, covariant objects. DFT coordinates lie in the fundamental representation of $O(D,D)$ whose dimension is $2D$. Then $X^{M}=(\tilde{x}_{\mu},x^{\mu})$ are coordinates of a double space. Here $x^{\mu}$ are the coordinates of the embedded supergravity, and $\tilde{x}_{\mu}$ are $D$ extra coordinates that we need for consistency. The dual coordinates are taken away considering that fields and parameters depend only on $x^{\mu}$ and therefore $\tilde \partial=0$. This is, in fact, the simplest solution to the ``strong constraint'',
\be
\partial_{M} (\partial^{M} \star) = (\partial_{M} \star) (\partial^{M} \star) = 0 \, ,
\label{SC}
\ee
where $\star$ means a product of arbitrary DFT fields/parameters. Contractions here are given by the $O(D,D)$ invariant metric, $\eta_{M N}$.

The DFT construction can be performed following four steps, 
\begin{enumerate}
\item {\bf Double geometry}: We consider a double geometry with coordinates $X^{M}$ with $M=0,\dots,2D-1$. We equip a group invariant metric $\eta_{MN}$ and its inverse $\eta^{MN}$. These metrics are used to lower and raise double curved indices.  On each point of the double space we consider a double tangent space, so we are able to define flat vectors $V^A$, with $A$ a flat index, $A=0,\dots,2D-1$. Then we consider two additional invariant and flat metrics $\eta_{A B}$ and ${\cal H}_{A B}$. The former is used to lower flat indices and both of them are used to construct the following flat projectors,
\bea
P_{AB} = \frac{1}{2}\left(\eta_{AB} - {\cal H}_{AB}\right) \, , \ \  \ \
\ov{P}_{AB} = \frac{1}{2}\left(\eta_{AB} + {\cal H}_{AB}\right)\ \, ,
\eea
which satisfy
\bea
&{\overline{P}}_{{A B}} {\overline{P}}^{ B}{}_{C}={\overline{P}}_{{A C}}\, , &\quad {P}_{{A B}} {P}^{B}{}_{C}={P}_{{A C}}, \nn\\
&{P}_{{A  B}}{\overline{P}}^{B}{}_{ C} = {\overline{P}}_{ {A B}}  {P}^{ B}{}_{C} = 0\, ,  &\quad {\overline{P}}_{{AB}} + {P}_{{A B}} = \eta_{{A B}}\,.
\eea
Thus, $P$ and $\bar P$ project onto complementary orthogonal subspaces and we can write any arbitrary vector as
\bea
V^{A} = P^{A}{}_{B} V^{B} + \ov P^{A}{}_{B} V^{B} = V^{\un A} + V^{\ov A} \, .
\label{projflat}
\eea
The notation underline and overline means that we use the projectors to lower/raise the indices, instead of using an invariant metric.

\item {\bf Symmetries:} DFT is a T-duality invariant formulation. Moreover, all the fields and parameters are written in representations of the duality group and, subsequently, duality invariance is always guaranteed. 
Infinitesimal $O(D,D)$ transformations acting on an arbitrary double vector reads
\bea
\delta_h V^M = V^{N} h_{N}{}^{M} \, ,
\label{duality}
\eea
where $h \in o(D, D)$ \footnote{We use the lower-case because $h$ belongs to the associated algebra.} is an arbitrary parameter. 

We can also define generalized diffeomorphisms. These are infinitesimal transformations acting on a generic double vector $V^M$ through a generalized Lie derivative, \textit{i.e.},
\bea
\delta_{\hat \xi} V^M = {\mathcal L}_{\hat \xi} V^M =\hat \xi^N \partial_{N} V^M + (\partial^M \hat \xi_{P} - \partial_P \hat \xi^M) V^{P} + \omega \partial_{N} \hat \xi^N V^{M}\, .
\eea
In the previous expression we consider a generic parameter $\hat \xi^{M}$ and a density weight factor $\omega$. The generalized Lie derivative differs from the ordinary one since we need to ensure ${\cal L}_{\hat \xi} \eta_{M N}=0$. The closure of the generalized diffeomorphism transformations on an arbitrary generalized tensor $V^{M}{}_{N}$,
\bea
\Big[\delta_{\hat \xi_1},\delta_{\hat \xi_2} \Big] V^{M}{}_{N} = \delta_{\hat \xi_{21}} V^{M}{}_{N} \,,
\eea
is provided by the C-bracket,
\bea
\hat \xi^{M}_{12} = \hat \xi^{P}_{1} \frac{\partial \hat \xi^{M}_{2}}{\partial X^{P}} - \frac12 \hat \xi^{P}_{1} \frac{\partial \hat \xi_{2P}}{\partial X_{M}} - (1 \leftrightarrow 2) \, .
\label{Cbra}
\eea
An important comment here is that the closure of the generalized diffeomorphisms requires the strong constraint (\ref{SC}). 

The partial derivative of a generic double vector does not transform as a tensor. Therefore we define a covariant derivative as follows
\bea
D_{M} V^{N} = \partial_{M} V^{N} - \Gamma_{M}{}^{N}{}_{P} V^{P} \, .
\label{DFTcovder}
\eea
The compatibility $D_{M}\eta_{NP}=0$ implies $\Gamma_{MNP}=-\Gamma_{MPN}$. It is usual to impose \bea
{\cal T}_{MNP}=3 \Gamma_{[MNP]}=0 \, ,
\eea
since ${\cal T}_{MNP}$ plays the role of generalized torsion. A curious aspect of the double geometry is that there are not enough compatibility conditions to  fully determine $\Gamma_{MNP}$, unlike in general relativity. Consequently, the generalized Riemann tensor cannot be fully determined but, nevertheless, a generalized Ricci tensor ${\cal R}_{MN}$ can be constructed as well as a generalized Ricci scalar ${\cal R}$, and these are fully determined in terms of the fundamental fields of DFT.

Another symmetry of DFT is the double Lorentz transformation that acts in the following way,
\bea
\delta_{\Lambda} V^{A} = V^{B} \Lambda_{B}{}^{A} \, ,
\eea
on an arbitrary flat vector $V^A$. Demanding $\delta_{\Lambda} \eta_{AB}=0$ we have $\Lambda_{AB}=-\Lambda_{BA}$. Moreover, using the decomposition/notation (\ref{projflat}), the condition $\delta_{\Lambda}{\cal H}_{AB}=0$ implies
\bea
\Lambda_{\un A \ov B} = \Lambda_{\ov A \un B} = 0 \, .
\eea
The partial derivative of an arbitrary flat vector $V^A$ does not transform as a tensor and therefore we introduce a flat covariant derivative,
\bea
D_{M} V^{A} = \partial_{M} V^{A} - W_{M}{}^{A}{}_{B} V^{B} \, ,
\eea
where $W_{MAB}=-W_{MBA}$  is the generalized spin connection. Imposing compatibility with the generalized frame we find that $W_{MAB}$ is not fully determined.

\item{\bf Fundamental fields:} The fundamental fields of DFT are a generalized frame $E_{MA}(X)=E_{MA}$ and a generalized dilaton $d(X)=d$. The former is equivalent to a vielbein for this double geometry and satisfies 
\bea
E_{M A} {\cal H}^{A B} E_{N B} & = & {\cal H}_{M N} \, , \\
E_{M A} \eta^{A B} E_{N B} & = & \eta_{M N} \, .
\eea
DFT can be written in terms of $d$ and ${\cal H}_{M N}$, the latter known as the generalized metric. In this case, the double Lorentz invariance is implicit. The generalized metric is an $O(D,D)$ element, \textit{i.e.}
\bea
{\cal H}_{M P} \eta^{P Q} {\cal H}_{Q N} = \eta_{M N} \, , 
\eea
and with the help of this dynamical metric and $\eta_{MN}$ one can define curved projectors,
\bea
P_{MN} = \frac{1}{2}\left(\eta_{MN} - {\cal  H}_{MN}\right) \, , \ \ \ \ \
\ov{P}_{MN} = \frac{1}{2}\left(\eta_{MN} + {\cal H}_{MN}\right)\ .
\label{proyc}
\eea
The previous projectors satisfy
\bea
&{\overline{P}}_{{M Q}} {\overline{P}}^{ Q}{}_{ N}={\overline{P}}_{{M N}}\, , &\quad {P}_{{M Q}} {P}^{Q}{}_{ N}={P}_{{M N}}, \nn\\
&{P}_{{M  Q}}{\overline{P}}^{Q}{}_{ N} = {\overline{P}}_{ {M Q}}  {P}^{ Q}{}_{ N} = 0\, ,  &\quad {\overline{P}}_{{MN}} + {P}_{{M N}} = \eta_{{M N}}\,,
\label{curveprojrel}
\eea
similar to the flat projectors. The main difference here is that $P_{AB}$ and $\ov P_{AB}$ are constants, while the curved ones are not for an arbitrary double background.

The fundamental fields transform with respect to generalized diffeomorphisms and double Lorentz transformations as follows,
\bea
\delta_{\hat \xi,\Lambda} E_{M A} & = & {\cal L}_{\hat \xi} E_{M A} + E_{M B} \Lambda^{B}{}_{A} \, ,  \\
\delta_{\hat \xi} d & = & \hat \xi^{N} \partial_N d - \frac12 \partial_{M} \hat \xi^{M} \, .
\eea
The generalized dilaton is a double Lorentz invariant, and its transformation under generalized diffeomorphisms is not covariant. However $e^{-2d}$ transforms as a generalized scalar density with $\omega=1$. On the other hand, both $E_{MA}$ and ${\cal H}_{MN}$ satisfy compatibility conditions 
\bea
D_{M}{\cal H}_{N P} & = & 0 \, , \quad
D_{M}{E}_{N P} = 0 \, .
\eea
The generalized fluxes are defined as 
\bea
F_{A B C} = 3 E_{[A} E^{M}{}_{B} E_{M C]} \, ,
\eea
with $E_{A}=\sqrt{2} E^M{}_{A} \partial_{M}$. It is possible to determine the totally antisymmetric part of the spin connection, $W_{ABC}\equiv \sqrt{2} W_{MBC} E^{M}{}_{A}$, in the following way
\bea
F_{A B C} = - 3 W_{[A B C]} \, .
\label{Fluxspin}
\eea
Finally, it is convenient to define 
\bea
F_{\un M \ov A \ov B} & = & \frac{1}{\sqrt{2}} E_{M}{}^{\un C} F_{\un C \ov A \ov B} \, , \\
F_{\ov M \un A \un B} & = & \frac{1}{\sqrt{2}} E_{M}{}^{\ov C} F_{\ov C \un A \un B} \, ,
\eea
for later use. The transformation rule of the previous objects is
\bea
\label{FluxM}
\delta_{\hat \xi, \Lambda} F_{\un M \ov A \ov B} & = & {\cal L}_{\hat \xi} F_{\un M \ov A \ov B} + \partial_{\un M} \Lambda_{\ov A \ov B} + 2 F_{\un M \ov C [\ov B} \Lambda^{\ov C}{}_{\ov A]} \, , \\
\delta_{\hat \xi, \Lambda} F_{\ov M \un A \un B} & = & {\cal L}_{\hat \xi} F_{\ov M \un A \un B} + \partial_{\ov M} \Lambda_{\un A \un B} + 2 F_{\ov M \un C [\un B} \Lambda^{\un C}{}_{\un A]} \, .
\eea
\item {\bf Action principle:} The action of DFT is given by
\bea
\int d^{2D}X e^{-2d} {\cal R}(E,d) \, ,
\eea
where ${\cal R}(E,d)$ is a two derivative scalar under generalized diffeomorphisms and it is invariant under Lorentz transformations. This object is known as the generalized Ricci scalar, and can be written in terms of the generalized fluxes,
\bea
{\cal R}(d,E_{MA}) & = & 2E_{\un{A}}F^{\un{A}} + F_{\un{A}}F^{\un{A}} - \frac16{F}_{\un{ABC}}F^{\un{ABC}} - \frac12{F}_{\ov{A}\un{BC}}F^{\ov{A}\un{BC}} \, ,
\label{GR}
\eea
where $F_{A} = \sqrt{2} \partial^{M} E_{M A} - 2 E_{A} d \, .$ 
\end{enumerate}

\subsubsection{Heterotic/Gauged Double Field Theory}
The duality group of the heterotic/gauged formulation of DFT is $O(D,D+n)$ with $n$ the dimension of the gauge group. Since this group differs from $O(D,D)$, the construction described in the previous part requires the following substantial modifications:
\begin{enumerate}
\item {\bf Double geometry:} We consider a double geometry with coordinates $X^{\cal M}=(\tilde{x}_{\mu}, x^{\mu}, x^{i})$ with ${\cal M}=0,\dots,2D-1+n$ and $i=1,\dots,n$. On each point of the extended double space we consider an extended double tangent space, so we are able to define flat vectors $V^{\cal A}$, with ${\cal A}$ a flat index, ${\cal A}=0,\dots,2D-1+n$. 

\item {\bf Symmetries:} The generalized diffeomorphisms now contain an extra term that depends on generalized structure constants $f_{\cal MNP}$,
\bea
\delta_{\hat \xi} V^{\cal M} = {\mathcal L}_{\hat \xi} V^{\cal M} =\hat \xi^{\cal N} \partial_{\cal N} V^{\cal M} + (\partial^{\cal M} \hat \xi_{\cal P} - \partial_{\cal P} \hat \xi^{\cal M}) V^{\cal P} + f^{\cal M}{}_{\cal NP} \hat \xi^{\cal N} V^{\cal P} + \omega \partial_{\cal N} \hat \xi^{\cal N} V^{\cal M} \, , \nn \\
\label{gaugeLie}
\eea
where $V^{\cal M}$ is an arbitrary double vector. The structure constants are fully antisymmetric and satisfy a Jacobi rule,
\bea
f_{{\cal MNP}}=f_{[{\cal MNP}]}\, , \qquad f_{[\cal MN}{}^{\cal R}f_{{\cal P}] {\cal R}}{}^{\cal Q}=0\, . \label{consf}
\eea

The closure of the transformations is given by a deformed bracket  
\bea
 [\hat \xi_1, \hat \xi_{2} ]^{ \cal M}_{(C_{f})}=2\hat \xi^{ \cal P}_{[1}\partial_{\cal P}\hat \xi_{2]}^{\cal M}-\hat \xi_{[1}^{ \cal N}\partial^{\cal M}\hat \xi_{2]\cal N}+f_{{\cal PQ}}{}^{\cal M} \hat \xi_{1}^{\cal P} \hat \xi_2^{\cal Q}\, ,
\eea
which reduces to the C-bracket when the structure constants vanish. Interestingly enough, the closure requires the strong constraint (\ref{SC}) plus an extra constraint,
\bea
f_{\cal M N P} \partial^{\cal M} \star = 0 \, .
\label{Strongf}
\eea
We solve the new constraint considering that the generalized structure constants are non-vanishing only when ${\cal M,N,P}=i,j,k$, \textit{i.e.} $f_{\cal MNP}=f_{ijk}$, and $\partial_{i}=0$. 

The covariant derivative (\ref{DFTcovder}) remains unchanged, but $\Gamma_{\cal MNP}$ receives an extra term in its transformation rule
\bea
\delta_{\hat \xi} \Gamma_{\cal MNP} = \delta^{\rm ungauged}_{\hat \xi} \Gamma_{\cal MNP} + f_{\cal N Q P} \partial_{\cal M} \hat \xi^{\cal Q} \, ,
\eea
where the explicit form of the first term corresponds to an exercise (see Exercises 3.3, part A, problem (9)). On the other hand, the double Lorentz symmetry remains unchanged.

\item{\bf Fundamental fields:} The fundamental fields of heterotic/gauged DFT are a generalized frame $E_{\cal M}{}^{\cal A}(X)$ and a generalized dilaton $d(X)$. The only modification at this point is related to the construction of the generalized fluxes $F_{\cal A B C}$, that now contain an extra term
\bea
F_{\cal A B C} = 3 E_{[\cal A} E^{\cal M}{}_{\cal B} E_{\cal M C]} + \sqrt{2} f_{\cal M N P} E^{\cal M}{}_{\cal A} E^{\cal N}{}_{\cal B} E^{\cal P}{}_{\cal C} \, .
\label{Gflux}
\eea
Similarly to the heterotic supergravity case we assume that $f_{\cal MNP}$ has the same units as derivatives.

\item {\bf Action principle:} The action of heterotic/gauged DFT is given by
\bea
\int d^{2D+n}X e^{-2d} {\cal R}(E,d) \, ,
\eea
where ${\cal R}(E,d)$ is a two derivative scalar under (\ref{gaugeLie}) and it is invariant under Lorentz transformations. This action principle now contains $f_{\cal MNP}$ contributions provided by the generalized fluxes (\ref{Gflux}) and, schematically, ${\cal R}$ has the same form as (\ref{GR}) but promoting the indices ${A}\rightarrow{\cal A}$.

\end{enumerate}

\subsubsection{Parametrization}
Here we present the parametrization of the different parameters, metrics and fields of heterotic/gauged DFT. The parametrization of the ungauged DFT is a particular case of this one. We follow the same structure of the previous part, step by step. 
\begin{enumerate}
\item {\bf Double geometry:} The parametrization of the coordinates is $X^{\cal M}=(\tilde{x}_{\mu}, x^{\mu}, x^{i})$. Since we solve the strong constraint using $\tilde \partial^{\mu} = \partial_{i}=0$, the parameters and fields of the resulting gauged supergravity do not depend on extra coordinates. The parametrization of the invariant metric is 
\be
{\eta}_{{\cal M N}}  = \left(\begin{matrix}\eta^{\mu\nu}&\eta^\mu{}_\nu&\eta^\mu{}_i\\ 
\eta_\mu{}^\nu&\eta_{\mu\nu}&\eta_{\mu i}\\\eta_{i}{}^\nu&\eta_{i\nu}&\eta_{ij}\end{matrix}\right)= \left(\begin{matrix}0&\delta^\mu{}_\nu&0\\ 
\delta_\mu{}^\nu&0&0\\0&0&\kappa_{ij}\end{matrix}\right) \ , \label{eta}
\ee
with $\mu, \nu=0,\dots, D-1$, $i,j=1,\dots, n$ and $\kappa_{ij}$ the Killing metric of the gauge group.  The flat indices can be split in the following way ${\cal A}=(\un a, \ov a, \ov i)$ where $\un a=1,\dots,D$, $\bar a=1,\dots,D$ and $\ov i$ runs from $1,\dots,n$. The latter is the flat version of the gauge index. The Lorentz invariant metrics are parametrized as 
\begin{equation}
\eta_{\cal A B} = \left(\begin{matrix} - \eta_{\un a \un b} & 0 & 0 \\ 0 & \eta_{\bar a \bar b} & 0 \\ 0 & 0 & \kappa_{\bar i \bar j} \end{matrix}\right) \ , \quad
{\cal H}_{\cal A B} = \left(\begin{matrix} \eta_{\un a \un b} & 0 & 0 \\ 0 & \eta_{\ov a \ov b} & 0 \\ 0 & 0 & \kappa_{\bar i \bar j} \end{matrix}\right) \ ,
\label{flatparam}
\end{equation}
where $\eta_{\un a \un b}$ and $\eta_{\ov a \ov b}$ can be identified with a flat and constant metric $\eta_{a b}$ 
\bea
\eta_{\bar a \bar b} \delta_{ab}^{\bar a \bar b} = \eta_{\un a \un b} \delta_{ab}^{\un a \un b} = \eta_{ab} \, , 
\eea
and $\kappa_{\bar i \bar j}$ is the flat version of the Killing metric. Defining the following constant object $e_{i}{}^{\bar i}$ we have a relation between the Cartan-Killing metric and its flat version,
\bea
\kappa_{i j} = e_{i}{}^{\bar i} \kappa_{\bar i \bar j} e_{j}{}^{\bar j} \, .
\eea
The same holds for the inverses of these metrics, defining $e^{i}{}_{\bar i}$ as the inverse of $e_{i}{}^{\bar i}$.

Before analyzing the symmetry rules, we include in this part the parametrization of the generalized frame,
\be
E^{\cal M}{}_{\cal A}  =\left(\begin{matrix}{ E}_{\mu \underline a}&  { E}^{\mu }{}_{\underline a} & E^i{}_{\underline a}\\ 
E_{\mu \overline  a}& E^\mu{}_{\overline  a}&E^i {}_{\overline a} \\
E_{\mu\overline i} &E^\mu{}_{\overline i} &E^i{}_{\overline i} \end{matrix}\right) \ = \
\frac{1}{\sqrt{2}}\left(\begin{matrix}-{ e}_{\mu \underline a}-C_{ \rho\mu} { e}^{\rho }{}_{\underline a} &  { e}^{\mu }{}_{\underline a} & -A_\rho{}^i { e}^{\rho }{}_{\underline{a}}\, , \\ 
e_{\mu \overline a}-C_{\rho \mu}{} e^{\rho }{}_{\overline{a}}& e^\mu{}_{\overline a}&-A_{\rho}{}^i  e^\rho{}_{\overline a} \\
\sqrt2 A_{\mu i}e^i{}_{\overline i} &0&\sqrt2 e^i{}_{\overline i} \end{matrix}\right)  \, ,
\label{Frame0}
\ee
where ${e}_{\mu \underline  a}$ and $e_{\mu \overline  a}$ are a pair of vielbeins for the same  metric tensor $g_{\mu \nu}$, i.e.,
\bea
e_{\mu}{}^{\ov a} \eta_{\ov a \ov b} e_{\nu}{}^{\ov b} & = & g_{\mu \nu} \, , \\
e_{\mu}{}^{\un a} \eta_{\un a \un b} e_{\nu}{}^{\un b} & = & g_{\mu \nu} \, .
\eea
We identify each vielbein (and their inverses) considering the following gauge fixing of the double Lorentz transformations to a single copy of Lorentz transformations,
\bea
{e}_{\mu \un a} \delta_{a}^{\un a} & = & {e}_{\mu \ov a} \delta_{a}^{\ov a} = e_{\mu a} \, , \\
{e}^{\mu}{}_{\un a} \delta_{a}^{\un a} & = & {e}^{\mu}{}_{\ov a} \delta_{a}^{\ov a} = e^{\mu}{}_{a} \, .
\eea
Finally in (\ref{Frame0}) we use the notation $C_{\mu \nu}=b_{\mu \nu} + \frac12 A_{\mu}{}^{i} A_{\nu i}$.

\item {\bf Symmetries:} The parametrization of the symmetry parameters is given by,
\bea
\hat \xi^{\cal M}& = & (\zeta_\mu, \xi^\mu, \lambda^i) \, , \label{diffeosP}\\
\Lambda_{\ov a \ov b} & = & \Lambda_{a b} \delta_{\ov a \ov b}^{a b} \, , \quad 
\Lambda_{\un a \un b} = - \Lambda_{a b} \delta_{\un a \un b}^{a b} \, , \\
\Lambda_{\ov a \ov{i}} & = & \label{gaugeuno} 0  \, , \quad
  \Lambda_{\ov{i} \ov{j}} =  f_{i j k} \lambda^{j} e^{k}{}_{\bar i} e^{i}{}_{\ov{j}}{} \, .
\eea
The components $\Lambda_{\ov a \ov{i}}$ and $\Lambda_{\ov i \ov{j}}$ are fixed to ensure $\delta E^{\mu}{}_{\ov i}=0$ and $\delta E^{i}{}_{\ov i}=0$ as required by (\ref{Frame0}). From (\ref{diffeosP}) we can understand how the generalized diffeomorphisms encode ordinary diffeomorphisms plus gauge transformations, while double Lorentz transformation encode the ordinary Lorentz transformations. For instance, from $\delta E^{\mu \ov a}$ we obtain 
\be
\delta E^{\mu}{}_{\ov a} \delta^{\ov a}_{a} = \delta e^{\mu}{}_{a} =  \xi^{\rho}\partial_{\rho}  e^{\mu}{}_{a}  - \partial_{\rho}\xi^{\mu} e^{\rho}{}_{a} + e^{\mu}{}_{b} \Lambda^{b}{}_{a} \, ,
\label{inverseparam}
\ee
in agreement with (\ref{inverse0}). 

\item{\bf Fundamental fields:} The parametrization of the generalized frame was given in (\ref{Frame0}), while the parametrization of the generalized metric, 
\bea
{\cal H}_{\cal M N} = \left(\begin{matrix} g^{\mu \nu} & - g^{\mu \rho} C_{\rho \nu} & - g^{\mu \rho} A_{\rho i} \\
- C_{\rho \mu} g^{\nu \rho}  & g_{\mu \nu} + C_{\rho \mu} C_{\sigma \nu} g^{\rho \sigma} + A_{\mu}{}^i \kappa_{ij} A_{\nu}{}^j &
C_{\rho \mu} g^{\rho \sigma} A_{\sigma i} + A_{\mu}{}^j \kappa_{ji} \\
- g^{\nu \rho} A_{\rho i} & C_{\rho \nu} g^{\rho \sigma} A_{\sigma i} +  A_{\nu}{}^j \kappa_{ij} & \kappa_{ij} + A_{\rho i} g^{\rho \sigma} A_{\sigma j}\end{matrix}\right) \ ,
\label{generalizedmetric}
\eea
can be easily obtained from the $E_{\cal M}{}^{\cal A} {\cal H}_{\cal A B} E_{\cal N}{}^{\cal B}={\cal H}_{\cal MN}$.  

The generalized dilaton is given by
\bea
e^{-2d} = \sqrt{-g} e^{-2\phi} \, .
\eea
Using the parametrization of the generalized frame and dilaton, it is straightforward to compute the parametrization of the different projections of the generalized fluxes,
\bea \label{f1}
F_{\overline a\underline{ bc}} & = & -
w^{(+)}_{ a bc} \delta_{\overline a\underline{ bc}}^{abc}\, ,\\
F_{\underline  a \overline{bc}} & = & w^{(-)}_{ a bc} \delta_{\underline  a \overline{bc}}^{abc}\, ,\\
F_{\overline{a bc}} & = & 3\left(w_{[{abc}]}-\frac16 \bar H_{{abc}}\right) \delta_{\overline{a bc}}^{abc}
\, ,\\
F_{\underline {a bc}}& = & -3\left(w_{[abc]}+\frac16 \bar H_{abc}\right) \delta_{\underline {a bc}}^{abc}\, ,\\
F_{\overline i \underline{ab}} & = & - \frac1{\sqrt2} e^ \mu{}_a e^\nu{}_b e{}_{i\overline i} F^i_{\mu\nu} \delta_{\underline{ab}}^{ab}\, , \label{Fluxg}\\
F_{\underline a \overline i \overline j} & = & -e^i{}_{\overline i} e^j{}_{\overline j} e^\mu{}_a A_{\mu}{}^{k} f_{ijk} \delta_{\underline a}^{a} \, ,\\
F_{\overline{ijk}} & = & \sqrt2e^i{}_{\overline i }e^j{}_{\overline j}e^k{}_{\overline k }f_{ijk}\, ,\\
F_{\underline a} & = & \left(\partial_\mu e_a^\mu+e_a^\mu e_b^\nu\partial_\mu e^b_\nu-2e^\mu{}_{ a}\partial_\mu\phi\right) \delta_{\un a}^{a} \label{f2}\, ,
\eea
where 
\bea
\bar H_{abc}&=& 3e^\mu{}_{a} e^\nu{}_b e^\rho{}_{c} \left(\partial_{[\mu}b_{\nu\rho]}-A_{[\mu}^i\partial_\nu A_{\rho]i}+\frac13
f_{ijk}A_{\mu}{}^i A_\nu{}^j A_\rho{}^k\right)\, .\label{flatH}
\eea

\item {\bf Action principle:} The action of heterotic/gauged DFT reduces to (\ref{het0}) after parametrization. For example, the last term of the heterotic/gauged DFT action principle, $-\frac12 F_{\bar i \un b \un c} F^{\bar i \un b \un c}$, reproduces the term $-\frac14 F_{\mu \nu}{}^{i} F^{\mu \nu}{}_{i}$ when the component $F_{\ov i \un b \un c}$ is replaced by (\ref{Fluxg}).

\end{enumerate}

\subsection{Part B: Single and Double Copy from the Double Field Theory Perspective}
\label{B2}
\subsubsection{Generalized Kerr-Schild Ansatz}
In this section we review DFT and the Generalized Kerr-Schild ansatz (GKSA).
The GKSA was originally formulated in \cite{PartGKS-1} as an exact and linear perturbation of the generalized background metric $H_{M N}$ ($M,N=0, \dots, 2D-1$) and an exact perturbation of the generalized background dilaton $d_{o}$ (see also \cite{PartGKS-2}-\cite{PartGKS-6}). We work with arbitrary $D$ until parametrization. 

Since the generalized metric is an $O(D,D)$ element, its perturbation has the following form
\bea
{\cal H}_{MN} = H_{MN} + \kappa (\bar{K}_{M} K_{N} + {K}_{M} \bar{K}_{N} ) \, ,
\label{DFTKS}
\eea
where $\bar{K}_{M}= \bar{P}_{M}{}^{N} \bar{K}_{N}$ and $K_{M}= {P}_{M}{}^{N} {K}_{N}$ are a pair of generalized null vectors 
\bea
\eta^{MN} \bar{K}_{M} \bar{K}_{N} & = & \eta^{MN} K_{M} K_{N} =  \eta^{MN} \bar{K}_{M} K_{N} = 0 \, .
\label{nulldft}
\eea 
According to (\ref{DFTKS}), the DFT projectors are
\bea
{\cal P}_{MN} &=& P_{MN} - \frac12 \kappa (\bar{K}_{M} K_{N} + {K}_{M} \bar{K}_{N}) \nn \\ 
\bar{\cal P}_{MN} &=& \bar{P}_{MN} + \frac12 \kappa (\bar{K}_{M} K_{N} + {K}_{M} \bar{K}_{N}) \, .
\eea
Each $O(D,D)$ multiplet can be written as a sum over its projections,
\bea
V_{M} = P_{M}{}^{N} V_{N} + {\bar P}_{M}{}^{N} V_{N} = V_{\underline M} + V_{\overline M} \, ,
\eea
where $V_{M}$ is a generic double vector. When we use the underline and overline notation, we consider the background projectors $P_{MN}$ and ${\bar P}_{MN}$.

The generalized background dilaton can be perturbed with a generic $\kappa$ expansion,
\be
d = d_{o} + \kappa f\, , \qquad f = \sum_{n=0}^{\infty}\kappa^{n}f_{n} \, ,
\ee
with $n\geq 0$.

Mimicking the ordinary Kerr-Schild ansatz, the generalized vectors $K_{M}$, $\bar{K}_{M}$ and $f$ obey some conditions in order to produce finite deformations in the DFT action and EOM's. If we consider a generic double vector $V_{N}$, the covariant derivative can be defined as
\bea
\nabla_{M} V_{N} = \partial_{M} V_{N} - \Gamma_{MN}{}^{P} V_{P} \, , 
\eea
where $\Gamma_{MNP}$ is the generalized affine connection. 
Demanding 
\bea
\nabla_{M}{\cal H}_{NP}&=&0 \, , \quad \nabla_{M}{\cal \eta}_{NP}=0 \, ,  
\eea
and a vanishing generalized torsion 
\bea
\Gamma_{[MNP]}=0 \, ,
\label{torsion}
\eea
the following projections of $\Gamma_{MNP}=-\Gamma_{MPN}$ are well-defined and can be perturbed,
\bea
\Gamma_{\underline M \underline N \overline Q} & = &  -\bar{\cal P}_{Q}{}^{R} {\cal P}_{M}{}^{S} \partial_{S}{\cal P}_{R N}  \, , \quad \Gamma_{\overline M \overline N \underline Q} =   \bar{\cal P}_{N}{}^{R} {\bar {\cal P}}_{M}{}^{S}\partial_{S} {\cal P}_{R Q}  \, ,  \nn \\
\Gamma_{\underline M \overline N \overline Q} & = &  2 \bar{\cal P}_{[N}{}^{R}  {\bar {\cal P}}_{Q]}{}^{S} \partial_{S} {\cal P}_{R M}  \, , \quad \Gamma_{\overline M \underline N \underline Q} =  2 \bar{\cal P}_{M}{}^{R} {\cal P}_{[N}{}^{S}\partial_{S} {\cal P}_{Q]R} \, .
\eea

Similarly to Riemannian geometry, the generalized Ricci scalar and the generalized Ricci tensor can be constructed from different (determined) projections of the generalized affine connection. Following the original construction of the GKSA we impose,
\bea
{\bar K}^{P} \partial_{P} K^{M} + K_{P} \partial^{M}{\bar K}^{P} - K^{P} \partial_{P}{\bar K}^{M} & = & 0 \, , \nn \\ {K}^{P} \partial_{P} {\bar K}^{M} + {\bar K}_{P} \partial^{M}{K}^{P} - {\bar K}^{P} \partial_{P}{K}^{M} & = & 0 \, , 
\label{geodesic1}
\eea
and
\bea
K^{M} \partial_{M}f = {\bar K}^{M} \partial_{M}f = 0 \, .
\label{geodesic2}
\eea
Using (\ref{torsion}), we can change $\partial \rightarrow \nabla$ in (\ref{geodesic1}) obtaining, 
\bea
{\bar K}^{P} \nabla_{P} K^{M} + K_{P} \nabla^{M}{\bar K}^{P} - K^{P} \nabla_{P}{\bar K}^{M} & = & 0 \, , \nn \\ {K}^{P} \nabla_{ P} {\bar K}^{M} + {\bar K}_{P} \nabla^{M}{K}^{P} - {\bar K}^{P} \nabla_{P}{K}^{M} & = & 0 \, . 
\label{geodesic11}
\eea

The generalized flux formulation is compatible with the GKSA if we consider perturbations of the form,
\bea
{\cal E}_{M}{}^{\ov A} = E_{M}{}^{\ov A} + \frac12 \kappa E_{M}{}^{\underline B} K_{\underline B} {\bar K}^{\overline A} \, , \nn \\ {\cal E}_{M}{}^{\underline A} = E_{M}{}^{\underline A} - \frac12 \kappa E_{M}{}^{\ov B} {\bar K}_{\overline B} K^{\underline A}\, ,
\label{GKSA}
\eea
where $K_{A} = {\cal E}^{M}{}_{\underline A} K_{M}={E}^{M}{}_{\underline A} K_{M}$ and $\bar{K}_{A} = {\cal E}^{M}{}_{\overline A} \bar{K}_{M}=E^{M}{}_{\overline A} \bar{K}_{M}$ and ${\cal E}_{MA}$ is an $O(D,D)/O(D-1,1)_{L} \times O(1,D-1)_{R}$ frame. In this formulation $\underline A= 0, \dots, D-1$ and $\overline A=0, \dots, D-1$ are $O(D-1,1)_{L}$ and $O(1,D-1)_{R}$ indices, respectively. 

We can define flat invariant projectors as follows,
\bea
{\cal P}_{A B} & = & {\cal E}_{M \underline A} {\cal E}^{M}{}_{\underline B} = P_{A B} \, , \nn \\ \bar{\cal P}_{{A B}} & = & {\cal E}_{M \ov A} {\cal E}^{M}{}_{\ov B} = \bar P_{{A B}} \, ,
\eea
where $P_{AB}=E_{M \underline A} E^{M}{}_{\underline B}$ and $\bar{P}_{AB}=E_{M \overline A} E^{M}{}_{\overline B}$ are the standard DFT flat projectors. Using these projectors we can construct two invariant metrics,
\bea
\eta_{AB} = {\cal E}_{M A}\eta^{MN} {\cal E}_{N B} = E_{M A}\eta^{MN} E_{N B} \, , \\
{H}_{AB} = {\cal E}_{MA} {\cal H}^{MN} {\cal E}_{N B} = E_{MA} {H}^{MN}E_{N B} \, .
\eea

The flat covariant derivative acting on a generic vector $V_{B}$ is
\bea
{\cal D}_{A} V_{B} = {\cal E}_{A} V_{B} + {\cal W}_{AB}{}^{C} V_{C} \, , 
\eea
where ${\cal E}_{A} = \sqrt{2} {\cal E}^{M}{}_{A} \partial_{M}$ and ${\cal W}_{AB}{}^{C}$ is the generalized spin connection that satisfies
\be
{\cal W}_{ABC} = - {\cal W}_{ACB}\, \qquad \mathrm{and} \qquad {\cal W}_{A\overline B \underline C} = {\cal W}_{A\underline B \overline C} = 0 \, .
\ee
With the help of the generalized frames we can construct the generalized fluxes, which are defined as
\bea
{\cal F}_{ABC} & = & 3 {\cal E}_{[A}({\cal E}^{M}{}_{B}){\cal E}_{M C]} \, , \nn \\ {\cal F}_{A} & = & \sqrt{2}e^{2d}\partial_{M}\left({\cal E}^{M}{}_{A}e^{-2d}\right)  \,.
\eea

In the flux formulation of DFT, conditions (\ref{geodesic1}) and (\ref{geodesic2}) become
\bea
K^{\underline A} E_{\underline A} {\bar K}^{\overline C} + {K}^{\un A} \bar K^{\ov B} F_{\un A \ov B}{}^{\ov C} & = & 0 \, , \nn \\
{\bar K}^{\overline A} E_{\overline A} {K}^{\underline C} + {\bar K}^{\ov A}  K^{\un B} F_{\ov A \un B}{}^{\un C} & = & 0 \, ,
\label{flatgeo1}
\eea
and
\bea
K^{\underline A} E_{\underline A} f = {\bar K}^{A} E_{\overline A} f = 0 \, . 
\label{flatgeo2}
\eea
It is straightforward to check that the previous conditions are double Lorentz invariant using
\bea
\delta_{\Gamma}{\cal E}_{M A} = {\cal E}_{M}{}^{B} \Gamma_{B A} \, , \quad \delta_{\Gamma}{E}_{M A} = {E}_{M}{}^{B} \Gamma_{B A}  
\label{Lorentz}
\eea 
where $\Gamma_{A B} = - \Gamma_{B A}$ is the double Lorentz parameter. 

Only the totally antisymmetric  and trace parts of ${\cal W}_{ABC}$ can be determined in terms of ${\cal E}_{M}{}^{A}$  and $d$,
\bea
{\cal W}_{[ABC]} & = & -\frac13{\cal F}_{ABC}\, , \\
{\cal W}_{BA}{}^{B} & = &  - {\cal F}_{A}\, ,
\eea
the latter arising from partial integration with the dilaton density. Using these identifications, conditions (\ref{flatgeo1}) and (\ref{flatgeo2}) can be written as 
\bea
K_{\underline A} D^{\underline A} {\bar K}^{\overline B}  & = &  \bar K_{\overline A} { D}^{\overline A} {K}^{\underline B} =  0 \, , \nn \\ K_{\underline A} {D}^{\underline A} f & = & K_{\overline A} {D}^{\overline A}f = 0 \, ,
\eea
where $D_{A}$ is the background covariant derivative. As we mentioned before, the generalized Ricci scalar and the generalized Ricci tensor are completely determined in terms of the degrees of freedom of DFT and, particularly, can be written in terms of different projections of the fluxes,
\bea
\label{GRicci_scalar}
{\cal R} & = & 2{\cal E}_{\underline{A}}{\cal F}^{\underline{A}} + {\cal F}_{\underline{A}}{\cal F}^{\underline{A}} - \frac16 {\cal F}_{\underline{ABC}} {\cal F}^{\underline{ABC}} - \frac12{\cal F}_{\ov{A}\underline{BC}}{\cal F}^{\ov{A}\underline{BC}} \, , \\
{\cal R}_{\ov{A}\un{B}} & = & {\cal E}_{\ov{A}}{\cal F}_{\un{B}} - {\cal E}_{\un{C}}{\cal F}_{\ov{A}\un{B}}{}^{\un{C}} + {\cal F}_{\un{C}\ov{DA}}{\cal F}^{\ov{D}}{}_{\un{B}}{}^{\un{C}} - {\cal F}_{\un{C}}{\cal F}_{\ov{A}\un{B}}{}^{\un{C}} \, .
\label{GRicci_tensor}
\eea

The previous projections of the fluxes can be computed using (\ref{GKSA}) and imposing (\ref{flatgeo1}) and (\ref{flatgeo2}), 
\bea
\label{constrained_fluxes0}
{\cal F}_{\un{ABC}} & = & F_{\un{ABC}} -\frac{3}{2}\kappa\ov{K}^{\ov{D}}K_{[\un{A}}F_{\un{BC}]\ov{D}}\, , \\
{\cal F}_{\un{A}\ov{BC}} & = & F_{\un{A}\ov{BC}} + \kappa\left(\ov{K}{}_{[\ov{C}}D{}_{\ov{B}]}K_{\un{A}} + K_{\un{A}}E_{[\ov{B}}\ov{K}_{\ov{C}]} - \frac{1}{2}\ov{K}^{\ov{D}}K_{\un{A}}F_{\ov{DBC}}\right)\, , \\
{\cal F}_{\ov{A}\un{BC}} & = & F_{\ov{A}\un{BC}} - \kappa\left(K_{[\un{C}}D_{\un{B}]}\ov{K}_{\ov{A}} + \ov{K}_{\ov{A}}E_{[\un{B}}K_{\un{C}]} - \frac{1}{2}K^{\un{D}}\ov{K}_{\ov{A}}F_{\un{DBC}}\right)\, , \\
{\cal F}_{\un{A}} & = & F_{\un{A}} - \frac{1}{2}\kappa\left(K_{\un{A}}D_{\ov{B}}\ov{K}^{\ov{B}} + F_{\ov{B}\un{AC}}\ov{K}^{\ov{B}}K^{\un{C}} + 4D_{\un{A}}f\right)\, .
\label{constrained_fluxes}
\eea
Replacing the previous expressions in (\ref{GRicci_scalar}) the generalized Ricci scalar can be written as
\bea
{\cal R} & = & R + \kappa\left[- K_{\un{A}}\ov{K}^{\ov{B}}E_{\ov{B}}F^{\un{A}} - D_{\un{A}}\left(K^{\un{A}}D_{\ov{B}}\ov{K}^{\ov{B}} + F_{\ov{B}}{}^{\un{AC}}\ov{K}^{\ov{B}}K_{\un{C}}\right) + F^{\ov{A}\un{BC}}K_{\un{C}}D_{\un{B}}\ov{K}_{\ov{A}}\right.\, \nn \\
& & \ \ \ \ \ \ \ \ \ \ \left. + F^{\ov{A}\un{BC}}\ov{K}_{\ov{A}}E_{\un{B}}K_{\un{C}} - 4D_{\un{A}}D^{\un{A}}f\right] + \kappa^{2}\left[4E_{\un{A}}f E^{\un{A}}f\right]\,,
\eea
and therefore in the case $f=\mathrm{const.}$, the generalized Ricci scalar can be linearized. 

With a similar procedure the generalized Ricci tensor can be written as,
\bea
{\cal R}_{\ov{A}\un{B}} = {R}_{\ov{A}\un{B}} + \kappa {\cal R}_{(\kappa)\ov{A}\un{B}} + \kappa^2 {\cal R}_{(\kappa^{2})\ov{A}\un{B}} \, , 
\eea
where 
\footnotesize
\bea
{\cal R}_{(\kappa)\ov{A}\un{B}} & = & - \frac{1}{2}D_{\ov{A}}\left(K_{\un{B}}D_{\ov{C}}\ov{K}^{\ov{C}}\right) + \frac{1}{2}E^{\un{C}}\left(K_{\un{C}}D_{\un{B}}\ov{K}_{\ov{A}}\right) - \frac{1}{2}E^{\un{C}}\left(K_{\un{B}}D_{\un{C}}\ov{K}_{\ov{A}}\right) + \frac{1}{2}E^{\un{C}}\left(\ov{K}_{\ov{A}}E_{\un{B}}K_{\un{C}}\right)\, \nn \\
& & - \frac{1}{2}E^{\un{C}}\left(\ov{K}_{\ov{A}}E_{\un{C}}K_{\un{B}}\right) - \frac{1}{2}\ov{K}^{\ov{D}}K^{\un{C}}E_{\ov{A}}F_{\ov{D}\un{BC}} - \frac{1}{2}E_{\ov{A}}\ov{K}^{\ov{D}}K^{\un{C}}F_{\ov{D}\un{BC}} - \frac{1}{2}\ov{K}^{\ov{D}}E_{\ov{A}}K^{\un{C}}F_{\ov{D}\un{BC}}\, \nn \\
& & - \frac{1}{2}E^{\un{C}}K^{\un{D}}\ov{K}_{\ov{A}}F_{\un{DBC}} - \frac{1}{2}K^{\un{D}}E^{\un{C}}\ov{K}_{\ov{A}}F_{\un{DBC}} - \frac{1}{2}K^{\un{D}}\ov{K}_{\ov{A}}E^{\un{C}}F_{\un{DBC}} + \frac{1}{2}\ov{K}^{\ov{D}}K^{\un{C}}E_{\ov{D}}F_{\ov{A}\un{BC}}\, \nn \\
& & - \frac{1}{2}K_{\un{C}}D_{\un{B}}\ov{K}_{\ov{E}}F^{\un{C}\ov{E}}{}_{\ov{A}} + \frac{1}{2}K_{\un{B}}D_{\un{C}}\ov{K}_{\ov{E}}F^{\un{C}\ov{E}}{}_{\ov{A}} - \frac{1}{2}\ov{K}_{\ov{E}}E_{\un{B}}K_{\un{C}}F^{\un{C}\ov{E}}{}_{\ov{A}} + \frac{1}{2}\ov{K}_{\ov{E}}E_{\un{C}}K_{\un{B}}F^{\un{C}\ov{E}}{}_{\ov{A}}\, \nn \\
& & + \frac{1}{2}\ov{K}{}_{\ov{A}}D{}_{\ov{D}}K_{\un{C}}F_{\un{B}}{}^{\un{C}\ov{D}} - \frac{1}{2}\ov{K}{}_{\ov{D}}D{}_{\ov{A}}K_{\un{C}}F_{\un{B}}{}^{\un{C}\ov{D}} + \frac{1}{2}K_{\un{C}}E_{\ov{D}}\ov{K}_{\ov{A}}F_{\un{B}}{}^{\un{C}\ov{D}} - \frac{1}{2}K_{\un{C}}E_{\ov{A}}\ov{K}_{\ov{D}}F_{\un{B}}{}^{\un{C}\ov{D}}\, \nn \\
& & + \frac{1}{2}K_{\un{C}}D_{\un{B}}\ov{K}_{\ov{A}}F^{\un{C}} + \frac{1}{2}K^{\un{C}}\ov{K}_{\ov{A}}E_{\un{C}}F_{\un{B}} - \frac{1}{2}K_{\un{B}}D_{\un{C}}\ov{K}_{\ov{A}}F^{\un{C}} + \frac{1}{2}\ov{K}_{\ov{A}}E_{\un{B}}K_{\un{C}}F^{\un{C}}\, \nn \\
& & - \frac{1}{2}\ov{K}_{\ov{A}}E_{\un{C}}K_{\un{B}}F^{\un{C}} - \frac{1}{2}\ov{K}^{\ov{E}}K_{\un{C}}F_{\ov{EDA}}F_{\un{B}}{}^{\un{C}\ov{D}} + \frac{1}{2}K^{\un{D}}\ov{K}_{\ov{E}}F_{\un{DBC}}F^{\un{C}\ov{E}}{}_{\ov{A}}\, \nn \\
& & - \frac{1}{2}K^{\un{D}}\ov{K}_{\ov{A}}F_{\un{DBC}}F^{\un{C}} - \frac{1}{2}\ov{K}^{\ov{E}}K^{\un{D}}F_{\ov{E}\un{D}}{}^{\un{C}}F_{\ov{A}\un{BC}} - 2D_{\ov{A}}D_{\un{B}}f\, ,
\eea
\normalsize
and
\bea
{\cal R}_{(\kappa^{2})\ov{A}\un{B}} & = & \frac{1}{2} \ov{K}^{\ov{D}}K^{\un{C}}\left(E_{\ov{D}}\ov{K}_{\ov{A}}\right)\left(E_{\un{C}}K_{\un{B}}\right) + \frac{1}{4} \ov{K}^{\ov{D}}K^{\un{C}}\ov{K}_{\ov{A}}\left(D_{\ov{D}}E_{\un{C}}K_{\un{B}}\right)\, \nn \\
& & - \frac{1}{4}K^{\un{C}}\ov{K}_{\ov{A}}K_{\un{B}}\left(E_{\un{C}}D_{\ov{D}}\ov{K}^{\ov{D}}\right) - \frac{1}{4}K^{\un{C}}\ov{K}_{\ov{A}}K^{\un{E}}E_{\un{C}}\left(\ov{K}^{\ov{D}}F_{\ov{D}\un{BE}}\right)\, \nn \\
& & - K^{\un{C}}\ov{K}_{\ov{A}}\left(E_{\un{C}}D_{\un{B}}f\right) + \left(K_{\un{B}}D_{\un{C}}\ov{K}_{\ov{A}}\right)\left(D^{\un{C}}f\right) - \left(\ov{K}_{\ov{A}}E_{\un{B}}K_{\un{C}}\right)\left(D^{\un{C}}f\right)\, \nn \\
& & + \left(\ov{K}_{\ov{A}}E_{\un{C}}K_{\un{B}}\right)\left(D^{\un{C}}f\right) + \left(K^{\un{D}}\ov{K}_{\ov{A}}F_{\un{DBC}}\right)\left(D^{\un{C}}f\right)\, .
\eea

As can be appreciated, the EOM of the generalized metric contains quadratic terms even if $f=0$, and unlike general relativity there no exist $\alpha_{1}$ and $\alpha_{2}$ such that the quadratic terms can be written as 
\bea
{{\cal R}_{(\kappa^2)}}_{\overline{A}\underline{B}} = \alpha_{1} \kappa \bar{K}_{\overline A} \bar{K}{}^{\overline C} {{\cal R}_{(\kappa)}}_{\overline{C}\underline{B}} + \alpha_{2} \kappa {K}_{\underline B} {K}^{\underline C} {{\cal R}_{(\kappa)}}_{\overline{A}\underline{C}} \, .
\nn
\eea

The previous equation shows that the equation of motion of the generalized metric cannot be linearized when the GKSA is considered. Nevertheless, upon breaking the global $O(D,D)$ invariance and using the equation of motion of $g_{\mu \nu}$ and $b_{\mu \nu}$, it is straightforward to probe that the quadratic contributions vanish when $f=0$.

\subsubsection{Generalized KS at the Supergravity Level}
Now we are interested in imposing the supergravity version of the GKSA on the previous formulation. One way is to directly parametrize the results of the previous section. Another way is to just perturb the supergravity Lagrangian with the GKSA. For simplicity, we do not consider perturbations of the gauge field, \textit{i.e.}, $A_{\mu i}=A_{o\mu i}$. The inverse of the 10-dimensional background metric is perturbed as
\bea
g^{\mu \nu} & = & g_{o}^{\mu \nu} + \kappa \phi l^{(\mu} \bar{l}^{\nu)} \, ,
\label{metricp}
\eea
where $l_{\mu}$ and $\bar{l}_{\mu}$ are null vectors with respect to $g^{\mu \nu}$ and $g_{o}^{\mu \nu}$, \textit{i.e.},
\bea
l_{\mu} l_{\nu} g^{\mu \nu} = l_{\mu} g_{o}^{\mu \nu} l_{\nu}  = 0 \, , \\
\bar l_{\mu} \bar l_{\nu} g^{\mu \nu} = \bar l_{\mu}  g_{o}{}^{\mu \nu} \bar l_{\nu} = 0 \, .
\eea

The previous objects also satisfy the following relations
\bea
\bar{l}^{\nu} \nabla_{o\nu}l_\mu =  0 \, , \quad {l}^{\nu} \nabla_{o\nu}{\bar l}_\mu =  0 \, , 
\label{geode}
\eea
which reduce to the standard geodesic conditions when $l$ and $\bar{l}$ are identified. The perturbation of the Kalb-Ramond field is given by
\bea
 b_{\mu \nu} & = & b_{o \mu \nu} - \wt{\kappa} \phi l_{[\mu} \bar{l}_{\nu]} \, ,
\label{bp}
\eea
where $\wt{\kappa}=\frac{2\kappa}{2+\kappa(l\cdot\bar l)}$. One can easily that after perturbing the supergravity action using these extended ansatz, the equations of motion are linear, as was shown from the DFT approach. We left this as an exercise for the reader.

\subsubsection{DFT Behind the Double Copy Maps of HJP}
The story of the double copy of Yang-Mills following HJP with the identifications,
 \be\label{id1}
  A_{\mu}{}^{i}(k) \; \rightarrow \; e_{\mu\bar{\mu}}(k,\bar{k})\;. 
 \ee
 \be\label{id2}
  \kappa_{ab} \; \rightarrow \; \frac{1}{2}\, \bar{\Pi}^{\bar{\mu}\bar{\nu}}(\bar{k})\,, 
 \ee
\begin{equation}\label{eq:precripcubic}
f_{abc} \; \to \;  \tfrac{i}{4} \,
\ov\Pi^{\ov\mu\ov\nu\ov\rho}(\ov k_{1},\ov k_{2},\ov k_{3})\; , 
\end{equation}
produce more that quadratic and cubic Einstein-Hilbert gravity. As the reader might suspect, the double copy produce the full quadratic and cubic NSNS supergravity, as we are going to show now. Particularly, after solving the section constraint imposing $x=\tilde x$, one should interpret the fields as
\bea
e_{\mu \nu} = h_{\mu \nu} + b_{\mu \nu} \, .
\eea
Let's explore the quadratic b-field contributions in the HJP double copy.
We already know that the quadratic action is given by
\begin{align}
     S_{\rm DC}^{(2)} = \frac{1}{4}& \int d^Dx \Big( e^{\mu{\nu}}\square e_{\mu{\nu}}+\partial^{\mu}  e_{\mu{\nu}}\,\partial^{\rho}  e_{\rho}{}^{{\nu}} \nn\\ &
     + {\partial}^{{\nu}} e_{\mu{\nu}}\, {\partial}^{{\sigma}}e^{\mu}{}_{{\sigma}}
          - \phi\square \phi +2\phi \partial^{\mu}{\partial}^{{\nu}} e_{\mu{\nu}}
     \Big) .
  \end{align}
Now we will focus in the extra contributions given by $b_{\mu \nu}$. It is easy to see that the terms with just one b cancel out, and we just have,
\begin{align}
     S_{\rm DC}^{(2)}|_{b} = \frac{1}{4}& \int d^Dx \Big( b^{\mu{\nu}}\square b_{\mu{\nu}} + 2 \partial^{\mu}  b_{\mu{\nu}}\,\partial^{\rho}  b_{\rho}{}^{{\nu}}
        \Big) .
  \end{align}
  The last terms are exactly the contributions coming from the $H^2$ contributions at quadratic order. Finally, the dilaton $\varphi$ can be easily incorporated at quadratic order by considering the shift
\bea
\phi = h + 4 \varphi \, .
\eea
Therefore, the field $\phi$ is usually known as the generalized dilaton, and $\varphi$ is the ordinary dilaton. 

Now we will explore the dilatonic and b-field contributions arising from the LR quadratic double copy.

\subsubsection{$b$-field Quadratic Contributions in LR Double Copy}

It is easy to show that at quadratic order terms mixing $b_{\mu \nu}$ and $h_{\mu \nu}$ from (\ref{DFDF_DFT}) cancel, and the remnant contributions are given by
\bea
{\cal L}^{(2)}_{b} = - \frac{a_1}{2}(\Box b^{\mu \nu} \Box b_{\mu \nu} - 2 \Box b^{\nu \mu} \partial_{\mu} \partial^{\rho} b_{\nu \rho}) \,  .
\eea
Defining $\bar h_{\mu \nu \rho}= 3 \partial_{[\mu} b_{\nu \rho]}$, the previous Lagrangian can be written as
\bea
{\cal L}^{(2)}_b = \frac{a_1}{6} \Box \bar h^{\mu \nu \rho} \bar h_{\mu \nu \rho} \, . 
\label{bcontribution}
\eea

Comparing with the four-derivative contributions to bosonic string theory in the Metsaev–Tseytlin formalism 
we see that there are no four-derivative quadratic contributions in the $b$-field. Therefore, the contribution obtained in \eqref{bcontribution} must correspond to a field redefinition. Indeed, let us consider 
\bea
\label{b-redef}
b_{\mu \nu} = b^{(0)}_{\mu \nu} + c_1 \nabla_{\rho} \bar h^{\rho}{}_{\mu \nu} 
\eea
with $c_1$ an arbitrary coefficient. Let's apply this field redefinition on the leading order term $T_{H}=-\frac{1}{12} H^2$ to quadratic order in perturbations. The induced four-derivative contribution is given by
\bea
T_{\textrm{induced}} = - \frac{c_1}{2} \partial_{\mu}(\nabla_{\lambda} \bar h^{\lambda}{}_{\nu \rho}) \bar h^{\mu \nu \rho} \, .
\eea 
Since we are considering only quadratic contributions, the covariant derivative reduces to an ordinary derivative $\nabla \rightarrow \partial$, and therefore we can write the previous expression as
\bea
T_{\textrm{induced}} = -\frac{c_1}{6} \Box \bar h_{\mu \nu \rho} h^{\mu \nu \rho} \, .
\label{redefb}
\eea
By fixing $c_1= a_1$, we can interpret the DC theory as Weyl gravity plus an ambiguous $b$-field contribution, which can be removed by the field redefinition \eqref{b-redef}.

\subsubsection{Dilatonic Quadratic Contributions in LR Double Copy}

A similar analysis applies to the dilaton. The quadratic dilatonic terms in (\ref{DFDF_DFT}) are given by
\bea
{\cal L}^{(2)}_{\phi}= - 8 a_2 (\Box h - \partial_{\mu} \partial_{\nu} h^{\mu \nu} + \Box \phi) \Box \phi \, .
\eea
Since the dilaton does not contribute to the quadratic structure of the $\alpha'$-corrections, these terms must also be removable by field redefinitions. Consider
\bea
\phi = \phi^{(0)} + c_2 (\Box h - \partial_{\mu} \partial_{\nu} h^{\mu \nu} + \Box \phi) \, .
\eea
Inspecting the two-derivative term $4 \partial_{\mu} \phi \partial^{\mu} \phi$, one finds that the desired contribution is induced by setting $c_2=- a_2$. Therefore, both the $b$-field and dilaton contributions in (\ref{DFDF_DFT}) are ambiguous at quadratic order, leaving Weyl gravity as the only unambiguous sector.

While the previous results might be seen as boring, indeed is very promising since the Lagrangian 
\bea
\int_{x} \sqrt{-g} (R + C^2)
\eea
can be turned into 
\bea
\int_{x} \sqrt{-g} (R + Riem^2)
\eea
 by performing field redefinitions. And this is excellent since the first quadratic correction coming from the $\alpha'$-corrections is exactly given by $Riem^2$. Indeed, the full four-derivative Lagrangian is given by
\bea
{\cal L}^{(1)} & = & \frac{\alpha'}{4} \Big[  \textrm{Riem}^2 - \frac12 H^{\rho \mu}{}_{\nu} H_{\mu}{}^{\alpha \lambda} R^{\nu}{}_{\rho \alpha \lambda} + \frac{1}{24} H^4 - \frac18 H^2_{\mu \nu} H^{2 \mu \nu} \Big]  \, \, . 
\label{MT}
\eea
 Therefore, combining the double copies of HJP and LR one can obtain the full quadratic bosonic Lagrangian up to quadratic order. While at first glance one might be worry about the structure of the DFDF Lagrangian, which is proportional to the leading order equations of motion, the point of the off-shell double copy is to not impose those covariant field redefinitions, since they produce a non-trivial gravitational result (in this sense, it is safe to combine both results).
 
 Before finishing this section it is important to mention that this line of research can be developed from the homotopy Lie algebra perspective \cite{PartHomotopy-1}-\cite{PartHomotopy-15}, which allows to extend the results of the double copy and explore further setups (supersymmetric extensions, higher-order contributions, etc.) in a much more mathematical/rigorous way. 

\newpage

\subsection{Exercises}
\subsubsection{Part A}
\begin{enumerate}
    \item Show that the curvature for the $b$-field can be written as $H_{\mu \nu \rho}=\nabla_{[\mu} b_{\nu \rho]}$ in bosonic supergravity. Then show its Bianchi identity $\nabla_{[\mu} H_{\nu \rho \sigma]}=0$.

\item Show the following identities: 
\bea
(\nabla_{\mu} H_{\nu \rho \sigma}) H^{\mu \rho}{}_{\tau} H^{\nu \sigma \tau} = 0 \, , \quad (\nabla_{\mu} H^{\nu \rho \sigma}) R^{\mu}{}_{\nu \rho \sigma} = 0 \, .
\eea

\item Prove that the parametrization of the generalized metric respects the $O(D,D)$ constrain ${\cal H}^2=1$.

\item Compute the symmetry transformation rule for $b_{\mu \nu}$ coming from DFT.

\item Show that the DFT action reduces to the NSNS Lagrangian after parametrization (include the non-Abelian gauge field to recover the Yang-Mills plus Chern-Simons terms of the heterotic supergravity).

\item Consider a generic double vector $V^{M}$ for heterotic DFT and show that $\Big[\delta_{\hat \xi_1},\delta_{\hat \xi_2} \Big] V^{M} = \delta_{\hat \xi_{21}} V^{M}$ holds if we impose the strong constraint.

\item Decide if the following projections are determined or undetermined: $\Gamma_{\un M \un N \un P}$, $\Gamma_{\un M \un N \ov P}$, $\Gamma_{\un M \ov N \ov P}$, $\Gamma_{\ov M \ov N \ov P }$, and write the determined projections in terms of ${\cal H}_{MN}$.

\item Demanding $
\int d^{2D}X e^{-2d} V \nabla_{M} V^{M}  = - \int d^{2D}X e^{-2d} V^{M} \nabla_{M} V $ show that the trace of the connection is $\Gamma_{M N}{}^{M} = -2\partial_{N} d$. 

\item Find $\delta_{\hat \xi} \Gamma_{MNP}$. Then show that $\Gamma_{[MNP]}$
transforms as a generalized tensor. 

\end{enumerate}

\subsubsection{Part B}
\begin{enumerate}
    \item Consider DFT in its generalized metric formulation. Compute the determined projections of $\nabla^{(0)}_{M} K_{N}$ and $\nabla^{(0)}_{M} \bar K_{N}$.

    \item Consider $R_{DFT}$ in its generalized metric formulation and perturb it with the GKSA. Use the connections of the previous exercise to write it in its covariant form.

    \item Consider $R_{\ov M \un N}=0$ under the generalized Kerr-Schild ansatz and prove that the equation contains the same information that $R_{\un M \ov N}=0$ under the same ansatz.

    \item Use the GKSA at the supergravity level for the metric and the b-field (the so-called extended ansatz, without perturbing the dilaton) and prove that the equations of motion are linear in the perturbations.

    \item Propose a list of gauge terms which can be potential candidates to reproduce the $Riem^3$ interactions of bosonic string supergravity. In order to do so, first do some research to understand what kind of pure gravitational terms does this theory have and then propose your gauge Lagrangian.    
\end{enumerate}

\newpage

\section{Summary and Outlook}
Throughout these lectures, we have reviewed several modern aspects of the double copy program and its role in connecting gauge and gravitational theories. This framework has emerged as a particularly promising direction for future research, especially in the study of higher-derivative corrections and their implications for quantum gravity. One of the central motivations behind the double copy is the possibility of reformulating gravitational dynamics in terms of structures that are better understood on the gauge theory side, where quantization is under significantly greater control.

In the final lecture, we showed how bosonic supergravity can emerge at quadratic order within this framework. Moreover, there are strong indications—particularly from the amplitudes community—that this correspondence may extend to higher orders, potentially providing a systematic way to construct increasingly refined gravitational theories. If such extensions can be fully established, they would offer a powerful bridge between perturbative gauge theory computations and nontrivial gravitational dynamics.

In parallel, new models incorporating higher-derivative corrections can be developed independently of traditional string-theoretic constructions. This opens the door to a broader landscape of theories that, while inspired by string theory, are not strictly constrained by it. A recent example of this line of research is the implementation of the double copy map for non-commutative gauge theories via the Seiberg–Witten map \cite{Part3-1}. Although this topic goes beyond the scope of the present course, it serves as a compelling illustration of how novel toy models for quantum gravity can naturally arise within the double copy framework \cite{Part3-2}.

The overarching idea is to exploit gauge theories—whose quantization is well understood—to construct corresponding gravitational theories that may inherit some of these desirable quantum properties. This perspective suggests a new strategy for approaching long-standing problems in quantum gravity \cite{Part3-3}-\cite{Part3-4}, shifting part of the difficulty to a domain where more computational and conceptual tools are available.

As you can imagine, there are many directions in which these ideas can be further developed, ranging from formal theoretical questions to concrete phenomenological applications. I hope that these lectures have provided a useful starting point for engaging with the current literature and for exploring the deep and fascinating connections between gauge and gravitational theories. 

\section{Acknowledgments}

The author is very grateful to A. Rodriguez and G. Menezes for collaboration in some works covered by these lectures. Although not explicitly discussed in these notes, the author thanks L. Jonke for collaboration in the (non-commutative) off-shell double copy program. The author is also in debt with N. Miron-Granese for his help in administrative aspects which made the course at UBA possible, and also preparing the webpage (asignaturas.df.uba.ar/eetdgytdg-lescano/). Last but not least, I would like to thank all the students who were attending these lectures in Universidad de Buenos Aires. I really enjoyed the experience, and I hope they keep their enthusiasm and motivation about hep-th. 

\vspace{0.3cm}
E.L. is supported by the SONATA BIS grant 2021/42/E/ST2/00304 from the National Science Centre (NCN), Poland.

\end{document}